\documentclass[a4paper,usenatbib,useAMS]{mnras}

\usepackage[T1]{fontenc}
\usepackage{ae,aecompl}

\usepackage{amsmath}
\usepackage{amsfonts}
\usepackage{amssymb}
\usepackage{gensymb}
\usepackage{graphicx}
\usepackage{mathptmx}
\usepackage{multirow}
\usepackage{subfig}
\usepackage{tabularx}
\usepackage{threeparttable}

\def\reff@jnl#1{{\rm#1\/}}

\def\aj{\reff@jnl{AJ}}                  
\def\araa{\reff@jnl{ARA\&A}}            
\def\apj{\reff@jnl{ApJ}}                        
\def\apjl{\reff@jnl{ApJ}}               
\def\apjs{\reff@jnl{ApJS}}              
\def\ao{\reff@jnl{Appl.Optics}}         
\def\apss{\reff@jnl{Ap\&SS}}            
\def\aap{\reff@jnl{A\&A}}                       
\def\apjl{\reff@jnl{ApJ}}               
\def\aapr{\reff@jnl{A\&A~Rev.}}         
\def\aaps{\reff@jnl{A\&AS}}             
\def\azh{\reff@jnl{AZh}}                        
\def\baas{\reff@jnl{BAAS}}              
\def\jrasc{\reff@jnl{JRASC}}            
\def\memras{\reff@jnl{MmRAS}}           
\def\mnras{\reff@jnl{MNRAS}}            
\def\nar{\reff@jnl{New A. Rev.}}          
\def\pra{\reff@jnl{Phys. Rev. A}}         
\def\prb{\reff@jnl{Phys. Rev. B}}         
\def\prc{\reff@jnl{Phys. Rev. C}}         
\def\prd{\reff@jnl{Phys. Rev. D}}         
\def\prl{\reff@jnl{Phys. Rev. Lett}}      
\def\pasp{\reff@jnl{PASP}}              
\def\pasj{\reff@jnl{PASJ}}              
\def\qjras{\reff@jnl{QJRAS}}            
\def\skytel{\reff@jnl{S\&T}}            
\def\solphys{\reff@jnl{Solar~Phys.}}    
\def\sovast{\reff@jnl{Soviet~Ast.}}     
\def\ssr{\reff@jnl{Space~Sci.Rev.}}     
\def\zap{\reff@jnl{ZAp}}                        
\def\nat{\reff@jnl{Nature}}             

\def\p#1by#2{{\partial{#1} \over \partial{#2}}}
\def\pp#1by#2#3{{\partial^2{#1} \over \partial{#2}\partial{#3}}}
\def\d#1by#2{{{\rm d}{#1} \over {\rm d}{#2}}}
\def\dd#1by#2#3{{{\rm d}^2{#1} \over {\rm d}{#2}{\rm d}{#3}}}

\title[325 MHz GMRT Observations of the Super-CLASS Field]{Deep observations of the Super-CLASS super-cluster at 325 MHz with the GMRT: the low-frequency source catalogue}

\label{firstpage}

\author[C.~J.~Riseley et al.]{C.~J.~Riseley$^{1,2}$\thanks{Corresponding author email: \href{mailto:c.j.riseley@manchester.ac.uk}{c.j.riseley@manchester.ac.uk}}, 
A.~M.~M.~Scaife$^{1}$, C.~A.~Hales$^{3}$\thanks{Jansky Fellow of the National Radio Astronomy Observatory.}, I.~Harrison$^1$,   M.~Birkinshaw$^{4,5}$,
\newauthor
R.~A.~Battye$^1$, R.~J.~Beswick$^{1}$,  M.~L.~Brown$^{1}$, C.~M.~Casey$^{6}$, S.~C.~Chapman$^{7}$,
\newauthor
 C.~Demetroullas$^1$, C.-L. Hung$^{6}$, N.~J.~Jackson$^1$, T.~Muxlow$^1$, B.~Watson$^1$
 \vspace{0.03in}\\
$^1$ Jodrell Bank Centre for Astrophysics, Alan Turing Building, School of Physics and Astronomy, The University of Manchester, Oxford Road, Manchester, \\
M13 9PL, U.K.\\
$^2$ School of Physics \& Astronomy, University of Southampton, Highfield, Southampton, SO17 1BJ, U.K.\\
$^3$ National Radio Astronomy Observatory, P.O. Box 0, Socorro NM 87801, USA.\\
$^4$ H.~H.~Wills Physics Laboratory, University of Bristol, Tyndall Avenue, Bristol BS8 1TL, U.K. \\
$^5$ Harvard-Smithsonian Center for Astrophysics, 60 Garden St., Cambridge, MA 02138, USA. \\
$^6$ Department of Astronomy, The University of Texas at Austin, 2515 Speedway Blvd Stop C1400, Austin, TX 78712, USA. \\
$^{7}$ Department of Physics and Atmospheric Science, Dalhousie University, Coburg Road, Halifax B3H1A6, Canada. \\
}

\date{Accepted ---; received ---; in original form \today}

\pubyear{2016}

\begin{document}
\maketitle

\begin{abstract}
We present the results of 325 MHz GMRT observations of a super-cluster field, known to contain five Abell clusters at redshift $z \sim 0.2$. We achieve a nominal sensitivity of $34\,\umu$Jy beam$^{-1}$ toward the phase centre. We compile a catalogue of 3257 sources with flux densities in the range $183\,\umu\rm{Jy}\,-\,1.5\,\rm{Jy}$ within the entire $\sim 6.5$ square degree field of view. Subsequently, we use available survey data at other frequencies to derive the spectral index distribution for a sub-sample of these sources, recovering two distinct populations -- a dominant population which exhibit spectral index trends typical of steep-spectrum synchrotron emission, and a smaller population of sources with typically flat or rising spectra. We identify a number of sources with ultra-steep spectra or rising spectra for further analysis, finding two candidate high-redshift radio galaxies and three gigahertz-peaked-spectrum radio sources. Finally, we derive the Euclidean-normalised differential source counts using the catalogue compiled in this work, for sources with flux densities in excess of $223 \, \umu$Jy. Our differential source counts are consistent with both previous observations at this frequency and models of the low-frequency source population. These represent the deepest source counts yet derived at 325 MHz. Our source counts exhibit the well-known flattening at mJy flux densities, consistent with an emerging population of star-forming galaxies; we also find marginal evidence of a downturn at flux densities below $308 \, \umu$Jy, a feature so far only seen at 1.4 GHz.
\end{abstract}


\section{Introduction}
Deep radio surveys of the extragalactic source population are powerful tools with which to probe a wide range of source populations across a variety of environments and redshifts. In previous decades, optical surveys have been preferred for the study of the formation, interactions and evolution of galaxies. However, radio emission is important for galaxy population studies, as the synchrotron emission is a clear indicator of magnetic fields from star formation and active galactic nuclei (AGN). Additionally, the radio emission is essentially unaffected by dust obscuration and as such provides a powerful tracer of the evolution of star-forming galaxies and AGN with redshift.

All-sky surveys -- for example the Very Large Array (VLA) Faint Images of the Radio Sky at Twenty-Centimetres (FIRST; \citealt{1994ASPC...61..165B}) survey and the NRAO VLA Sky Survey (NVSS; \citealt{1998AJ....115.1693C}) at 1.4 GHz, and the 325 MHz Westerbork Northern Sky Survey (WENSS; \citealt{1997A&AS..124..259R}) -- have been effective in identifying large numbers of bright radio sources and have led to studies of the populations they represent. At higher frequencies ($\nu \gtrsim 1.4$ GHz) and/or higher flux densities ($S \gtrsim 1-10$ mJy at 1.4 GHz) the dominant population of sources are radio-loud AGN (for example \citealt{1984ApJ...287..461C}, \citealt{1994ASPC...61..165B}, \citealt{1998AJ....115.1693C}, \citealt{1999MNRAS.305..297G}, \citealt{2005ApJ...624..135A}, \citealt{2008ApJ...681.1129B}).

Moving to fainter flux densities, the contribution from star-forming galaxies (SFG) and radio-quiet (RQ) AGN become increasingly important, and these sources are believed to dominate at $S \lesssim 0.1$ mJy (for example \citealt{1999ApJ...526L..73R}, \citealt{2005MNRAS.358.1159M}, \citealt{2007ASPC..380..205P}, \citealt{2008ApJS..179...95M}, \citealt{2009MNRAS.397..281I}, \citealt{2009ApJ...694..235P}, \citealt{2013MNRAS.436.3759B}, \citealt{2015MNRAS.452.1263P}). The physics of low-luminosity SFG is still poorly understood; all-sky surveys are too shallow to recover sufficient numbers of these faint sources to infer much detail. Increasingly deep surveys such as the VLA-COSMOS survey (for example \citealt{2004AJ....128.1974S}) as well as smaller, deep fields (e.g. \citealt{2004A&A...424..371M}, \citealt{2013ApJS..205...13M}) have recovered sources down to $\sim \umu$Jy flux densities at 1.4 GHz. Radio emission from SFG is comprised of two components: synchrotron emission dominates at low frequencies, whereas thermal bremsstrahlung (free-free emission) from the ionized interstellar medium dominates at higher frequencies (for example \citealt{1992ARA&A..30..575C}, \citealt{2002A&A...392..377B}, \citealt{2008A&A...477...95C}). Whilst synchrotron emission presents itself with typical spectral index\footnote{Adopting the convention $S \propto \nu^{\alpha}$.} of $\alpha \simeq -0.8$, free-free emission has a flatter spectrum (typically $\alpha \simeq -0.1$). In addition to these features at high frequency, SFG spectra exhibit a number of other features below 1 GHz, with bends and inversions detected in some spectra \citep{2010MNRAS.405..887C}.

Surveys at low frequencies (for example \citealt{1991PhDT.......241W}, \citealt{2008MNRAS.383...75G}, \citealt{2008MNRAS.387.1037G}, \citealt{2009MNRAS.395..269S}, \citealt{2009AJ....137.4846O}, \citealt{2010iska.meetE..51S}, \citealt{2013MNRAS.435..650M}, \citealt{2014MNRAS.443.2590S}) open a new window for study, offering a number of advantages over their higher-frequency counterparts. Low-frequency observations are powerful at detecting (ultra-)steep spectrum sources, which are often galaxies at high redshift (\citealt{2008A&ARv..15...67M} and references therein). Low-frequency observations also enable detailed studies of the radio synchrotron spectral index, which allows for more precise characterisation of the source properties. 

In this work, we present the results of a deep 325 MHz continuum survey of a super-cluster field, performed using the Giant Metrewave Radio Telescope (GMRT) and carried out as part of the Super-Cluster Assisted Shear Survey (Super-CLASS). The remainder of this paper is divided as follows: we firstly introduce the Super-CLASS project in \S\ref{sec:proj}; subsequently we detail the observations and data reduction methodology in \S\ref{sec:obs}. We present our results in \S\ref{sec:RES}, including a sample from our source catalogue; we verify the catalogue and analyse the statistical properties in \S\ref{sec:AN}. In \S\ref{sec:DISC} we derive the spectral index distribution and identify sources with steep and/or inverted spectra for further study, which may constitute ultra-steep spectrum radio sources or gigahertz-peaked spectrum radio sources. We derive the source counts distribution from our catalogue, as well as evaluate the various bias and incompleteness corrections that must be applied, in \S\ref{sec:src}. Finally, we draw our conclusions in \S\ref{sec:CONC}. All errors are quoted to $1\sigma$. We adopt the spectral index convention that $S \propto \nu^{\alpha}$ where the radio spectral index $\alpha < 0$. We assume a concordance cosmology of H$_0 = 73 \rm{km} \rm{s}^{-1}$ Mpc$^{-1}$, $\Omega_{\rm{m}} = 0.27$, $\Omega_{\rm{\Lambda}} = 0.73$. At a redshift of $z = 0.2$, representative of the constituent clusters of the Super-CLASS super-cluster, an angular size of 1 arcsecond corresponds to a physical size of 3.2 kpc.

\subsection{The Super-Cluster Assisted Shear Survey}\label{sec:proj}
Weak lensing in the radio regime is emerging as promising cosmological probe, as many of the issues which strongly affect optical lensing studies (such as atmospheric effects and anisotropic PSFs) are negated by shifting to radio wavelengths. While measurements of cosmic shear on scales of $1-4$ degrees have been made at radio wavelengths (for example \citealt{2004ApJ...617..794C}, and see also \citealt{2010MNRAS.401.2572P}) previous shear surveys have been severely limited by resolution, field of view, and low source counts. A large, high-resolution catalogue is required to disentangle the shear signal (a factor $\sim0.01$) from intrinsic source ellipticity (typically $\sim0.3$). Recent work by \cite{2016MNRAS.456.3100D} has demonstrated that biases in shear measurements can be mitigated by cross-correlating both radio and optical survey data

Of the current generation of instruments, perhaps the best suited to studies of cosmic shear is the expanded Multi-Element Remote-Linked Interferometer Network (e-MERLIN). A number of legacy surveys are currently underway with e-MERLIN, including the Super-CLASS project\footnote{For more details, see \url{http://www.e-merlin.ac.uk/legacy/projects/superclass.html}}.

Super-CLASS is a wide-area, deep e-MERLIN legacy survey at L-band (reference frequency 1.4 GHz) targeting a region of sky known to contain five moderate-redshift $(z\sim0.2)$ Abell clusters -- A968, A981, A998, A1005 and A1006 \citep{1989ApJS...70....1A}. Some observational properties of these clusters are listed in Table \ref{tab:clusters}. Hereafter this region is referred to as the Super-CLASS field. The principal goal of the project is to detect the effects of cosmic shear in a super-cluster environment, where the increased level of structure should allow for a statistically significant detection of shear over a wide range of scales. However, a number of ancillary science goals exist, such as investigation of polarization properties of AGN and SFG, studies of cosmic magnetism in cluster- and super-cluster environments, classifying the galaxy population in the super-cluster, and detailed studies of the radio source population at $\umu$Jy flux densities. 

A wide range of other instruments are involved, including the Karl G. Jansky Very Large Array (JVLA) and LOw-Frequency Array (LOFAR; \citealt{2013A&A...556A...2V}) in the radio band, and a number of optical and mm-/sub-mm wavelength telescopes. The weak lensing component of the survey is similar in manner (although with lower source densities and on a smaller field) to those that may ultimately be conducted by the Square Kilometre Array (SKA; \url{http://skatelescope.org}). For more details on weak lensing with the SKA see for example \citet{2015aska.confE..23B}. See \citet{harrison2016arxiv} for cosmology forecasts from weak lensing with the SKA; for simulated catalogues see \citet{bonaldi2016arxiv}. 

\begin{table}
\begin{center}
\caption{Properties of galaxy clusters constituting the Super-CLASS super-cluster.}
\label{tab:clusters}
\begin{threeparttable}
\begin{tabular}{cccccc}
\hline
 & & & \\ 
Name & RA & Dec & $z$ & $L_x$ (0.1-2.4 keV) 	\\
 & (J2000) & (J2000) & & $[\times10^{44}$ erg s$^{-1}]$\\
\hline
Abell 968   & 10$^{\rm{h}}$21$^{\rm{m}}$09.5$^{\rm{s}}$ 	& +68$\degree$15$^{\prime}$53$^{\prime\prime}$ & 0.195 & 0.401 \\
Abell 981   & 10$^{\rm{h}}$24$^{\rm{m}}$24.8$^{\rm{s}}$ 	& +68$\degree$06$^{\prime}$47$^{\prime\prime}$ & 0.202 & 1.670 \\
Abell 998   & 10$^{\rm{h}}$26$^{\rm{m}}$17.0$^{\rm{s}}$ 	& +67$\degree$57$^{\prime}$44$^{\prime\prime}$ & 0.203 & 0.411 \\
Abell 1005 & 10$^{\rm{h}}$27$^{\rm{m}}$29.1$^{\rm{s}}$ 	& +68$\degree$13$^{\prime}$42$^{\prime\prime}$ & 0.200 & 0.268 \\
Abell 1006 & 10$^{\rm{h}}$27$^{\rm{m}}$37.2$^{\rm{s}}$ 	& +67$\degree$02$^{\prime}$41$^{\prime\prime}$ & 0.204 & 1.320 \\
\hline
\end{tabular}
References:
\begin{tablenotes}
	\item{Redshift, $z$: \citet{1990ApJ...365...66H}}
	\item{X-ray luminosity, $L_x$: BAX database; \citet{2004A&A...424.1097S}}
\end{tablenotes}
\end{threeparttable}

\end{center}
\end{table}

\begin{table}
\begin{center}
\caption{Summary of GMRT pointings on the Super-CLASS field. For each pointing, we list the integration time $(\tau_{\rm{int}})$ as well as the percentage of data that was used (i.e. unflagged) in the final image, as well as the predicted thermal noise.}
\label{tab:pointing_summary}
\scalebox{0.85}{
\begin{tabular}{ccccccc}
\hline
 & & & & \multicolumn{2}{c}{$\sigma_{\rm{rms}} [\umu$Jy beam$^{-1}]$ } \\
RA & Dec & $\tau_{\rm{int}}$ & Unflagged & Thermal & Measured \\
(J2000) & (J2000) & $[$s$]$ & $[$\%$]$ &  &  \\
\hline
10$^{\rm{h}}$21$^{\rm{m}}$56.13$^{\rm{s}}$ & 68$\degree$09$^{\prime}$44.8$^{\prime\prime}$ & $14.8\times10^3$ & 66.0 & 33.9 & 46.4 \\
10$^{\rm{h}}$24$^{\rm{m}}$17.12$^{\rm{s}}$ & 67$\degree$35$^{\prime}$26.7$^{\prime\prime}$ & $15.3\times10^3$ & 62.9 & 34.2 & 42.9 \\
10$^{\rm{h}}$26$^{\rm{m}}$31.43$^{\rm{s}}$ & 67$\degree$01$^{\prime}$01.4$^{\prime\prime}$ & $15.0\times10^3$ & 63.8 & 34.3 & 42.8 \\
10$^{\rm{h}}$28$^{\rm{m}}$50.28$^{\rm{s}}$ & 67$\degree$35$^{\prime}$25.0$^{\prime\prime}$ & $14.5\times10^3$ & 64.7 & 34.6 & 44.0 \\
10$^{\rm{h}}$26$^{\rm{m}}$32.15$^{\rm{s}}$ & 68$\degree$09$^{\prime}$57.5$^{\prime\prime}$ & $14.5\times10^3$ & 68.0 & 33.8 & 41.5 \\
10$^{\rm{h}}$24$^{\rm{m}}$15.01$^{\rm{s}}$ & 68$\degree$44$^{\prime}$23.0$^{\prime\prime}$ & $14.6\times10^3$ & 66.8 & 34.0 & 44.8 \\
\hline
\end{tabular}
}
\end{center}
\end{table}

\begin{figure}
	\centering
		\includegraphics[width=0.4\textwidth]{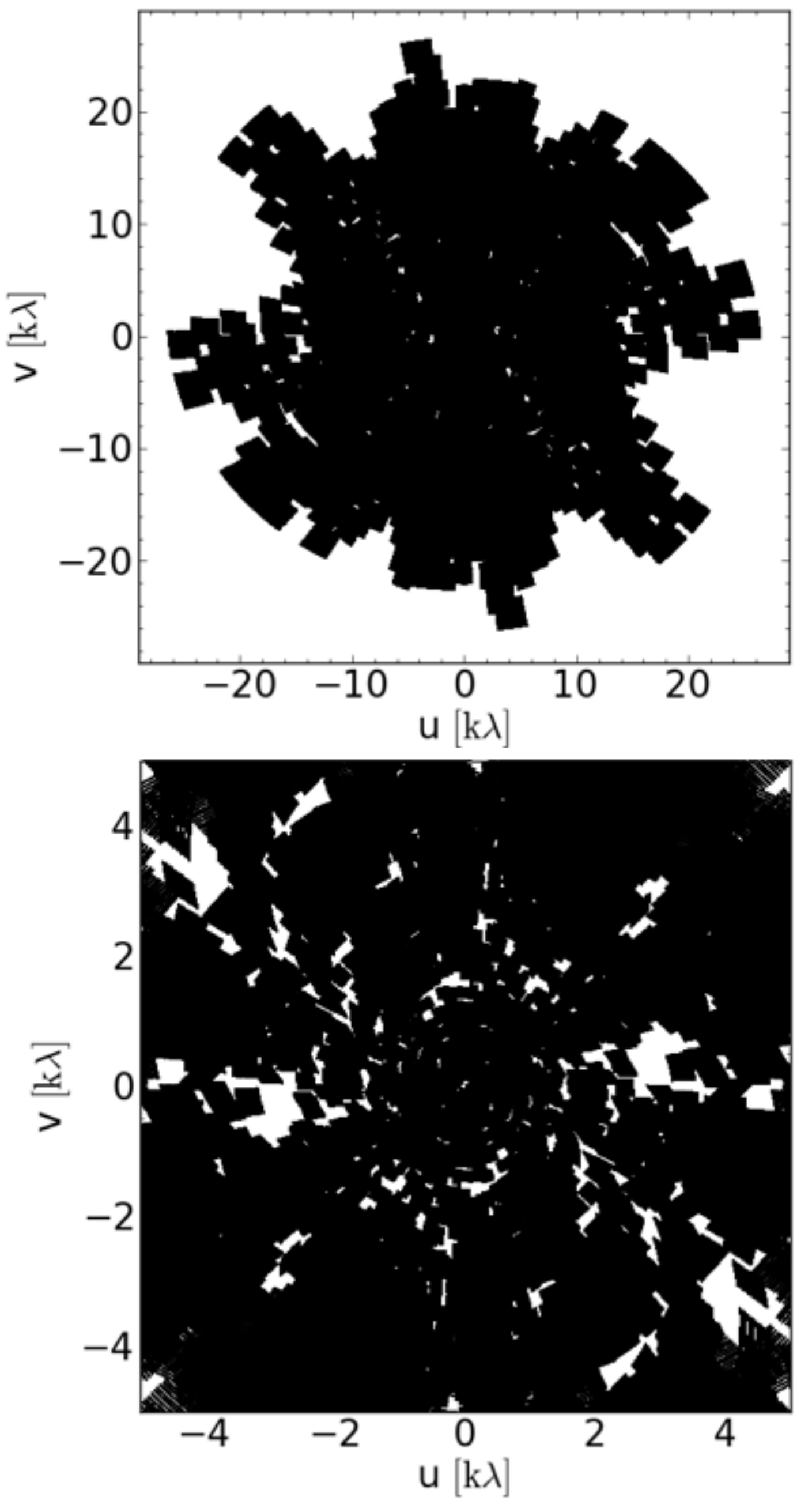}
		\label{fig:uvcovg_full}
\caption{Typical \emph{uv}-coverage of a single pointing on the Super-CLASS field. \emph{Top panel:} Plot showing coverage of the full \emph{uv-}plane. \emph{Bottom panel:} Zoom on the central portion of the \emph{uv-}plane, illustrating the coverage on short baselines.}
\label{fig:uvcovg}
\end{figure}

\section{Observations \& Data Reduction}\label{sec:obs}
\subsection{Observing Details}
The Super-CLASS field was observed in 2014 January using the GMRT over the course of five nights, under project code 25\_052 (P.I. Scaife, A.). At 325 MHz, the GMRT has a large primary beam full width at half-maximum (FWHM) of 84 arcminutes. To achieve as close to uniform coverage as possible across the e-MERLIN survey area while retaining efficiency, the Super-CLASS field was covered in six close-packed pointings, summarised in Table \ref{tab:pointing_summary}. 

\begin{table*}
\begin{center}
\caption{Summary of the GMRT observations of the Super-CLASS field.}
\label{tab:obs_summary}
\begin{tabular}{ccccp{2cm}p{6.5cm}}
\hline
 & & \\ 
Date 	& Start time & Hours observed & \# Antennas & Antennas missing throughout run & Comments     \\
 		& (IST) & & & & \\
\hline 
2014 Jan 09 & 23:00 & 10.3 & 29 & S02 & W02 stopped at 01:36. MESB power failures at 23:26 for 6 minutes and at 01:33 for 20 minutes. \\
2014 Jan 11 & 02:30 & 7.5 & 27 & W02, C04, S02 & C09 out 05:12--05:44. GSB problem at 07:44 forced reset. \\
2014 Jan 12 & 04:45 & 5.3 & 30 & -- & W02 (C10) stopped at 08:47 (09:25). \\
2014 Jan 13 & 05:00 & 4.3 & 30 & -- & -- \\
2014 Jan 14 & 21:00 & 12.5 & 30 & -- & W02 (C10) stopped at 08:47 (09:10). \\
\hline
\end{tabular}
\end{center}
\end{table*}

The GMRT was used with the GMRT Software Backend (GSB; \citealt{2010ExA....28...25R}) in default configuration in total intensity mode, with an acquisition bandwidth of 33 MHz, 256 channels, and an integration time of 8.05 seconds. Calibrator sources 3C\,286 and 3C\,48 were scanned for 10--15 minutes at the beginning and end of each observing run. 3C\,286 was typically available for both initial and final calibrator scans; 3C\,48 was only observed on 2 nights when 3C\,286 was unavailable for one or other of the calibrator scans.

Throughout the observing runs, a minimum of 26 of the 30 antennas were functional at all times; those missing may have been offline for a number of reasons including being taken for painting, problems with the hardware, or experiencing particularly strong radio frequency interference (RFI). On one night (Jan 9th) power failures forced a temporary halt in data collection, and on one night (Jan 11th) issues with the GSB forced a restart. Data taken during these periods were inspected for quality, and compromised data were discarded. In Table \ref{tab:obs_summary} we present a summary of the observations, including integration time, number of antennas, any missing antennas, and any general comments on each night's observing.

Scans were conducted in a 23 minute // 4 minute (target // calibrator) ratio; to maximise \emph{uv-}coverage for each pointing, scans on the same field were repeated only after a full cycle through all fields. The interleaved calibrator was observed after each pointing. The total integration time per pointing was of the order of $14.8\times10^3$ seconds, for a total integration time on target of 24.6h. The theoretical noise was derived using the following expression:

\begin{equation}
	\sigma_{\rm{th}} = \frac{ \sqrt{2} T_{\rm{sys}} }{ G \sqrt{n (n-1) f_{\rm{d}} \Delta\nu \tau_{\rm{int}} }}
\end{equation}
where the system temperature $T_{\rm{sys}} = 106$~K, the antenna gain $G = 0.32$~K Jy$^{-1}$ Antenna$^{-1}$, $n$ is the number of working antennas (taken as being 26, the minimum number functioning at any given time). $\Delta\nu$ is the bandwidth available; after removal of edge channels where the sensitivity drops off rapidly, $\Delta\nu$ reduces from $33$ to $30$ MHz. $\tau_{\rm{int}}$ is the integration time on target, and $f_d$ is the fraction of data used (i.e. the unflagged fraction). In Table \ref{tab:pointing_summary} we also list the integration time, the fraction of the data that was used and the expected noise for each pointing. The measured noise per pointing is also listed in Table \ref{tab:pointing_summary}, from which it is clear that we achieve a sensitivity of approximately $1.3\times\sigma_{\rm{th}}$. The final \emph{uv-}coverage of a single pointing is shown in Figure~\ref{fig:uvcovg}; it is clear that the \emph{uv-}plane is well-filled out to approximately $10-15 \,\rm{k}\lambda$.

\subsection{Data Reduction}
\subsubsection{Calibration}
The data were reduced using the Source Peeling and Atmospheric Modelling (\texttt{SPAM}) software \citep{2009A&A...501.1185I} which employs NRAO Astronomical Image Processing Software (\texttt{AIPS}) tasks through the \texttt{ParselTongue} interface \citep{2006ASPC..351..497K}. \texttt{SPAM} employs the \citet{2012MNRAS.423L..30S} flux density scale, which yields a flux density of 24.138 Jy and spectral index $\alpha=-0.197$ for 3C\,286 at 325 MHz; for 3C\,48 the flux density and spectral index are 43.742 Jy and $\alpha=-0.607$, respectively. \cite{2009A&A...501.1185I} describe data reduction with \texttt{SPAM} in detail; however here we will summarise the process.

Following removal of edge channels, the data were averaged by a factor 4 in frequency (yielding 64 channels of width 502.8 kHz) and 2 in time, as a compromise between improving data processing speed and mitigating bandwidth-/time-smearing effects. Prior to calibration, the data were visually inspected for strong RFI and bad antennas/baselines. Calibration solutions were derived for 3C\,286 and 3C\,48 using standard techniques in \texttt{SPAM} and applied to the target field. The interleaved calibrators are not used during the reduction process\footnote{The interleaved calibrators (0949+662 and 1101+724) were used to track atmospheric effects and data quality during the observing run itself.}. Instead, \texttt{SPAM} performs an initial phase calibration and astrometry correction using a sky model derived from the NVSS \citep{1998AJ....115.1693C} before proceeding with three rounds of direction-independent phase-only self-calibration.

Following the self-calibration, \texttt{SPAM} identifies strong sources within the primary beam FWHM that are suitable for direction-dependent calibration \& ionospheric correction. Only sources with well-defined astrometry are selected, yielding a catalogue of approximately 20 sources. Subsequently, \texttt{SPAM} performs direction-dependent calibration on a per-facet basis, using the solutions to fit a global ionospheric model, as described by \cite{2009A&A...501.1185I}. Throughout the reduction process, multiple automated flagging routines are used between cycles of imaging and self-calibration in order to reduce residual RFI and clip outliers. Across the six fields, around 32--37 per cent of the data were flagged out; this is not uncommon for GMRT data at this frequency -- for example, approximately 60 per cent of the data were flagged in the work of \citet{2009MNRAS.395..269S}. Additionally, the average flagged fraction was 40 per cent in \citet{2013MNRAS.435..650M}. Flagging statistics for each pointing are also listed in Table \ref{tab:pointing_summary}.

\subsubsection{Imaging}
During the self-calibration and imaging cycles, images were made using an \texttt{AIPS} \texttt{ROBUST} of $-1.0$ in order to achieve a compromise between sensitivity and resolution\footnote{An \texttt{AIPS} \texttt{ROBUST} = $-5.0$ indicates pure uniform weighting, for maximum resolution; \texttt{ROBUST} = $+5.0$ indicates pure natural weighting, for maximum sensitivity.}. All imaging was performed with facet-based wide-field imaging as implemented in \texttt{SPAM}. 


The direction-dependent and ionospheric calibration routine described above was repeated on each field separately. Following this calibration routine, the final images of all six fields were convolved to a common circular synthesised beam of FWHM 13 arcsec. These final images were then corrected for primary beam attenuation (with a cutoff of 30 per cent) and mosaicked in the image plane, weighted by the inverse of the local rms, using \texttt{AIPS} tasks as employed by \texttt{SPAM}.

\section{Results}\label{sec:RES}
\begin{figure}
	\centering
	\includegraphics[width=0.45\textwidth]{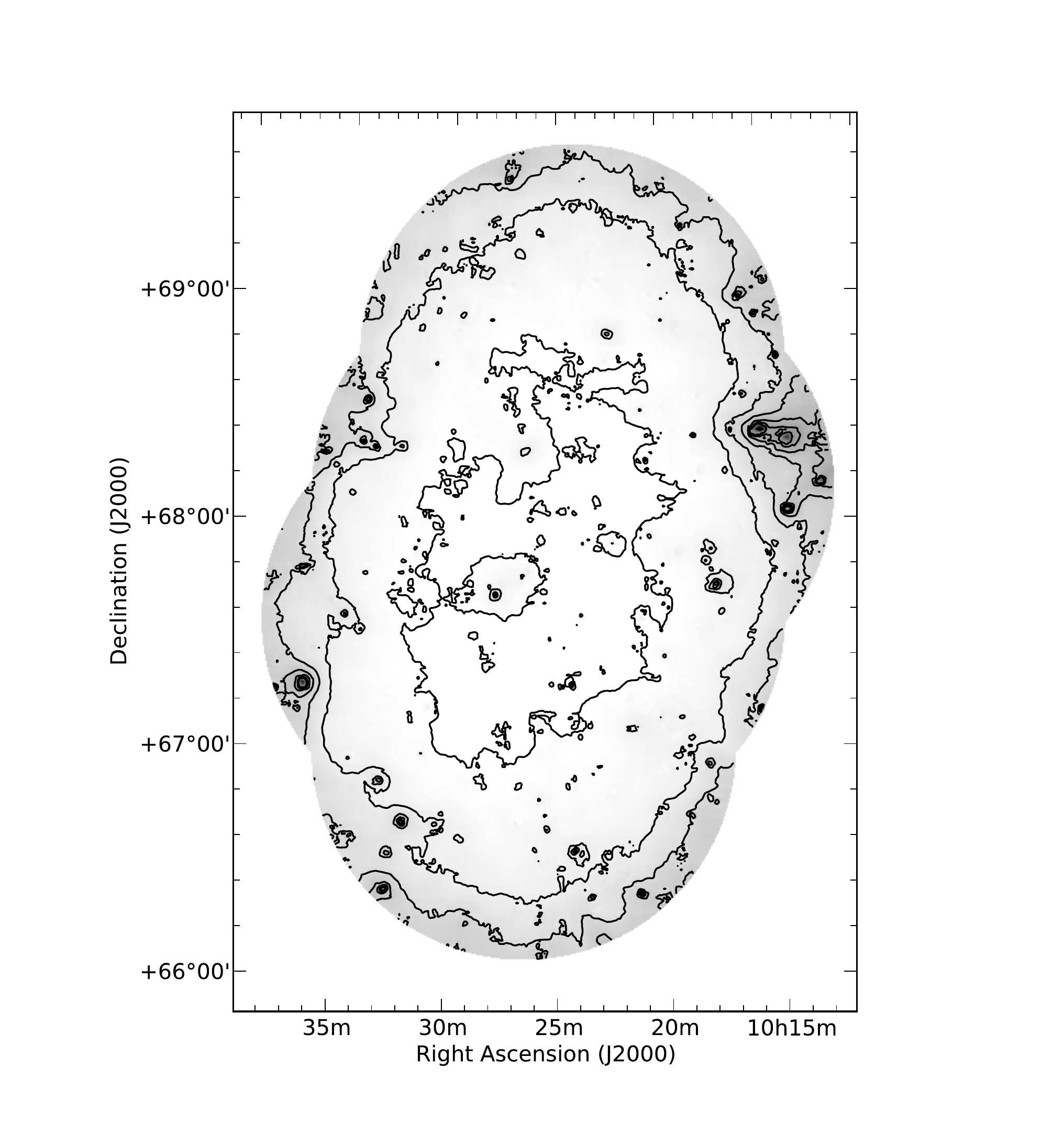}
\caption{Grayscale image showing the local rms noise of the final mosaicked GMRT image, derived using \texttt{PyBDSM} (see \S\ref{sec:cat}). The colour scale ranges from 50--750 $\umu$Jy beam$^{-1}$; contours are $[1,2,3,4,5,6,7] \,\times \, 50\,\umu$Jy beam$^{-1}$.}
\label{fig:mos_sens}
\end{figure}

\begin{figure*}
	\centering
	\includegraphics[width=1.\textwidth]{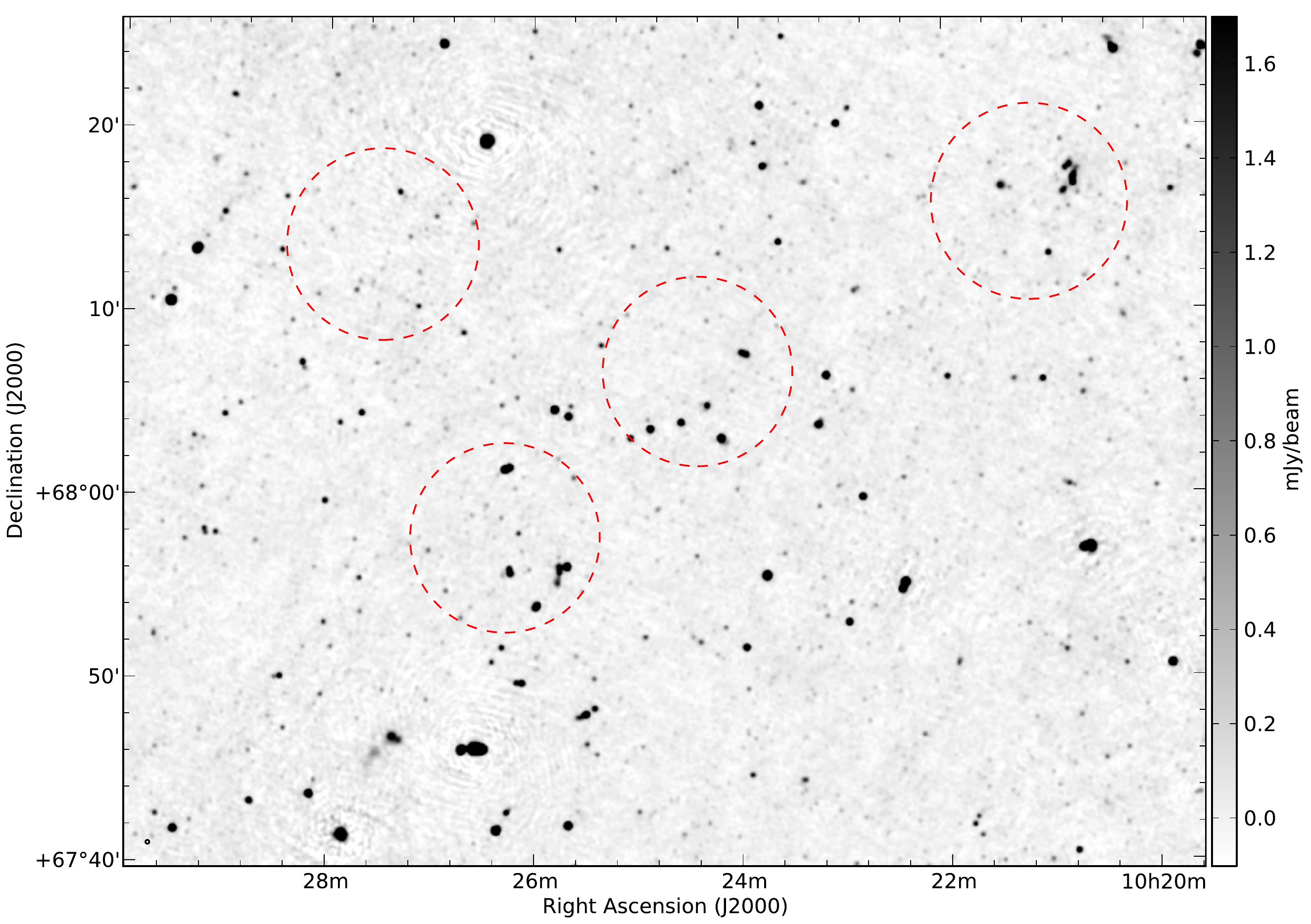}
\caption{Example region from the Super-CLASS field as observed with the GMRT. The grayscale ranges from $-3\sigma_{\rm{nom}}$ to $50\sigma_{\rm{nom}}$, where the nominal noise level is $\sigma_{\rm{nom}} = 34 \,\, \umu$Jy beam$^{-1}$. The restoring beam FWHM is 13 arcsec. The mosaicked image is corrected for the primary beam response of the GMRT. We note that while the noise in this region is generally low, and varies smoothly, some residual errors remain around some bright and/or complex sources. Dashed red circles trace a radius of 1 Mpc, centred on the reference coordinates of the clusters A968, A981, A998 and A1005 (see Table \ref{tab:clusters}).}
\label{fig:example_gmrt}
\end{figure*}

Figure~\ref{fig:mos_sens} shows the sensitivity of the final mosaicked image, derived using the Python Blob Detection and Source Measurement (\texttt{PyBDSM}\footnote{\texttt{PyBDSM:} \url{http://www.astron.nl/citt/pybdsm/}}; \citealt{2015ascl.soft02007M}) software. From Figure~\ref{fig:mos_sens}, the typical noise in the image is low, apart from a number of regions near bright and/or complex sources: the image rms is below $50\,\umu$Jy beam$^{-1}$ for the inner portion of the mosaic, and below $100\,\umu$Jy beam$^{-1}$ for the majority of the mosaic. We present a cutout region of the GMRT survey area in Figure~\ref{fig:example_gmrt} as an example of the image quality recovered from this work. The GMRT image has a nominal off-source noise of $\sigma_{\rm{nom}} = 34 \umu$Jy beam$^{-1}$ in this region; from Figure~\ref{fig:example_gmrt}, it appears that the quality is generally very high, although artefacts caused by residual phase and/or amplitude errors remain around some of the most complex sources, and there are some dynamic range issues associated with some of the brighter sources in the field.

\begin{table*}
\begin{center}
\caption{Excerpt from the 325 MHz SCG catalogue. The full catalogue is available through CDS (\url{http://cds.u-strasbg.fr}). Column (0): Source name, following the nomenclature SCG\_J\emph{hhmmss+ddmmss}. Columns (1) and (2): Right ascension and declination in sexagesimal format, J2000 reference. Column (3): Peak flux density at 325 MHz  and associated error. Column (4): Integrated flux density at 325 MHz and associated error. For unresolved sources, this is taken as being equal to the peak flux density, see \S\ref{sec:size}. Columns (5), (6) and (7): Deconvolved source major axis, minor axis, and position angle. Columns (8), (9) and (10): Uncertainty in the deconvolved major axis, minor axis and position angle. Columns (5) to (10) list a dash (-) if the source is unresolved.}
\label{tab:src_cat}
\begin{tabular}{cccrrrrrrrrrrr}
\hline
(0) & (1) & (2) & (3)\hspace{0.5cm} & (4)\hspace{0.5cm} & (5) & (6) & (7) & (8) & (9) & (10) \\
Source name & RA (J2000) & DEC (J2000) & $S_{\rm{peak}}$\hspace{0.4cm} & $S_{\rm{int}}$\hspace{0.46cm} & $\theta_{\rm{maj.}}$ & $\theta_{\rm{min.}}$ & PA & $\Delta(\theta_{\rm{maj.}})$ & $\Delta(\theta_{\rm{min.}})$ & $\Delta(\rm{PA})$ \\
& \emph{hh mm ss.ss} & \emph{dd mm ss.s} & $[$mJy beam$^{-1}]$ & $[$mJy$]$\hspace{0.35cm} & $[$arcsec$]$ & $[$arcsec$]$ & $[$deg$]$ & $[$arcsec$]$ & $[$arcsec$]$ & $[$deg$]$  \\
\hline
SCG\_J103820+672701 & 10 38 20.09 & 67 27 01.2 & $2.52\pm0.22$ & $3.43\pm0.32$ & 9.13 & 6.42 & 58.45 & 1.15 & 0.97 & 35.37 \\
SCG\_J103820+673505 & 10 38 20.80 & 67 35 05.8 & $0.89\pm0.18$ & $1.08\pm0.29$ & - & - & - & - & - & - \\
SCG\_J103809+672129 & 10 38 09.58 & 67 21 29.8 & $1.42\pm0.22$ & $1.91\pm0.34$ & 11.59 & 1.31 & 73.44 & 2.77 & 1.63 & 25.97 \\
SCG\_J103809+672449 & 10 38 09.32 & 67 24 49.4 & $1.44\pm0.20$ & $1.35\pm0.32$ & - & - & - & - & - & - \\
SCG\_J103805+672256 & 10 38 05.68 & 67 22 56.6 & $4.68\pm0.38$ & $6.45\pm0.55$ & 10.84 & 4.46 & 70.49 & 0.77 & 0.52 & 11.91 \\
SCG\_J103807+673552 & 10 38 07.42 & 67 35 52.3 & $0.91\pm0.18$ & $1.07\pm0.29$ & - & - & - & - & - & - \\
SCG\_J103804+674543 & 10 38 04.88 & 67 45 43.7 & $11.38\pm0.83$ & $20.77\pm1.55$ & - & - & - & - & - & - \\
SCG\_J103747+671514 & 10 37 47.72 & 67 15 14.1 & $5.05\pm0.46$ & $7.88\pm0.71$ & 13.34 & 5.59 & 92.72 & 1.20 & 0.73 & 12.77 \\
SCG\_J103746+671551 & 10 37 46.58 & 67 15 51.5 & $1.79\pm0.26$ & $2.25\pm0.42$ & - & - & - & - & - & - \\
SCG\_J103749+672539 & 10 37 49.18 & 67 25 39.5 & $1.68\pm0.19$ & $1.93\pm0.30$ & - & - & - & - & - & - \\
\hline
\end{tabular}

\end{center}
\end{table*}

\subsection{Source Catalogue}\label{sec:cat}
A catalogue of sources was compiled with \texttt{PyBDSM}. We compiled a catalogue of sources above 5$\sigma_{\rm{local}}$ in significance, where $\sigma_{\rm{local}}$ is the local noise level. \texttt{PyBDSM} derives a sensitivity map from the data and determines the local noise iteratively in a moving box. Subsequent `islands' are isolated above a user-defined threshold (set to $4\sigma_{\rm{local}}$) and Gaussians fitted to regions above a user-defined peak (set to $5\sigma_{\rm{local}}$). In order to reduce the number of spurious sources fitted to deconvolution artefacts, we defined two moving boxes: a `large' box of 200 pixels was used across the entire image, and a `small' box of 50 pixels was used in regions close to sources of high signal-to-noise. To better model the extended emission in the field, we also used the wavelet functionality of \texttt{PyBDSM} to decompose the residual image into wavelet images on a small number of scales, fitting to islands initially identified by \texttt{PyBDSM} during the Gaussian fitting routine.

The catalogue produced by \texttt{PyBDSM} was visually inspected; we removed any false detections arising from artefacts near complex/bright sources, and any sources that were cut by the primary beam cutoff (a total of ten sources were removed). The final number of sources in our catalogue is 3257, and we present a sample from the full catalogue in Table \ref{tab:src_cat}. The full catalogue is available online through CDS (\url{http://cds.u-strasbg.fr}). Hereafter we refer to our catalogue as the Super-CLASS-GMRT (SCG) catalogue. A short description of the catalogue is as follows:

Column (0): Source name, following the nomenclature SCG\_J\emph{hhmmss+ddmmss}.

Columns (1) and (2): Source right ascension (RA) and declination (DEC) in J2000 coordinates in sexagesimal format.

Column (3): Fitted peak flux density at 325 MHz in mJy beam$^{-1}$, with its associated uncertainty. The measurement uncertainty is taken as the fitted error plus five per cent of the peak flux density, added in quadrature.

Column (4): Fitted integrated flux density at 325 MHz in mJy, with its associated uncertainty. The measurement uncertainty is taken as sum of the fitted error plus five per cent of the integrated flux density, again added in quadrature. 

Columns (5), (6) and (7): Deconvolved major- and minor-axis FWHM (in arcsec) and position angle (PA) of the elliptical Gaussian (in degrees East of North). Sizes are only quoted for resolved sources; if unresolved, these columns are marked with a dash (-). See \S\ref{sec:size} for our definition of resolved and unresolved sources.

Columns (8), (9) and (10): Error in the deconvolved major- and minor-axis FWHM (in arcsec) and position angle (PA) of the elliptical Gaussian (in degrees East of North). These are quoted only for sources that are resolved; if unresolved these columns are marked with a dash (-).

\subsection{Completeness}\label{sec:completeness}
To quantify the efficiency with which \texttt{PyBDSM} detects sources in our field, we established a number of Monte-Carlo simulations for sources with flux densities at log-spaced intervals between $160\,\umu$Jy and 1.5 Jy. For each flux density, 10 catalogues of 160 point sources were generated, with positions randomised for each catalogue. These were then inserted into the residual map generated by \texttt{PyBDSM}, and catalogued in the same manner as the real data. Subsequently, the recovered sources were cross-referenced with the known simulated population to establish the fraction of sources missed as a function of flux density.

\begin{figure}
	\centering
	\includegraphics[width=0.45\textwidth]{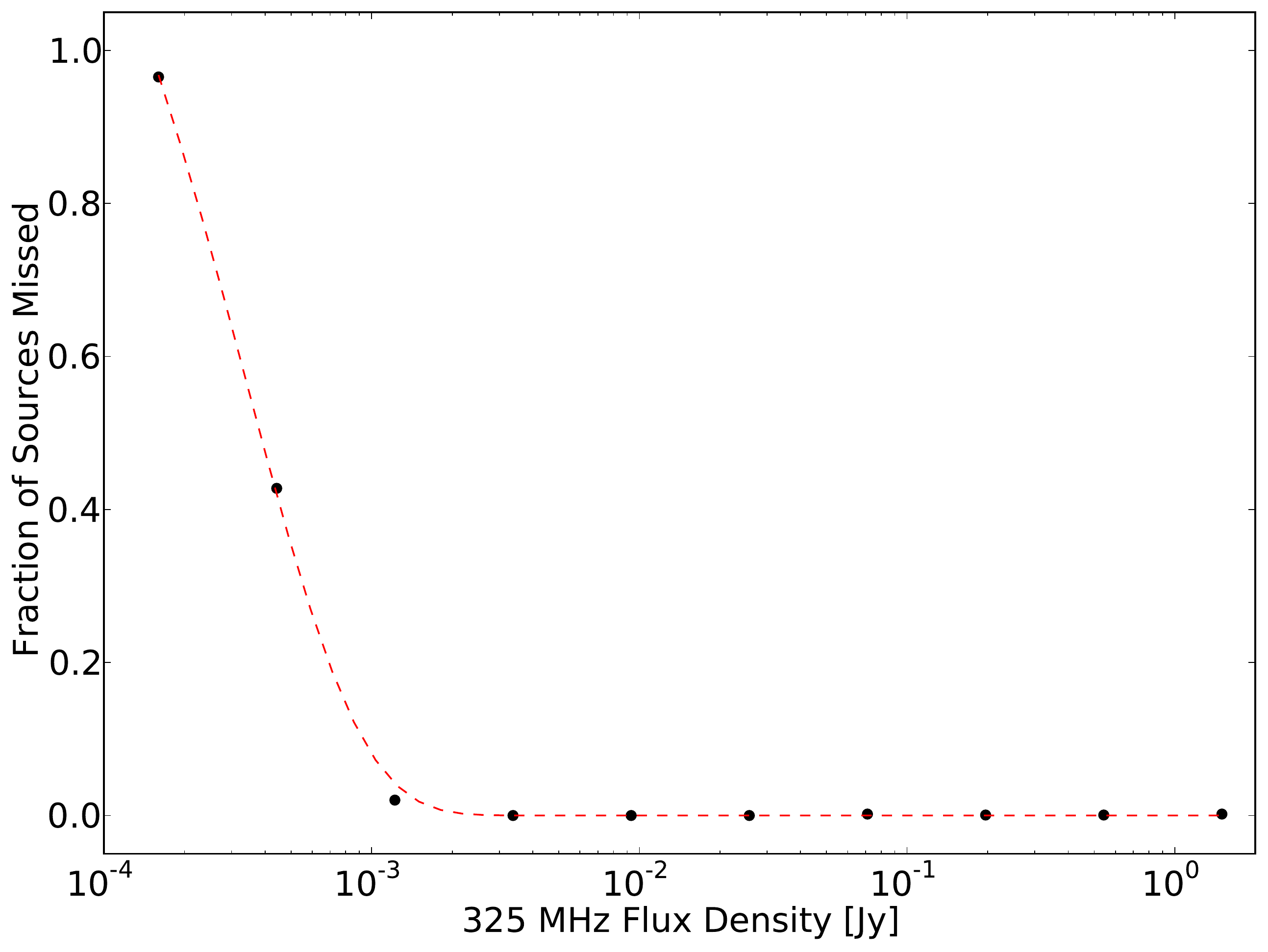}
	\caption{Fraction of sources missed as a function of 325 MHz flux density. Black points mark flux densities where simulated catalogues were analysed, dashed line marks the exponential fit.}
	\label{fig:completeness}
\end{figure}

This missed fraction as a function of flux density is presented in Figure~\ref{fig:completeness}, with the fitted exponential function shown as a dashed curve. The fit suggests our catalogue is 95 per cent complete at a flux density $S = 1.25$ mJy, and the completeness drops rapidly below 1 mJy.

\section{Analysis}\label{sec:AN}
\subsection{Verification}
The Super-CLASS region studied in this work has been covered by a number of surveys in the radio band. Catalogues exist at 74 MHz from the VLA Low-frequency Sky Survey (VLSS; \citealt{2007AJ....134.1245C}) and VLSS-Redux (VLSSR; \citealt{2014MNRAS.440..327L}), at 325 MHz from the WENSS \citep{1997A&AS..124..259R} and at 1.4 GHz from the NVSS \citep{1998AJ....115.1693C}. The high declination of the field puts this region outside the coverage of the VLA-FIRST survey \citep{1994ASPC...61..165B}. In this section we use the available survey data to validate our GMRT source catalogue.

\defcitealias{2012MNRAS.423L..30S}{SH12} 

It is important to note that in all our comparisons, the flux density measurements from the literature have been adjusted to bring them in line with the \citet{2012MNRAS.423L..30S} flux scale adopted in this work (hereafter \citetalias{2012MNRAS.423L..30S}). The flux density scale of WENSS is complex, having been set using observations of the sources 3C\,48, 3C\,147, 3C\,286 and 3C\,295. The correction factor to scale the WENSS survey to the \citetalias{2012MNRAS.423L..30S} flux scale is an average correction across the discrete set of WENSS calibrators. It would be incorrect to apply this factor to any localised area of the sky as the difference between the flux densities of the WENSS calibrators and the \citetalias{2012MNRAS.423L..30S} source models range from $\sim1$ per cent to $\sim18$ per cent (see \citealt{1997A&AS..124..259R} and \citetalias{2012MNRAS.423L..30S}). We note that, given the small difference between the native WENSS flux densities in this region and the GMRT values measured here (see \S\ref{sec:flux}) as well as the RA range, this field was likely calibrated using either 3C\,147 or 3C\,48. These calibrators both have ratios to the \citetalias{2012MNRAS.423L..30S} scale within the measurement uncertainty. Hence, we do not apply a correction factor to the flux densities from WENSS during catalogue verification.

The \citetalias{2012MNRAS.423L..30S} flux scale is calibrated on the RCB flux density scale \citep{1973AJ.....78.1030R}; at frequencies greater than 300 MHz, the RCB flux density scale is consistent with the KPW flux density scale \citep{1969ApJ...157....1K}. Therefore we can use Table 7 from \citet{1977A&A....61...99B} to adjust the flux density of sources on the Baars scale before comparing them with these GMRT data; hence the NVSS flux densities have been adjusted by a factor 0.972. The VLSS and VLSSR were also calibrated on the Baars flux density scale; measurements from the VLSS and VLSSR catalogues have been adjusted by a mean factor 1.10 to align with the \citetalias{2012MNRAS.423L..30S} scale, see \citet{2014MNRAS.440..327L}. 

\subsubsection{Match Criteria}
We attempted to match all sources in our catalogue to their equivalent in other catalogues. We matched sources from the SCG catalogue to reference catalogues (NVSS/WENSS) using a maximum offset of twice the synthesised beam FWHM of the GMRT image (i.e. 26 arcsec). Given the difference in resolution between the SCG catalogue and the references -- 13 arcsec for the SCG catalogue compared to 45 arcsec ($54 \, {\rm{arcsec}} \,\, \rm{cosec}\delta$) arcsec for NVSS (WENSS) -- we would expect a number of previously-unresolved sources to become resolved in the GMRT data. Where this was the case, flux densities of the SCG sources were summed for the purposes of flux density verification and spectral index derivation.

The SCG catalogue consists of a total of 3257 sources. We find 335 sources in the NVSS catalogue \citep{2002yCat.8065....0C} with matches in our catalogue, and 137 sources in the WENSS catalogue \citep{2000yCat.8062....0D} with SCG counterparts. From visual inspection, we note a total of 13 sources present in the NVSS catalogue that do not have counterparts in the SCG catalogue; all have integrated flux densities in the range 2.2--3.2 mJy at 1.4 GHz, and peak flux densities below 2.5 mJy beam$^{-1}$ (based on the NVSS mosaic images). In the GMRT image, the peak flux density at these positions ranged from $-80 \, \umu$Jy beam$^{-1}$ to $240 \, \umu$Jy beam$^{-1}$. In two cases, potential matches were nearby but separated by more than three times the GMRT restoring beam FWHM. For the remaining 11, no nearby SCG catalogue sources exist, and manual fits to these positions recovered no significant flux density. The positions of these sources and their integrated flux densities from the NVSS are presented in Table \ref{tab:unmatched}.

\begin{table}
\begin{center}
\caption{Sources in the NVSS catalogue that do not have SCG counterparts.}
\label{tab:unmatched}

\begin{threeparttable}
\begin{tabular}{ccccccc}
\hline
RA & Dec & $S_{\rm{int}}$ (1.4 GHz) & Comments \\
(J2000) & (J2000) & $[$mJy$]$ &  \\
\emph{hh mm ss.ss} & \emph{dd mm ss.s} & & \\
\hline
10 16 37.25 & 67 26 34.7 & $2.3\pm0.4$ & - \\
10 20 41.93 & 68 14 40.5 & $2.6\pm0.5$ & - \\
10 23 54.07 & 66 36 29.5 & $2.6\pm0.5$ & - \\
10 24 39.65 & 68 05 54.2 & $2.3\pm0.5$ & - \\
10 26 23.13 & 67 51 21.5 & $2.8\pm0.5$ & \tnote{a}\\
10 27 32.25 & 67 45 16.8 & $2.9\pm0.5$ & \tnote{b} \\
10 29 23.90 & 68 06 11.3 & $2.4\pm0.4$ & - \\
10 29 35.18 & 69 10 04.3 & $2.3\pm0.4$ & - \\
10 29 51.30 & 67 38 50.5 & $2.2\pm0.4$ & - \\
10 29 55.05 & 68 24 28.0 & $2.7\pm0.4$ & - \\
10 33 12.42 & 68 19 24.6 & $2.2\pm0.4$ & - \\
10 35 12.61 & 67 35 08.7 & $3.2\pm0.5$ & - \\
10 37 43.52 & 67 38 43.6 & $2.8\pm0.5$ & - \\
\hline
\end{tabular}
\begin{tablenotes}
	\item[a] {Two SCG sources are nearby, but separated by more than three times the GMRT synthesised beam FWHM. }
	\item[b] {An extended SCG source is nearby, but separated by more than three times the GMRT synthesised beam FWHM. }
\end{tablenotes}
\end{threeparttable}

\end{center}
\end{table}

\subsubsection{Astrometry}
\begin{figure}
	\centering
	\includegraphics[width=0.45\textwidth]{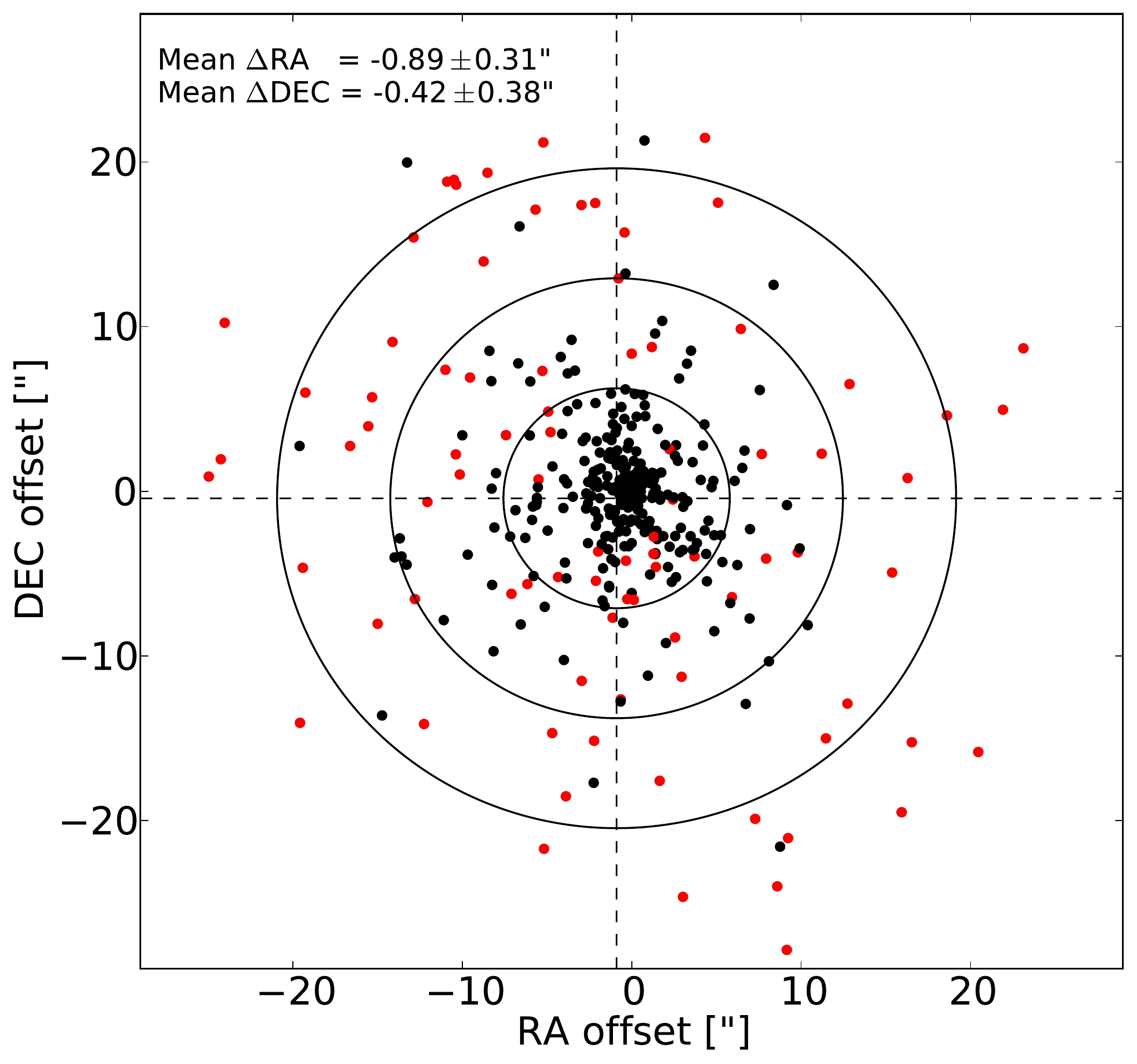}
\caption{Offset in RA and DEC of the sources in the GMRT field of view that have matches in the NVSS catalogue. The mean $\Delta$RA ($\Delta$DEC) is $-0.89\pm0.31$ ($-0.42\pm0.38$) arcsec, and is indicated by the dashed lines. Circles correspond to $[1,2,3]\sigma_{\rm{pos}}$ where the standard deviation in position offset, $\sigma_{\rm{pos}} = 6.68$ arcsec. Red points mark multiple SCG sources which have a single association in the NVSS catalogue.}
\label{fig:astrometry}
\end{figure}

Sources in the NVSS catalogue are known to have positional accuracies better than around an arcsecond for sources with integrated flux densities greater than 15 mJy, and accuracies better than 7 arcsec for fainter sources \citep{1998AJ....115.1693C}. Using the NVSS catalogue RA and DEC for reference, we determined the position offsets for the 335 sources present in the GMRT data which have matches in the NVSS catalogue. The offsets in RA and DEC are presented in Figure~\ref{fig:astrometry}. 

From Figure~\ref{fig:astrometry}, it is clear that the vast majority of sources in the GMRT catalogue have very little offset with respect to the NVSS catalogue. However, a number of sources lie more than $3\sigma$ away from the NVSS reference position. From inspection, these high offsets are caused by one of a number of factors. Foremost, the high offset may be due to single NVSS sources becoming resolved into multiple, separate sources as a result of the superior resolution. This is the case for the vast majority of outliers; we have indicated these cases in Figure~\ref{fig:astrometry}. For the remaining sources of high offset, it is typically the case that single NVSS sources become complex or extended at 325 MHz; in these situations the position offset is due to differences in the location of the emission peak.

The mean offsets in right ascension and declination are $\Delta\rm{RA} \, (\Delta\rm{DEC}) = -0.89\pm0.31$ ($-0.42\pm0.38$) arcsec. The synthesised beam FWHM of these GMRT observations is 13 arcsec. Hence, within the uncertainties, the positions of the SCG sources are consistent with the NVSS reference positions.

\subsubsection{Flux Density}\label{sec:flux}
\begin{figure}
	\centering
	\includegraphics[width=0.48\textwidth]{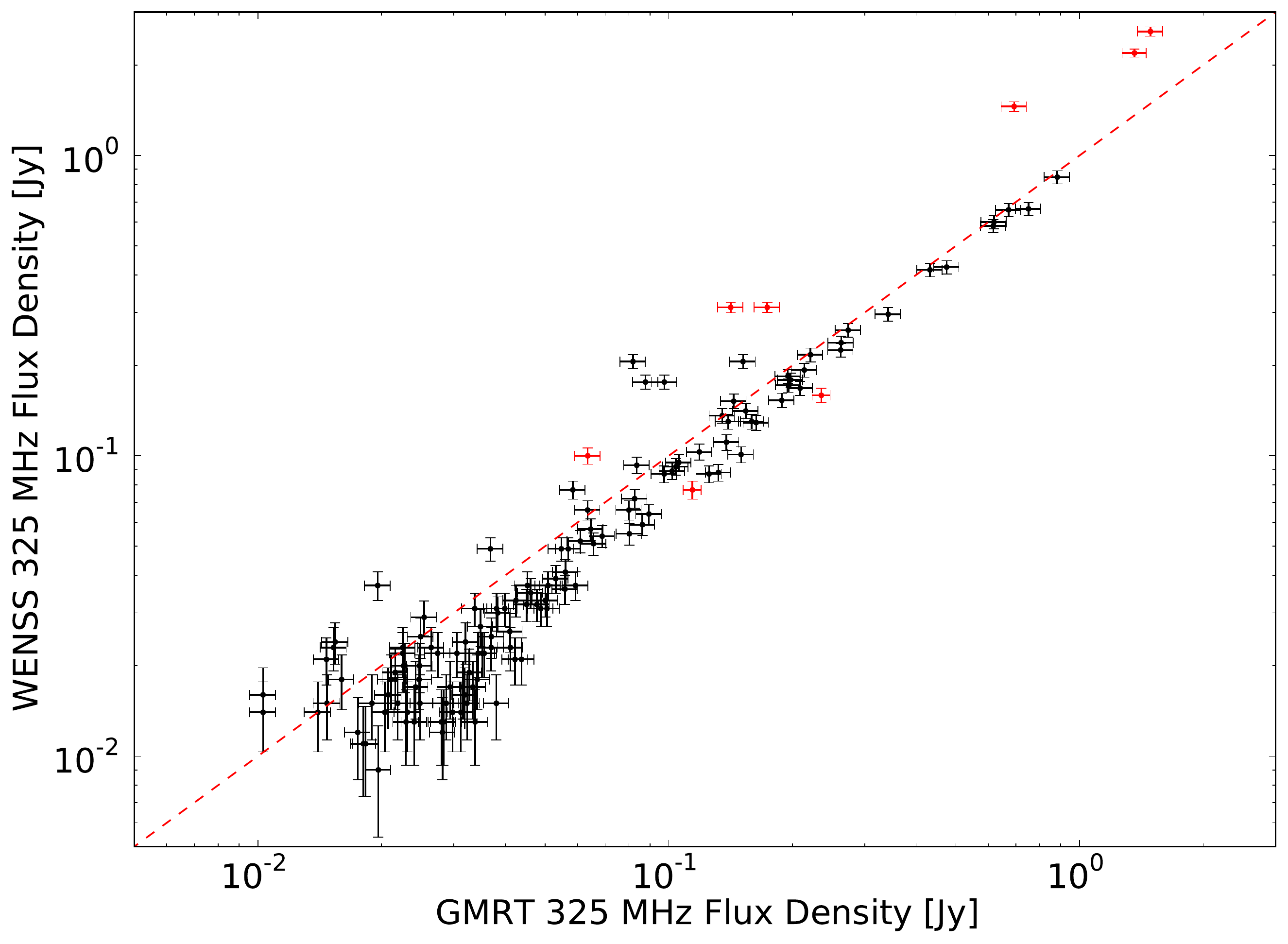}
\caption{Integrated flux densities of sources common to both the SCG catalogue and the WENSS catalogue. The dashed line indicates a flux density ratio equal to unity. The WENSS flux densities have been corrected to bring them in line with the \protect\citet{2012MNRAS.423L..30S} flux scale adopted in this work. Red points mark sources resolved by WENSS, black points mark unresolved WENSS sources.}
\label{fig:gmrt_wenss}
\end{figure}

The WENSS catalogue contains sources above a $5\sigma$ limiting peak flux density of 18 mJy beam$^{-1}$. Given that the WENSS observations were conducted at approximately the same frequency as this work, we are able to directly compare the integrated flux densities of sources common to both catalogues. Within the mosaicked GMRT area, we find 137 sources in the WENSS catalogue, of which 6 are resolved by WENSS. In Figure~\ref{fig:gmrt_wenss} we present the integrated flux densities for sources common to both the WENSS and SCG catalogues. We take the error in the WENSS flux density measurements to be five per cent of the integrated flux density plus the nominal image noise, added in quadrature. 

From Figure~\ref{fig:gmrt_wenss} it appears that the flux densities of sources recovered by the GMRT are generally consistent with their counterparts in WENSS. There is a slight excess at high flux densities, although it is possible that this can be attributed to improved ionospheric calibration. At lower flux densities there is broad scatter in the flux density ratio. Several effects may contribute to this difference -- firstly, the improved calibration routine may allow us to recover more flux from these sources. Secondly the superior sensitivity and resolution of these GMRT observations may play a role: of the SCG sources with point-source counterparts in WENSS, around half become extended at the resolution of these GMRT data, particularly those sources of lower integrated flux density. The high sensitivity of these GMRT observations may go some way to explaining this scatter, recovering fainter emission from these resolved sources.

We have performed two additional tests to verify the flux density scale of our observations. Firstly, comparing our GMRT data with the reprocessed 408 MHz all-sky survey of \citet{1981A&A...100..209H} -- for the reprocessed data see \citet{2015MNRAS.451.4311R} -- we find the flux scale of our map to be consistent to within around 2 per cent. Secondly, we have analysed the data from our main secondary calibrator source, 0949+662.

\subsubsection{Secondary Calibrator: 0949+662}
\begin{figure}
	\centering
	\includegraphics[width=0.48\textwidth]{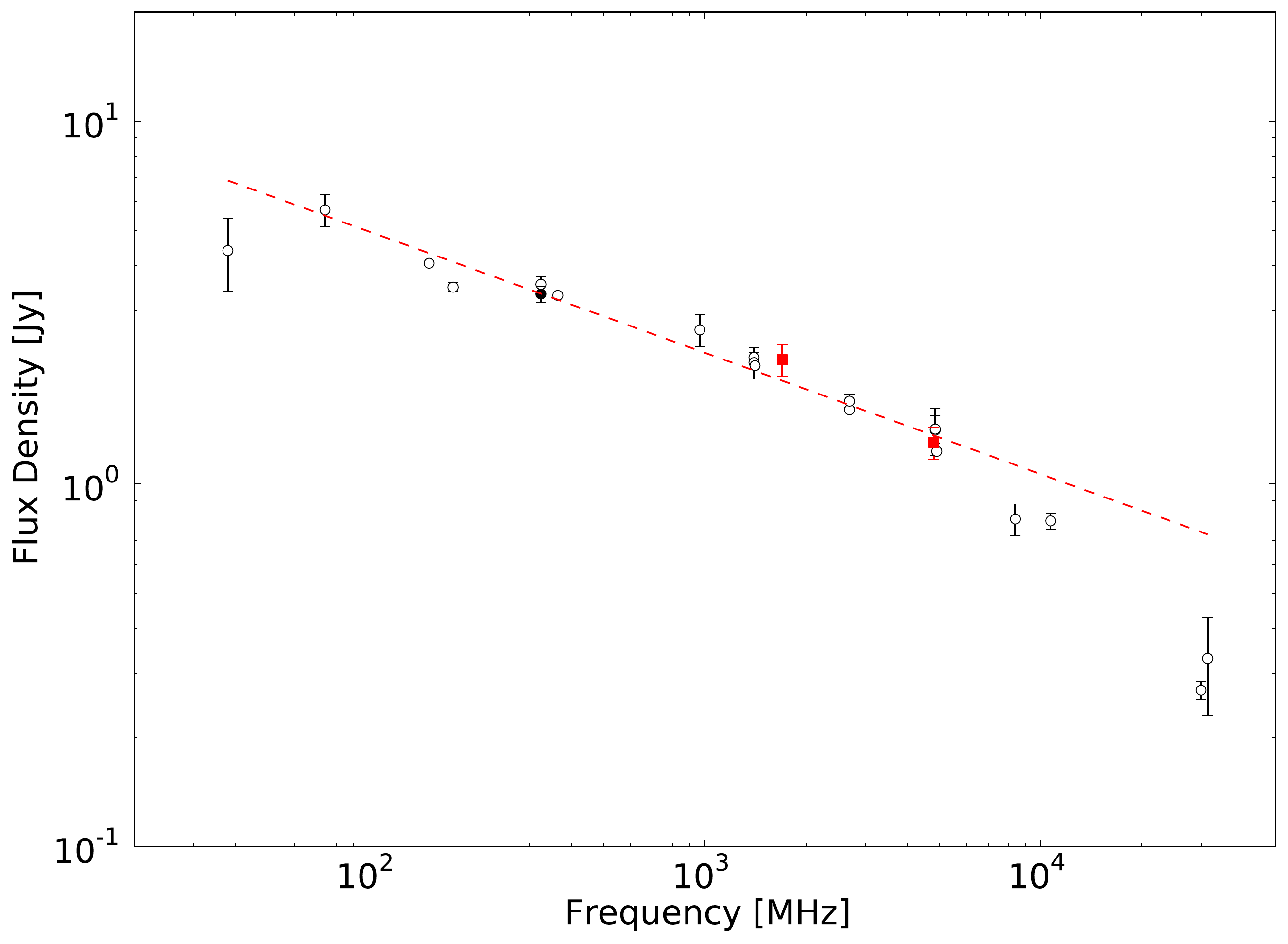}
\caption{Flux density as a function of frequency for the principal secondary calibrator used in this work, 0949+662 (4C +66.09) between 38 MHz and 31.4 GHz. Empty circles mark flux density measurements from the literature; filled squares mark flux densities measured with the VLBA, multiplied by a factor 5 to show the total source flux density (see \protect\citealt{2005A&A...434..449R}). Filled circle indicates the flux density recovered by the GMRT. Dashed line indicates power-law fit to the spectral index between 38 MHz and 31.4 GHz, where $\alpha = -0.33\pm0.03$.}
\label{fig:phasecalspix}
\end{figure}

The principal secondary calibrator source\footnote{Selected from the VLA Calibrator Manual, available at \url{http://www.aoc.nrao.edu/~gtaylor/csource.html}} observed during this work was 0949+662 (J2000 right ascension and declination $09^{\rm{h}}49^{\rm{m}}12.1^{\rm{s}} +66\degree15^{\prime}00^{\prime\prime}$, respectively). This source is also known in the literature as 4C\,+66.09, and has been studied extensively in the radio band between 38 MHz and 30 GHz. We use previous measurements of the flux density to model the spectral energy distribution (SED) for this source and examine how the flux density recovered by the GMRT compares with established measurements. Flux density measurements from the literature are listed in Table \ref{tab:phasecal}, as well as the scaling factor required to bring the measurement into line with the \citet{2012MNRAS.423L..30S} flux density scale. Figure~\ref{fig:phasecalspix} presents the flux density as a function of frequency for 0949+662. 

\begin{table}
\begin{center}
\caption{Flux density measurements from the literature for 0949+662 (4C +66.09), the principal secondary calibration source observed during this work, along with the native flux scale and the conversion factor required to bring the measurement onto the \citetalias{2012MNRAS.423L..30S} flux scale.}
\label{tab:phasecal}
\begin{threeparttable}
\scalebox{0.95}{
\begin{tabular}{ccccc}
\hline
 & & \\ 
Frequency & $S_{\rm{int}}$ & Flux scale & Factor & Catalogue reference \\
$[$MHz$]$ & $[$Jy$]$& & & \\ 
\hline 
31400 & $0.33\pm0.10$ & B77 & - \tnote{a} & W78 \\
30000 & $0.27\pm0.02$ & - \tnote{b} & - \tnote{b} & L07 \\
10695 & $0.77\pm0.04$ & B77 & 1.027 & K81 \\
8400 & $0.78\pm0.08$\tnote{c} & B77 & 1.021 & P92 \\
4900 & $1.23\pm0.02$ & KPW & - & PT78 \\
4850 & $1.41\pm0.12$ & B77 & 1.007 & BWE91 \\
4850 & $1.40\pm0.21$\tnote{d} & B77 & 1.007 & GC91 \\
4800 & $0.26\pm0.03$\tnote{c,e} & VLBA & - & R05 \\
2695 & $1.71\pm0.08$ & B77 & 0.989 & K81 \\
2695 & $1.62\pm0.02$ & B77 & 0.989 & K81 \\
1700 & $0.44\pm0.04$\tnote{c,e} & VLBA & - &  R05 \\
1410 & $2.18\pm0.01$ & B77 & 0.972 & K81 \\
1400 & $2.22\pm0.22$\tnote{c} & B77 & 0.972 & WB92 \\
1400 & $2.30\pm0.07$ & B77 & 0.972 & C98 \\
966 & $2.66\pm0.27$ & KPW & - & C77 \\
365 & $3.39\pm0.05$ & TXS & 0.977 & D96 \\
325 & $3.34\pm0.17$ & SH12 & - & SCG \\
325 & $3.55\pm0.18$ & WENSS\tnote{f} & - & R97 \\
178 & $3.32\pm0.48$\tnote{d} & KPW & 1.09 & GSW67 \\
151 & $4.06\pm0.09$ & RBC73 & - & H90 \\
74 & $5.18\pm0.52$ & B77 & 1.10 & C07 \\ 
38 & $4.40\pm1.00$\tnote{g} & RBC73 & - & C95 \\
\hline
\multicolumn{5}{l}{Flux scale notation:} \\
\multicolumn{5}{l}{B77: \protect{\citet{1977A&A....61...99B}} } \\
\multicolumn{5}{l}{KPW: \protect\citet{1969ApJ...157....1K} } \\
\multicolumn{5}{l}{RCB73: \protect\citet{1973AJ.....78.1030R} } \\
\multicolumn{5}{l}{SH12: \protect\citet{2012MNRAS.423L..30S}} \\
\multicolumn{5}{p{8cm}}{TXS: \protect\citet{1996AJ....111.1945D}. Sources in the Texas radio survey were found to have flux densities that were consistent with a factor $0.9607\times$B77. } \\
\multicolumn{5}{l}{References:}\\
\multicolumn{5}{p{8cm}}{BWE91: \citet{1991ApJS...75....1B}; C77: \citet{1977MmRAS..84....1C}; C95: \citet{1995MNRAS.274..447H}; C98: \citet{1998AJ....115.1693C}; C07: \citet{2007AJ....134.1245C}; D96: \citet{1996AJ....111.1945D}; GC91: \citet{1991ApJS...75.1011G}; GSW67: \citet{1967MmRAS..71...49G}; H90: \citet{1990MNRAS.246..256H}; K81: \protect{\citet{1981A&AS...45..367K}}; L07: \protect{\citet{2007A&A...474.1093L}}; P92: {\citet{1992MNRAS.254..655P}}; PT78: \citet{1978AJ.....83..451P}; R97: {\citet{1997A&AS..124..259R}}; R05: {\citet{2005A&A...434..449R}}; SCG: this work; W78: \protect{\citet{1978AJ.....83..475W}}; WB92: \citet{1992ApJS...79..331W}; }
\end{tabular}
}
\begin{tablenotes}
	\item[  ]{\hspace{-0.3cm} Table notes:}
	\item[a] 30 GHz lies outside the frequency range covered by Table 7 of
	\item[  ] \protect{\citet{1977A&A....61...99B}}. We do not apply a correction factor.
	\item[b] The flux scale used by the observations of \protect{\citet{2007A&A...474.1093L}} was set using 
	\item[  ] the calibrator source NGC 7027. We do not apply a correction factor. 
	\item[c] We assume an uncertainty of 10 per cent.
	\item[d] The original uncertainty was quoted as 15 per cent of the integrated flux
	\item[  ] density. 
	\item[e] The flux densities measured by the VLBA are 20 per cent of the total flux 
	\item[  ] density at that frequency; here we quote the initial value, but in Figure~\ref{fig:phasecalspix} 
	\item[  ] we use the full flux density (i.e. $5\times$ this value). See \protect{\citet{2005A&A...434..449R}} 
	\item[  ] for details. We apply no correction factor to the VLBA measurements. 
	\item[f] The WENSS flux scale is complex, depending on the sky position. We do 
	\item[  ] not  apply a correction factor to this source, see \S\ref{sec:flux}.
	\item[g] Multiple components exist at 38 MHz \protect{\citep{1995MNRAS.274..447H}}.
\end{tablenotes}
\end{threeparttable}
\end{center}
\end{table}

The SED is modelled using a simple power-law spectral index between 38 MHz and 31.4 GHz. For 0949+662, the best-fit spectral index is $\alpha = -0.33\pm0.03$. From Figure~\ref{fig:phasecalspix}, it is clear that there is some departure from a power-law in the high frequency regime $(\nu \gtrsim 10 \,  \rm{GHz})$. There also appears to be departure from power-law behaviour at low frequencies $(\nu \lesssim 74 \, \rm{MHz})$ which may indicate self-absorption effects. Additional low-frequency observations would be required to investigate this further. The flux density recovered by the GMRT for 0949+662 is consistent with the power-law fit, and is also in agreement with the flux density measurement from the WENSS catalogue. 

\subsubsection{Source Sizes}\label{sec:size}
The observed spatial extent of a source may be estimated using the ratio of integrated flux density to peak flux density (for example  \citealt{2000A&AS..146...41P}, \citealt{2010ApJS..188..384S}, \citealt{hales2014a}) via
\begin{equation}
	\frac{S_{\rm{int}}}{S_{\rm{peak}}} = \frac{\theta_{\rm{maj}}\theta_{\rm{min}}}{B_{\rm{maj}}B_{\rm{min}}}
\end{equation}
where $\theta_{\rm{maj}}$ and $\theta_{\rm{min}}$ are the \emph{observed} (i.e. not deconvolved) major and minor axes, respectively, and $B_{\rm{maj}}$ and $B_{\rm{min}}$ are the restoring beam major and minor axis FWHM. In the absence of image noise, unresolved sources have an integrated flux density equal to the peak flux density. However, image noise may affect the fit, and some unresolved sources may appear to have an integrated flux density that differs from the peak value.

\begin{figure}
	\centering
	\includegraphics[width=0.48\textwidth]{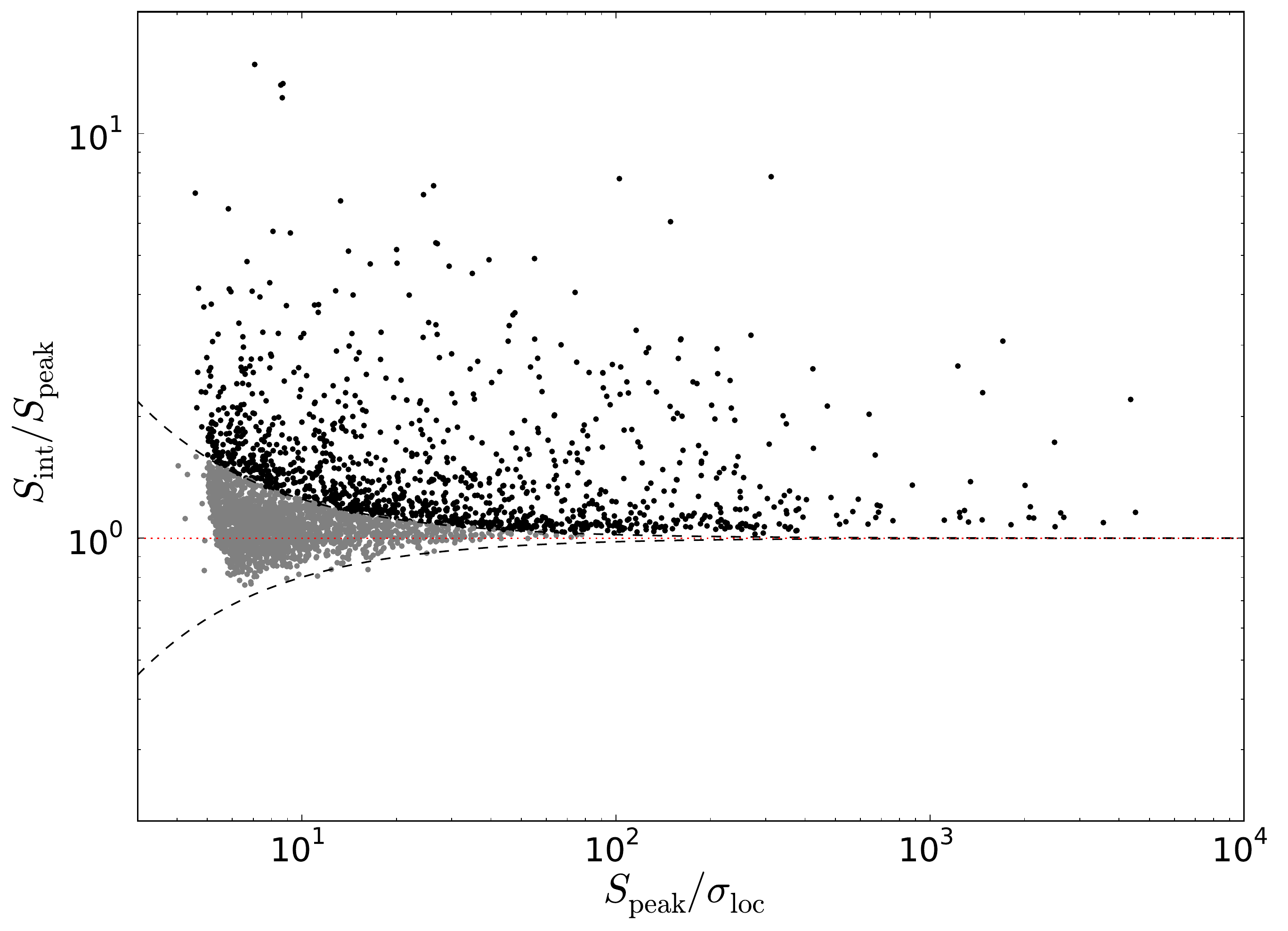}
\caption{Ratio of integrated to peak flux density $(S_{\rm{peak}}/S_{\rm{int}})$ as a function of detection significance $S_{\rm{peak}} / \sigma_{\rm{loc}})$ for sources in the SCG catalogue. Dotted line marks $(S_{\rm{peak}}/S_{\rm{int}})$ equal to unity. Dashed lines mark the locus used to differentiate resolved (black) and unresolved (grey) sources, defined by Equation \ref{eq:locus} and mirrored below $S_{\rm{peak}} = S_{\rm{int}}$. }
\label{fig:peakvint}
\end{figure}

In order to identify which sources are resolved and which are unresolved, we used the ratio of peak $(S_{\rm{peak}})$ and integrated $(S_{\rm{int}})$ flux density as a function of detection significance (i.e. $S_{\rm{peak}} / \sigma_{\rm{loc}}$). This is shown in Figure~\ref{fig:peakvint}. We define a locus that envelops 99.5 per cent of the sources with $S_{\rm{peak}}/S_{\rm{int}} > 1$, and mirrored this above $S_{\rm{peak}} = S_{\rm{int}}$, following the assumption that a similar number of unresolved sources will be scattered to $S_{\rm{peak}}/S_{\rm{int}} < 1$ by noise as those scattered to $S_{\rm{peak}}/S_{\rm{int}} > 1$. The locus was defined using the function
\begin{equation}\label{eq:locus}
	\frac{S_{\rm{peak}}}{S_{\rm{int}}} = k^{(S_{\rm{peak}} / \sigma_{\rm{loc}})^{-c}}
\end{equation}
where $k = 11.517$ and $c=1.042$ provide the best fit to our data. All sources above this locus are considered to be resolved, and we use the integrated flux densities as recovered by \texttt{PyBDSM} in the SCG catalogue. Sources below this locus are considered to be unresolved; we use the peak flux density in place of the integrated flux density when deriving the differential source counts in \S\ref{sec:src}, and the angular size is undetermined. Following Equation \ref{eq:locus}, 1207 sources in the SCG catalogue are considered to be resolved, and 2050 are unresolved. For resolved sources, we can estimate the deconvolved angular size via
\begin{equation}\label{eq:angsize}
	\Theta \simeq \sqrt{ \theta_{\rm{maj}} \theta_{\rm{min}} - B_{\rm{maj}} B_{\rm{min}}}
\end{equation}
and we set the size of unresolved sources to zero. Unresolved sources have major and minor axis FWHM and position angles, as well as their associated uncertainties, denoted by `-' in Table \ref{tab:src_cat}.

\section{Discussion}\label{sec:DISC}
\subsection{Spectral Index Distribution}
\begin{figure}
	\centering
		\includegraphics[width=0.47\textwidth]{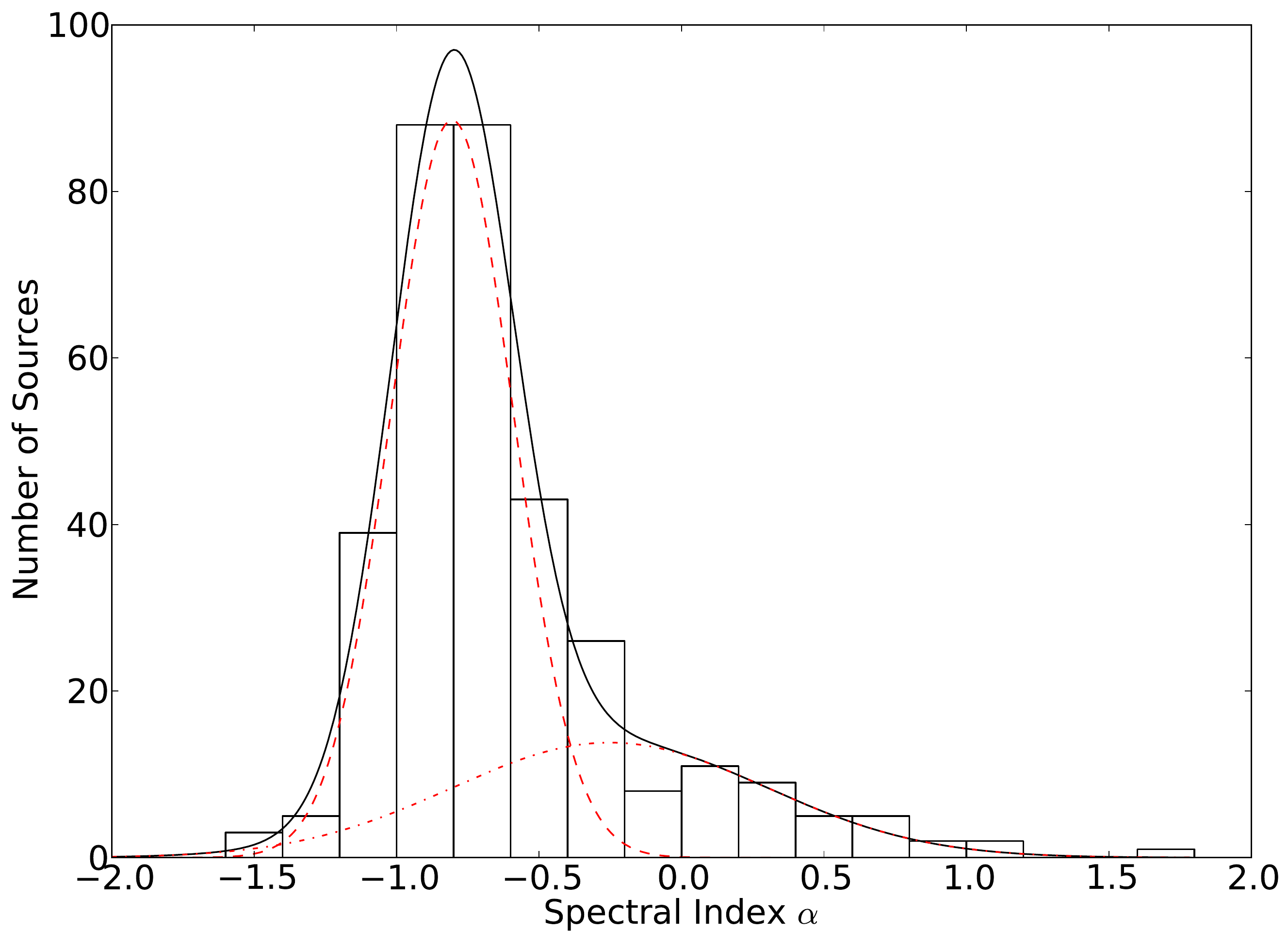}
		\includegraphics[width=0.47\textwidth]{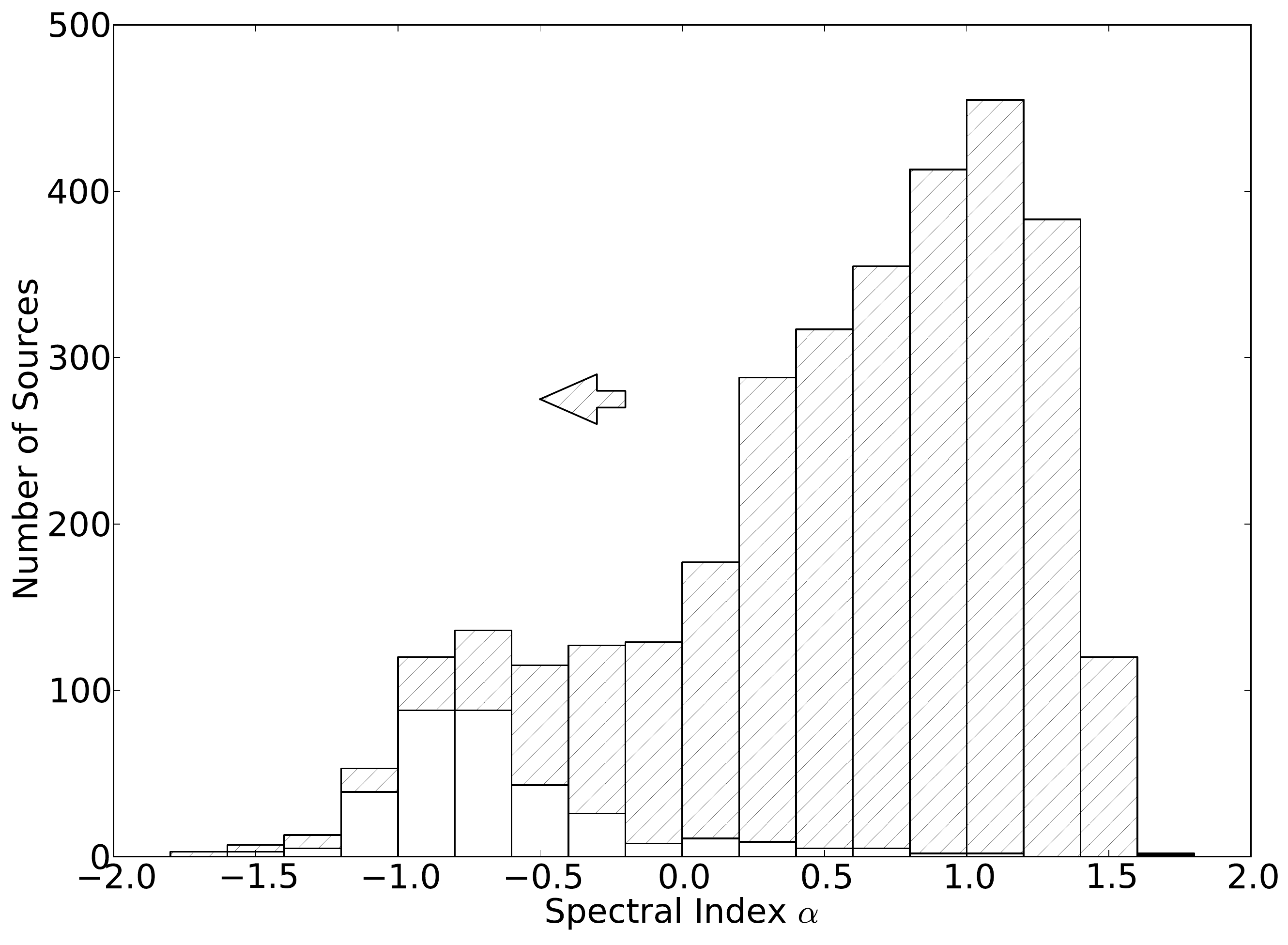}
\caption{Source spectral index distribution between 1.4 GHz (from the NVSS catalogue) and 325 MHz (from this work). \emph{Top panel:} spectral index estimates for sources with matches in the NVSS. The solid line indicates the combined best-fit distribution, dashed (dot-dashed) lines indicate the Gaussians fitted to the steep-spectrum (flat-spectrum) population. The best-fit models for these populations have peaks centred on $\alpha_1 =  -0.81$ (dashed) and $\alpha_2 = -0.25$ (dot-dashed). \emph{Bottom panel:} measured spectral index values for sources detected in the NVSS (no fill) and upper limits on the spectral index for sources without NVSS counterparts (hatched, and indicated by the arrow).}
\label{fig:spix_dist}
\end{figure}

For all sources with NVSS counterparts, we derive spectral index values using a simple power-law fit. We present the distribution of spectral index values between 325 and 1400 MHz in Figure~\ref{fig:spix_dist}. From inspection, it is clear that the distribution is not well-described by a single Gaussian. This suggests that two populations exist within this distribution: a larger population of sources centred around a spectral index of $-0.8$ (typical of synchrotron-dominated emission) and a smaller population of sources with flat or rising spectra. We note that this distribution is biased due to the limited sensitivity of the NVSS observations used as the higher frequency reference; as such we also include upper limits to the spectral index for sources without NVSS counterparts in the lower panel of Figure~\ref{fig:spix_dist}.

Subsequently, we fit two Gaussian functions to this distribution. The best fit yields distributions with mean spectral index $\alpha_1 = -0.81\pm0.21$ for the steep-spectrum population and $\alpha_2 = -0.25\pm0.55$ for the flat-spectrum population. These population distributions are also indicated in Figure~\ref{fig:spix_dist}.

The mean spectral index of the dominant source population within the SCG catalogue is consistent with that derived in previous work (for example \citealt{2013MNRAS.435..650M}, \citealt{2014MNRAS.443.2590S}). Figure~11 of \citet{2009MNRAS.395..269S} presents the histogram of spectral index values for the sources in the ELAIS-N1 field, which appears to exhibit marginal evidence of a small, rising-spectrum population alongside the dominant population. \citet{2013MNRAS.435..650M} find a spectral index distribution that is well-described by a single Gaussian centred on $\alpha = -0.71$. This difference is naturally explained by the difference in sensitivity between the observations in this work (nominal sensitivity $34 \umu$Jy beam$^{-1}$) and the work of \citeauthor{2013MNRAS.435..650M} (where the nominal sensitivity was $\sim1-7$ mJy beam$^{-1}$). The deeper observations presented here recover sources with flux densities far below 1 mJy; a regime where the contribution from SFG and RQ-AGN become increasingly important (for example \citealt{2001A&A...369..787P}, \citealt{2007ASPC..380..205P}, \citealt{2008ApJS..177...14S}, \citealt{2009MNRAS.397..281I}, \citealt{2009ApJ...694..235P}, \citealt{2015MNRAS.452.1263P}). 

However, the spectral index distribution presented in Figure~\ref{fig:spix_dist} is unlikely to be probing the population of SFG in this field. With only 335 spectral index estimates out of a catalogue of 3257 sources, the spectral index distribution we have derived is severely limited by the sensitivity of the NVSS catalogue; the NVSS catalogue is limited to sources of flux densities in excess of $\sim2-3$ mJy, whereas SFG typically dominate below $0.1-1$ mJy at 1.4 GHz. Hence, the population of flat-/rising-spectrum sources in the SCG catalogue may be comprised of a number of different source types -- the brighter sources may be flat-spectrum radio quasars (FSRQ) or blazars; some of the fainter flat-spectrum sources may be core-dominated AGN \citep{2012MNRAS.421.1644R} or hybrid sources exhibiting both star-formation and a central engine (for example \citealt{1999AJ....117..111H}, \citealt{2001AJ....121..128H}). Alternatively some of these flat spectra may be attributed to absorption effects in the local environment. Further observations at other frequencies are required to discriminate between scenarios. A detailed investigation of the nature of this flat-spectrum population will form the subject of future work, as the highly-sensitive 1.4~GHz data required are currently being taken with e-MERLIN and the JVLA.

\subsubsection{Relation between spectral index and flux density}

\begin{figure}
	\centering
	\includegraphics[width=0.475\textwidth]{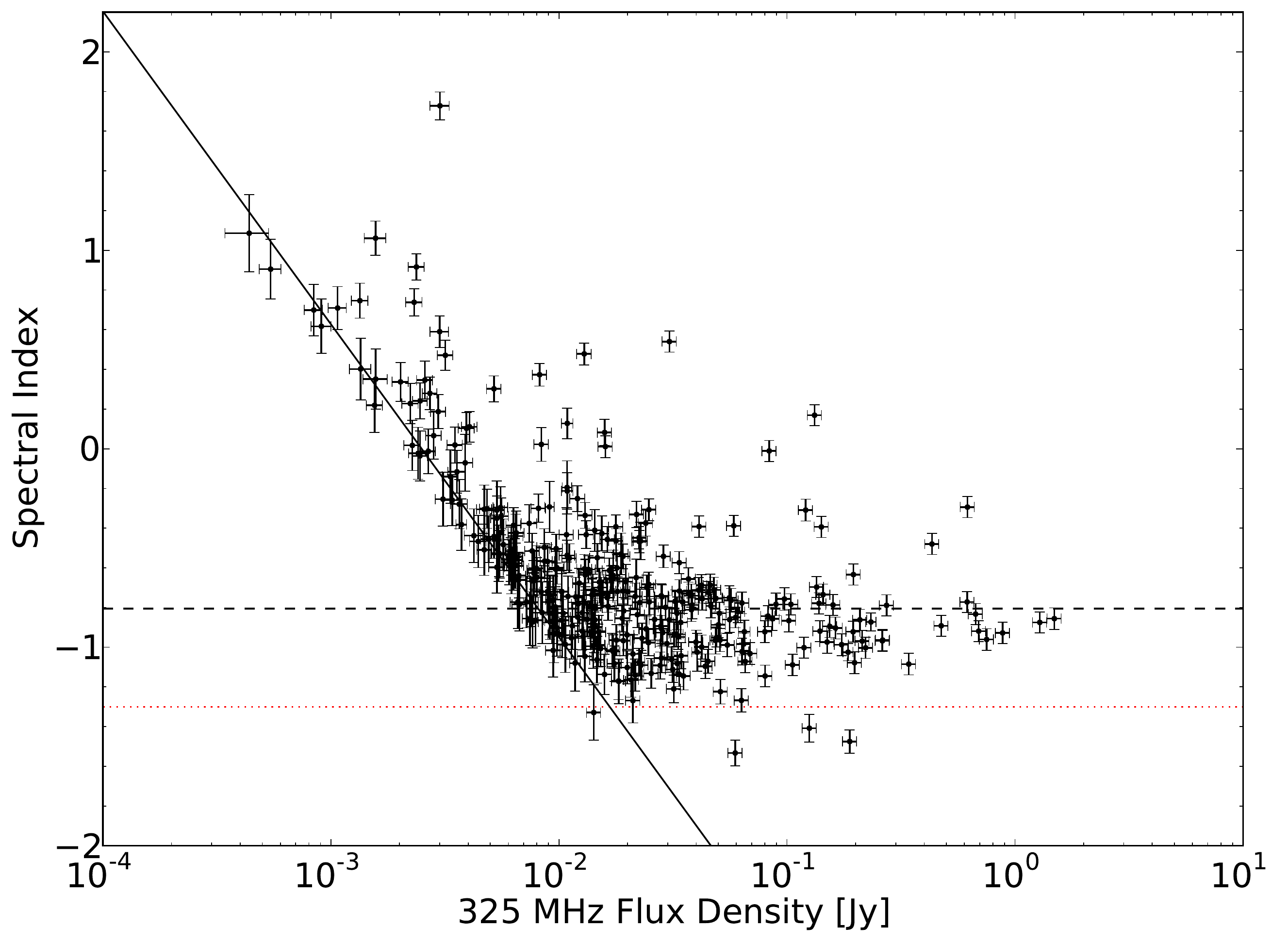}
	\caption{Spectral index as a function of 325 MHz flux density for the 335 sources in the SCG catalogue which have counterparts detected in the NVSS. Solid sloping line marks the flux density traced by the nominal NVSS detection limit of 2.5 mJy. Dashed line marks mean spectral index of the dominant population. Dotted line marks nominal `steep spectrum' index of $\alpha = -1.3$.}
	\label{fig:fluxvspix}
\end{figure}

\begin{figure*}
	\centering
	\begin{tabular}{cc}
	\subfloat{
		\includegraphics[width=0.63\textwidth]{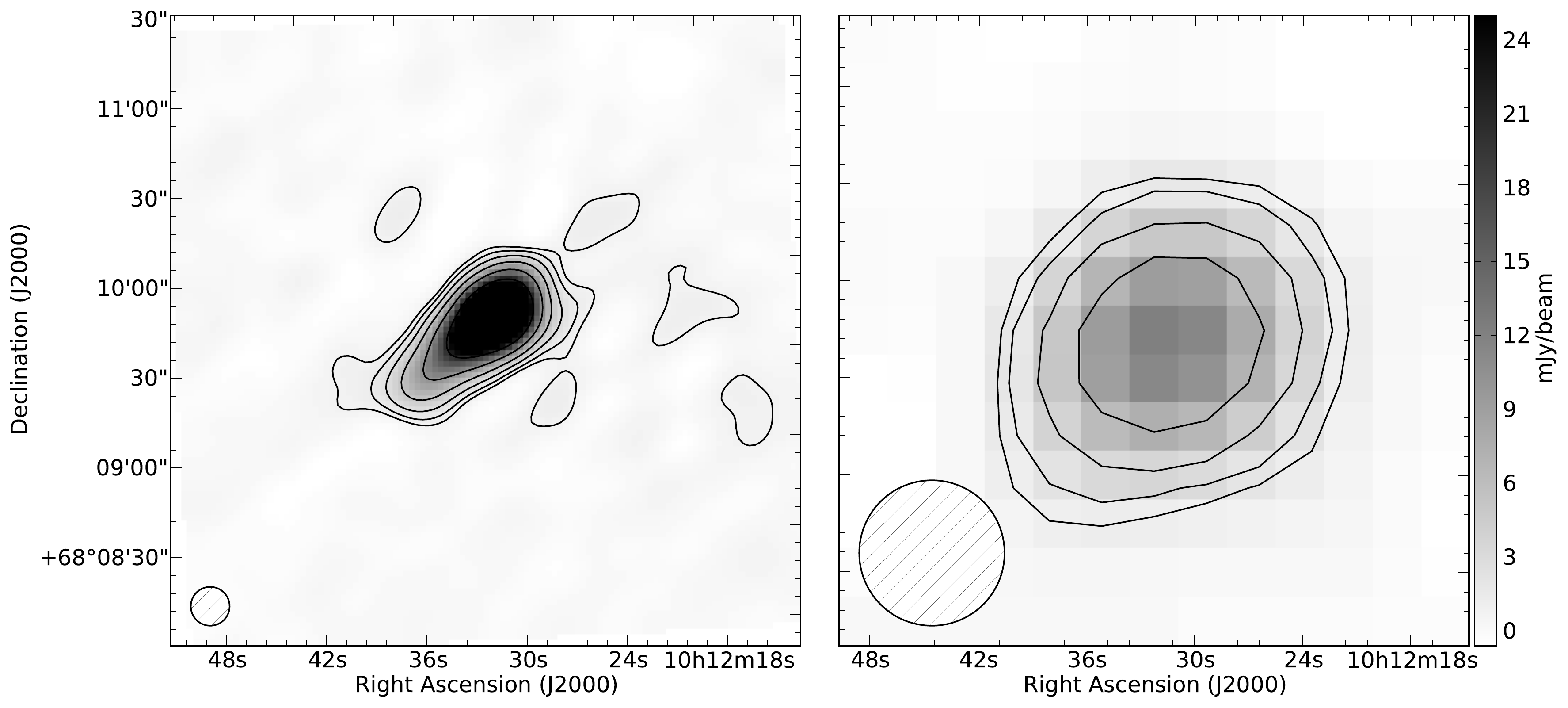}
	} \\
	\subfloat{
		\includegraphics[width=0.63\textwidth]{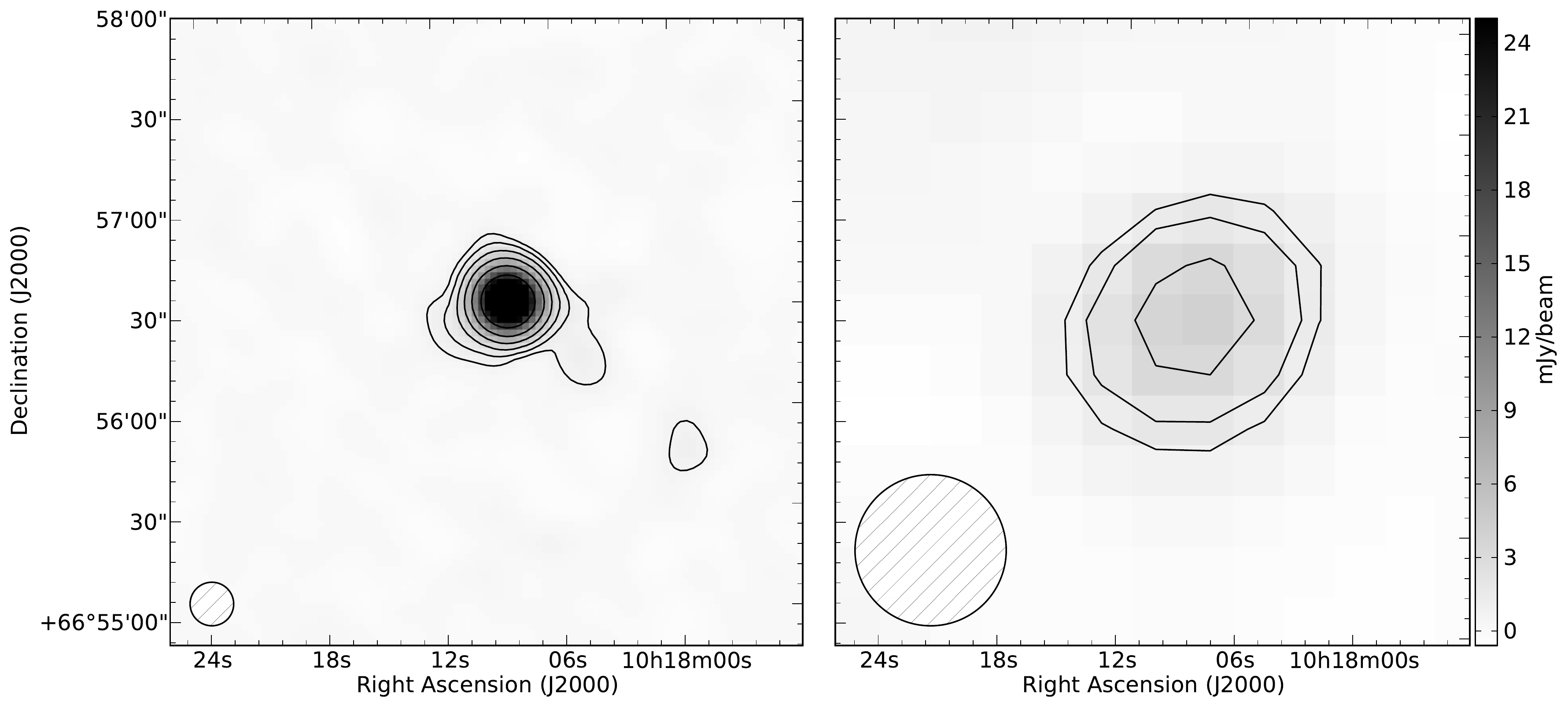}
	} \\
	\subfloat{
		\includegraphics[width=0.63\textwidth]{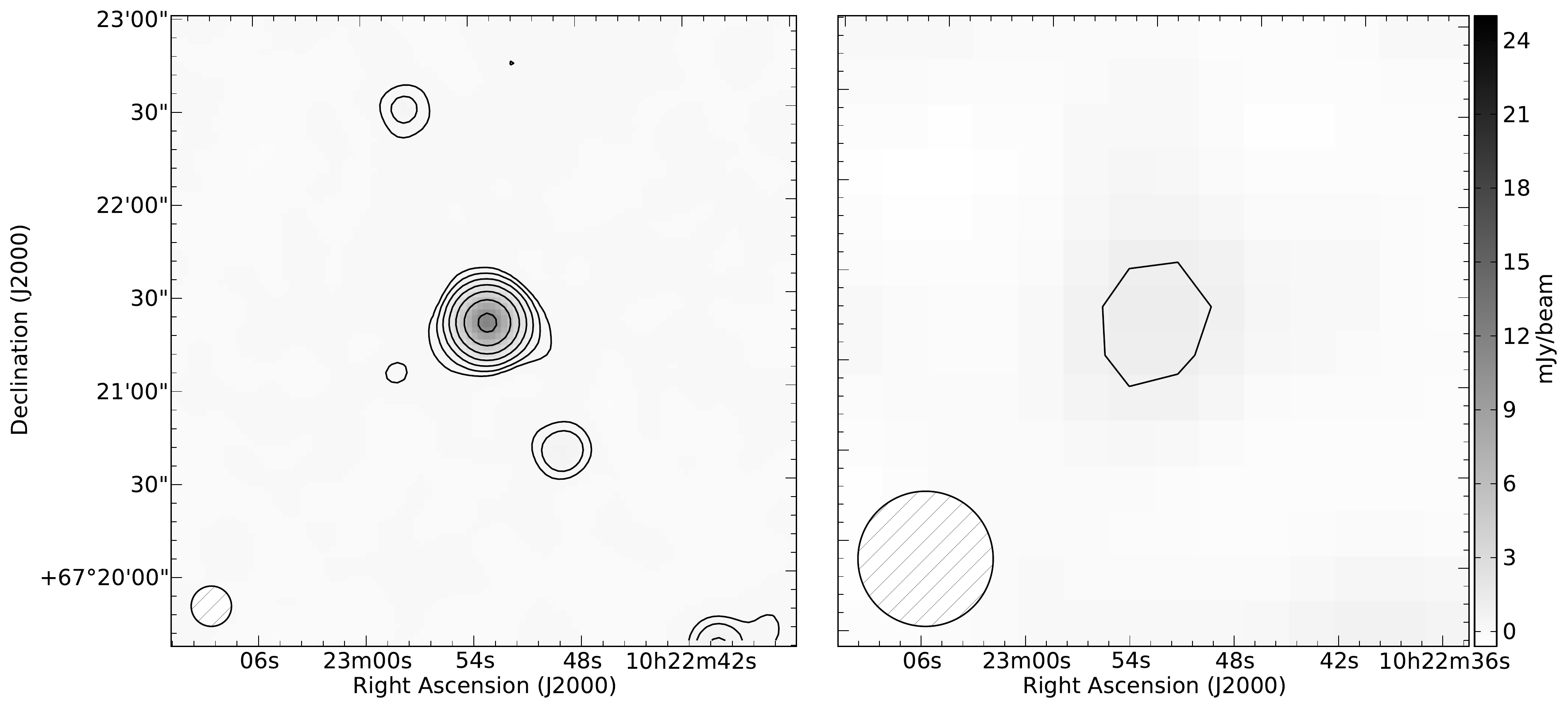}
	} \\
	\subfloat{
		\includegraphics[width=0.63\textwidth]{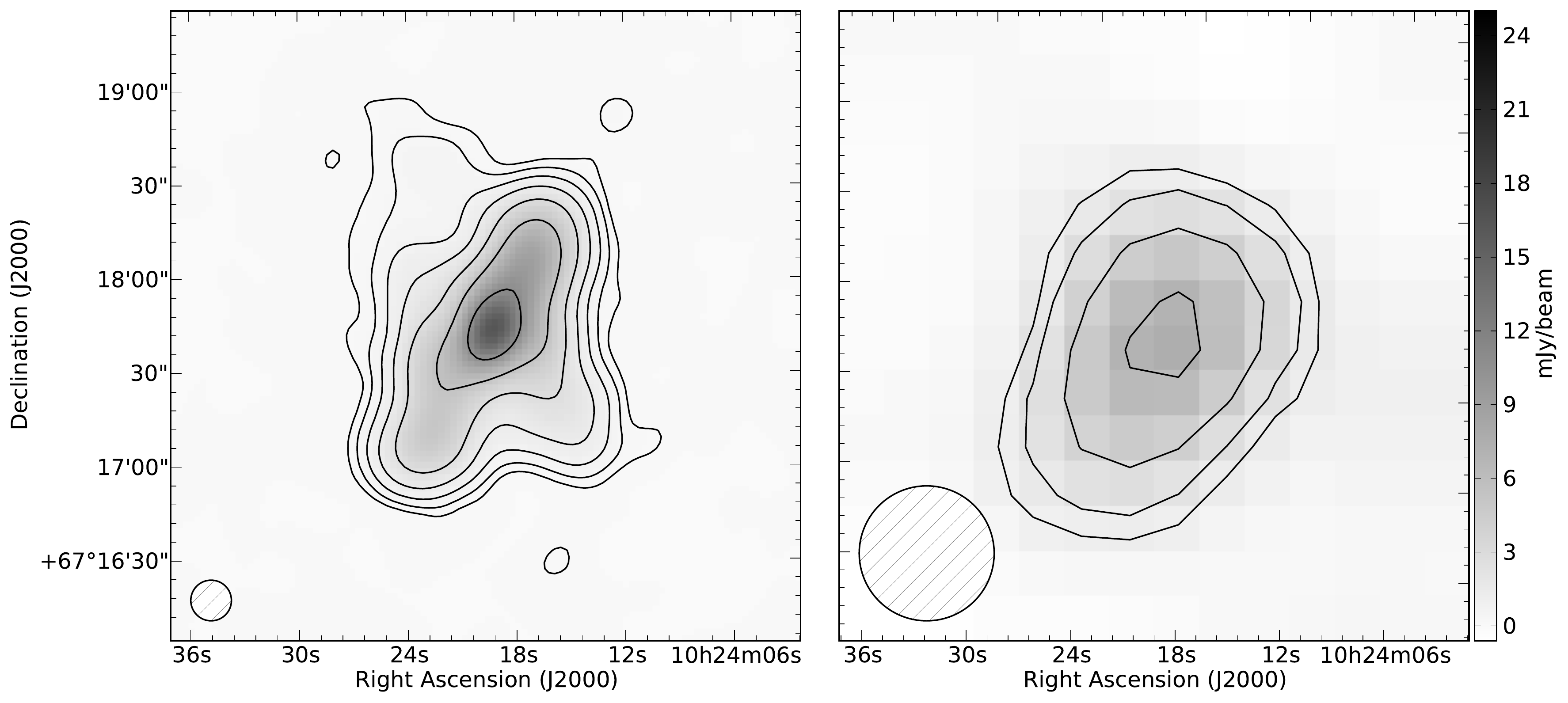}
	} 
	\end{tabular}
\caption{Candidate ultra-steep-spectrum radio sources from the SCG catalogue. \emph{From top to bottom:} SCG\_J101231+680942 , SCG\_J101808+665632, SCG\_J102253+672121, SCG\_J102419+671742. \emph{Left panels:} GMRT images. Contours start at $5\sigma_{\rm{local}}$ and scale by a factor 2, where the local noise $\sigma_{\rm{local}} = 140 \, /  \, 125 \, / \, 36 \, / \, 35 \, \umu$Jy beam$^{-1}$ from top to bottom. \emph{Right panels:} postage stamp from the NVSS at 1.4 GHz. First contour is at $3\sigma_{\rm{nom}}$, then scale by a factor 2 from $5\sigma_{\rm{nom}}$ where $\sigma_{\rm{nom}}$ is the nominal NVSS image noise of 0.5 mJy beam$^{-1}$. Images are set to matching colour scales and saturate at 25 mJy beam$^{-1}$. Beam sizes are indicated by the hatched circle in the lower left corner.}
\label{fig:uss_src}
\end{figure*}

\begin{table*}
\begin{center}
\caption{Ultra-steep-spectrum (USS) sources and gigahertz-peaked-spectrum (GPS) sources from the SCG catalogue that have been selected for further study. Integrated flux densities at 1.4 GHz are from the NVSS, and have been corrected by a factor 0.972 to bring them onto the Scaife \& Heald (2012) scale. We quote alternative names from the literature where they exist; the NVSS source identification if no others exist.}
\label{tab:uss_src}
\begin{tabular}{cccrrcc}
\hline
Name & RA (J2000) & DEC (J2000) & $S_{\rm{int}}$(325 MHz) & $S_{\rm{int}}$(1.4 GHz) & $\alpha$ & Alternate Name \\
& \emph{hh mm ss.ss} & \emph{dd mm ss.s} & $[$mJy$]$\hspace{0.4cm} & $[$mJy$]$\hspace{0.4cm} & &  \\
\hline
SCG\_J101231+680942 & 10 12 31.47 & 68 09 42.5 & $188.44\pm13.35$ & $21.87\pm1.07$ & $-1.48\pm0.06$ & NVSS J101231+680940 \\
SCG\_J101808+665632 & 10 18 08.51 & 66 56 32.8 & $59.30\pm4.23$ & $6.32\pm0.39$ & $-1.53\pm0.06$ & NVSS J101808+665632 \\
SCG\_J102253+672121 & 10 22 53.05 & 67 21 21.1 & $14.21\pm1.01$ & $2.04\pm0.39$ & $-1.33\pm0.14$ & NVSS J102253+672118 \\
SCG\_J102419+671742 & 10 24 19.15 & 67 17 42.5 & $125.37\pm8.88$ & $16.04\pm1.17$ & $-1.41\pm0.07$ & 8C~1020+675 \\
\hline
SCG\_J101418+683826 & 10 14 18.30 & 68 38 26.5 & $12.91\pm0.95$ & $25.95\pm0.87$ & $+0.48\pm0.06$ & NVSS J101418+683826 \\
SCG\_J101538+672844 & 10 15 38.12 & 67 28 44.8 &   $3.01\pm0.29$ & $37.52\pm1.17$ & $+1.73\pm0.07$ & CGRaBS J1015+6728 \\
SCG\_J101723+673633 & 10 17 23.75 & 67 36 33.3 & $30.51\pm2.17$ & $67.17\pm2.04$ & $+0.54\pm0.05$ & 87GB 101339.1+675144 \\
SCG\_J103401+683226 & 10 34 01.15 & 68 32 26.6 & $132.14\pm9.37$ & $169.03\pm5.05$ & $+0.17\pm0.05$ & 87GB 103023.1+684757 \\
\hline
\end{tabular}
\end{center}
\end{table*}

We present the dependence of the spectral index on the 325 MHz integrated flux density in Figure~\ref{fig:fluxvspix}. There is significant bias in Figure~\ref{fig:fluxvspix} due to the sensitivity limit of the NVSS; this is indicated by the solid line in Figure~\ref{fig:fluxvspix}, and cuts out a large region of the parameter space. In some previous works, there has been a detection of a `flattening' in the spectral index distribution below $S_{\rm{1.4\,GHz}} = 10$\,mJy (for example \citealt{2006A&A...457..517P}, \citealt{2008A&A...477..459M}) whereas other authors do not detect such a feature (for example \citealt{2009MNRAS.397..281I}, \citealt{2012MNRAS.421.1644R}). \citet{2012MNRAS.421.1644R} suggest that this flattening may be associated with a population of core-dominated AGN at the faintest flux densities; as such, the mean spectral index flattens to $\alpha > -0.7$.  

The picture from studies of the low-frequency spectral index is also mixed. Figure~13 of \citet{2009MNRAS.395..269S} exhibits marginal evidence of a flattening in the region 1--10 mJy, whereas \citet{2006A&A...456..791T} and \citet{2013MNRAS.435..650M} find `little-to-no evidence' of this. Again, sensitivity differences provide a natural explanation for this. From Figure~\ref{fig:fluxvspix} there is also marginal evidence of this flattening in the spectral index/flux density relation in the 1--10 mJy region. With a sample size of 335 out of a total catalogue of 3257 sources, more sensitive observations of this region are required at higher frequency to better examine any relationship between flux density and spectral index and confirm whether this flattening is a real feature or simply due to sample bias.

\subsection{Ultra Steep Spectrum Sources}
Ultra steep spectrum (USS) radio sources are often associated with radio galaxies at high redshift (HzRGs). Studies have shown these are among the most luminous and massive galaxies (for example \citealt{2005A&A...430L...1D}, \citealt{2007ApJS..171..353S}, \citealt{2014A&A...569A..52S}) and are believed to be progenitors of the massive elliptical galaxies in the local Universe. These have also been shown to be associated with overly dense regions of space: galaxy clusters and proto-clusters at redshifts $z \sim 2 - 5$ (for example \citealt{2003Natur.425..264S}, \citealt{2007A&A...461..823V}, \citealt{2015ApJ...802...31D}, \citealt{2015ApJ...808L..33C}). Identifying and studying HzRGs enables us to better understand the mechanisms by which these sources form and evolve in dense environments and at high-$z$. Most HzRG sources at $z \gtrsim 3$ have been identified through detection of USS radio sources, which are bright at low frequencies and often appear compact (for example \citealt{2007ASPC..380..213K}, \citealt{2008A&ARv..15...67M}). 

We find two sources with spectral index more than $3\sigma_1$ below the mean spectral index of the dominant population (i.e. $\alpha < -1.45$) which we investigate as USS sources. In the literature, USS radio sources are commonly defined as those with spectral index values $\alpha < -1.3$. We find a further two sources which satisfy this condition, for a total of four ultra steep spectrum candidates (USSc). Postage stamp images of these sources are presented in Figure~\ref{fig:uss_src}. In Table \ref{tab:uss_src} we list the integrated flux densities at 325 MHz (from this work) and 1400 MHz (from the NVSS) and the two-point spectral index values. In this section, we discuss these sources and cross-reference them with the literature.

\subsubsection{SCG\_J101231+680942}
\begin{figure}
	\centering
	\begin{tabular}{l}
	\subfloat{
		\includegraphics[width=0.46\textwidth]{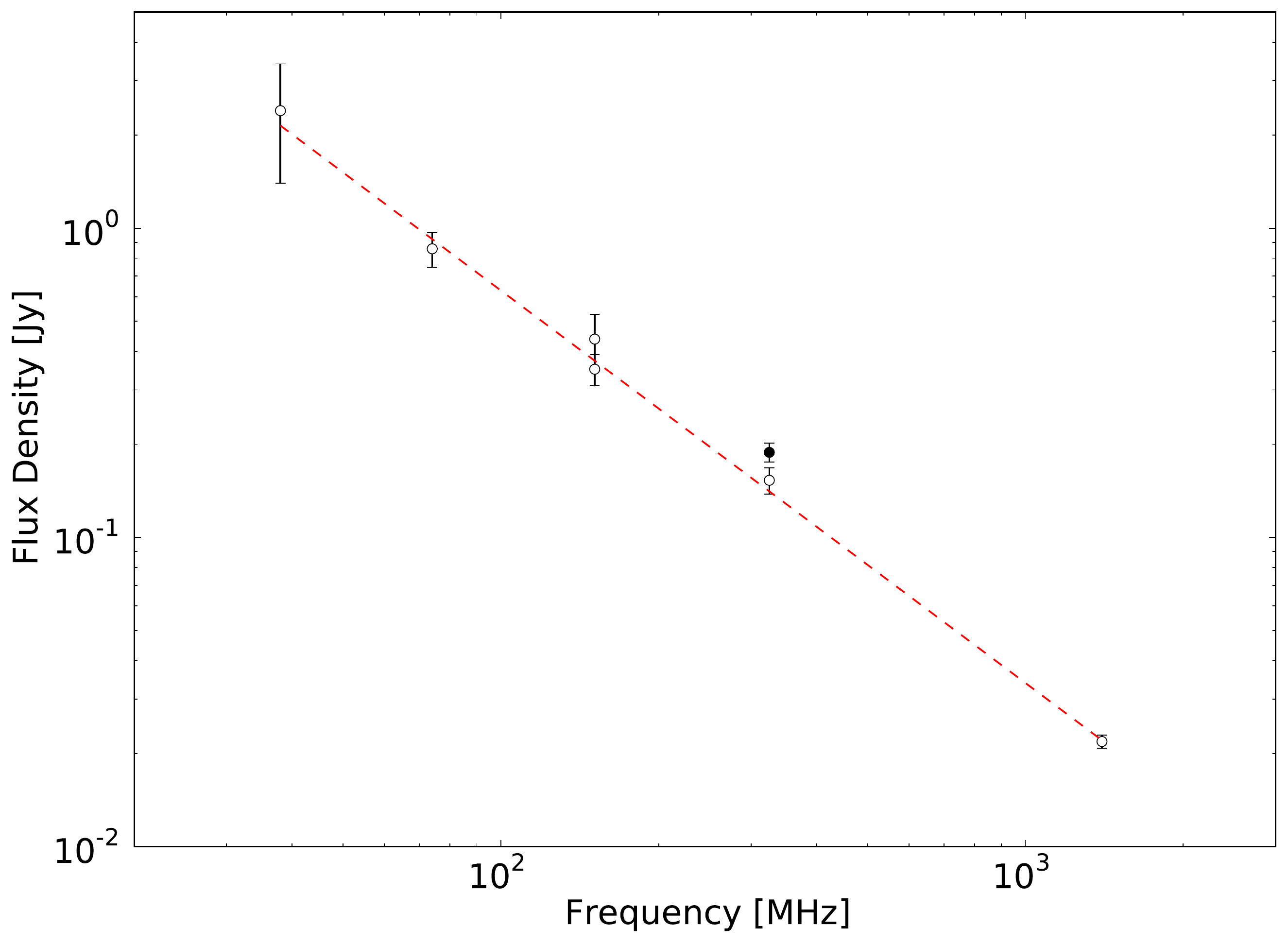}
	} \\
	\subfloat{
		\includegraphics[width=0.46\textwidth]{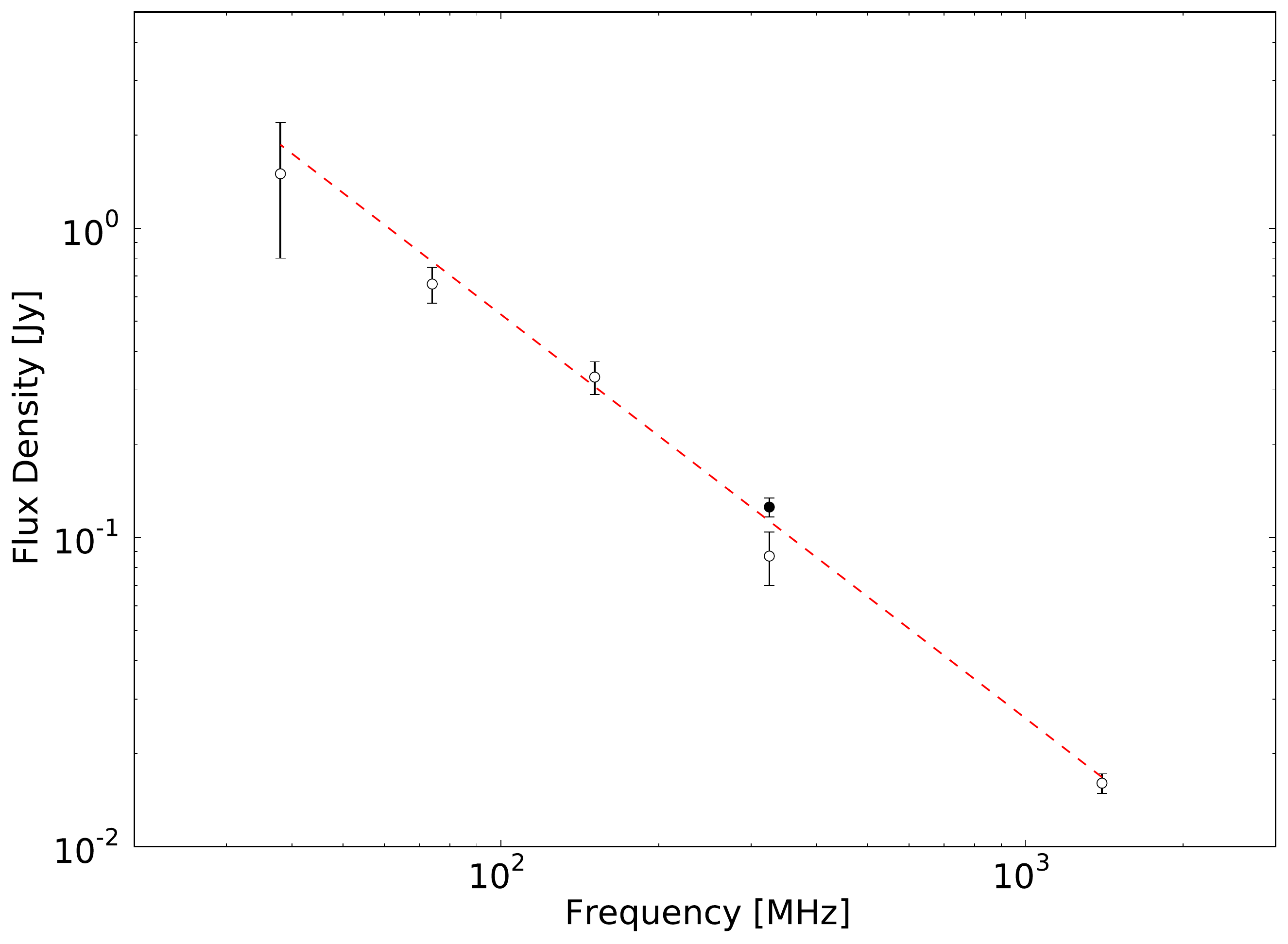}
	}
	\end{tabular}
	\caption{Flux density as a function of frequency for SCG\_J101231+680942 (NVSS J101231+680940; \emph{top panel}) and SCG\_J102419+671742 (8C~1020+675; \emph{bottom panel}) between 38 MHz and 1.4 GHz. Filled (empty) circles indicate flux density measurements from this work (the literature). Dashed lines indicate the fitted multifrequency spectral index $\alpha = -1.27\pm0.22$ $(-1.31\pm0.26)$ in the top (bottom) panel.}
	\label{fig:uss_spix}
\end{figure}

This source has been identified in a number of previous radio surveys; for example in the NVSS its identifier is NVSS J101231+680940. Additional flux density measurements for this source exist at 38, 74, 151 and 325 MHz, from a number of historic radio surveys. These flux density measurements are listed in Table \ref{tab:ussc}, with the appropriate correction factor required to bring them into line with the \citetalias{2012MNRAS.423L..30S} flux density scale. Subsequently, we use the corrected flux densities to fit a spectral index in the form of a power-law between 38 MHz and 1.4 GHz. We present the flux density as a function of frequency in the top panel of Figure~\ref{fig:uss_spix}, as well as the spectral index fitted to the data from the literature, $\alpha = -1.27\pm0.22$. This fit suggests a spectral index that is marginally shallower than the two-point spectral index indicated in Table \ref{tab:uss_src}. 

Based on the multi-frequency spectral index, we cannot conclusively say whether this source has an ultra-steep-spectrum; from Figure~\ref{fig:uss_spix}, the integrated flux density recovered by the GMRT appears to be in excess of that expected from the flux density measurements in the literature. However, from Figure~\ref{fig:uss_src} (top panel) it appears that this source is extended at the resolution of the GMRT, so it is possible that this excess is a result of the superior sensitivity to faint emission compared to previous surveys.

\begin{table*}
\begin{center}
\caption{Flux density measurements from this work and the literature for ultra-steep spectrum (USS) and gigahertz-peaked-spectrum (GPS) radio sources identified in this work (see Table \ref{tab:uss_src}). The `Factor' column indicates the correction assumed to bring the flux density measurement onto the {\protect\citet{2012MNRAS.423L..30S}} flux scale.}
\label{tab:ussc}
\scalebox{0.95}{
\begin{threeparttable}
\begin{tabular}{ccccccc}
\hline
 & & \\ 
Source & Source type & Frequency & $S_{\rm{int}}$ & Flux scale & Factor & Catalogue \\
& & $[$MHz$]$ & $[$mJy$]$ & & & \\ 
\hline 
\multirow{7}{3cm}{SCG\_J101231+680942 (NVSS J101231+680940)} & \multirow{7}{*}{USS} & 1400 & $22.5\pm1.1$ & B77 & 0.972 & NVSS {\citep{1998AJ....115.1693C}} \\
							& & 325 & $153\pm31$ & WENSS & - & WENSS {\citep{1997A&AS..124..259R}} \\
							& & 325 & $188.4\pm13.4$ & SH12 & - & SCG (this work) \\
							& & 151 & $350\pm40$ & RCB73 & - & 6C {\citep{1990MNRAS.246..256H}} \\
							& & 151 & $438\pm88$ & RCB73 & - & 7C {\citep{2007MNRAS.382.1639H}} \\
							& & 74 & $\,\,\,780\pm100$ & B77 & 1.10 & VLSS {\citep{2007AJ....134.1245C}} \\
							& & 38 & $\,\,\,2400\pm1000$ & RCB73 & - & 8C {\citep{1995MNRAS.274..447H}} \\
\hline
\multirow{7}{3cm}{SCG\_J102419+671742 (8C~1020+675)} & \multirow{7}{*}{USS} & 1400 & $16.5\pm1.2$ & B77 & 0.972 & NVSS {\citep{1998AJ....115.1693C}}  \\
							& & 325 & $132.1\pm9.4\,\,\,\,$ & SH12 & - & SCG (this work) \\
							& & 325 & $\,\,\,\,87\pm17$ & WENSS & - & WENSS {\citep{1997A&AS..124..259R}} \\
							& & 232 & $310$\tnote{a} & RCB73 & - & Miyun {\citep{1997A&AS..121...59Z}} \\
							& & 151 & $330\pm40$ & RCB73 & - & 6C {\citep{1990MNRAS.246..256H}} \\
							& & 74 & $600\pm80$ & B77 & 1.10 & VLSS {\citep{2007AJ....134.1245C}} \\
							& & 38 & $1500\pm700$ & RCB73 & - & 8C {\citep{1995MNRAS.274..447H}} \\
\hline
\multirow{6}{3cm}{SCG\_J101538+672844 (CGRaBS J1015+6728)} & \multirow{6}{*}{GPS} &15000 & $105.0\pm1.0\,\,\,\,$ & B77 & 1.035 & OVRO Blazar Monitoring Program; {\citep{2011ApJS..194...29R}}  \\
							& & 8400 & $150.1\pm15.0$\tnote{b} & B77\tnote{c} & 1.021 & CGRaBS {\citep{2008ApJS..175...97H}} \\
							& & 8400 & $146.1\pm14.6$\tnote{b} & B77\tnote{c} & 1.021 & {\citet{1991ApJS...75....1B}} \\
							& & 4850 & $117.0\pm11.0$ & B77\tnote{d} & 1.007 & {\citet{1991ApJS...75.1011G}} \\
							& & 1400 & $39.1\pm1.3$ & B77 & 0.972 & NVSS {\citep{1998AJ....115.1693C}}  \\
							& & 325 & $\,\,\,3.1\pm0.3$ & SH12 & - & SCG (this work) \\
\hline

\multirow{8}{3cm}{SCG\_J101723+673633 (87GB 101339.1+675144)} & \multirow{8}{*}{GPS} & 8400 & $19.3\pm0.2$ & B77\tnote{c} & 1.021 & JVAS/CLASS {\citep{2007MNRAS.376..371J}}  \\
							& & 8400 & $20.4\pm2.0$\tnote{b} & B77 & 1.021 & VLA-CLASS {\citep{2003MNRAS.341....1M}} \\
							& & 4850 & $38.0\pm3.8$\tnote{b} & B77 & 1.007 & {\citet{1991ApJS...75....1B}} \\
							& & 4850 & $41.0\pm6.0$ & B77 & 1.007 & 87GB {\citep{1991ApJS...75.1011G}} \\
							& & 4850 & $35.0\pm4.0$ & B77 & 1.007 & GB6 {\citep{1996ApJS..103..427G}} \\
							& & 1400 & $69.1\pm2.1$ & B77 & 0.972 & NVSS {\citep{1998AJ....115.1693C}}  \\
							& & 325 & $22.0\pm3.8$ & WENSS & - & WENSS {\citep{1997A&AS..124..259R}} \\
							& & 325 & $30.5\pm2.2$ & SH12 & - & SCG (this work) \\ 
\hline
\multirow{9}{3cm}{SCG\_J103401+683226 (87GB 103023.1+684757)} & \multirow{8}{*}{GPS} & 8400 & $103.7\pm1.0\,\,\,$ & B77\tnote{c} & 1.021 & JVAS/CLASS {\citep{2007MNRAS.376..371J}}  \\
							& & 8400 & $110.9\pm11.1$\tnote{b} & B77\tnote{c} & 1.021 & CRATES {\citep{2007ApJS..171...61H}} \\
							& & 8400 & $111.0\pm11.1$\tnote{b} & B77 & 1.021 & {\citep{1992MNRAS.254..655P}} \\
							& & 5000 & $105.5\pm10.6$\tnote{b,e} & B77 & 1.007 & MASIV {\citep{2008ApJ...689..108L}} \\
							& & 4850 & $151.0\pm13.0$ & B77 & 1.007 & 87GB {\citep{1991ApJS...75.1011G}} \\
							& & 4850 & $154.0\pm14.0$ & B77 & 1.007 & GB6 {\citep{1996ApJS..103..427G}} \\
							& & 1400 & $173.9\pm5.2\,\,\,$ & B77 & 0.972 & NVSS {\citep{1998AJ....115.1693C}}  \\
							& & 325 & $88.0\pm5.7$ & WENSS & - & WENSS {\citep{1997A&AS..124..259R}} \\
							& & 325 & $132.1\pm9.4\,\,\,\,$ & SH12 & - & SCG (this work) \\

\hline

\multicolumn{7}{l}{Flux scale notation:} \\
\multicolumn{7}{p{18cm}}{B77: \protect{\citet{1977A&A....61...99B}} } \\
\multicolumn{7}{l}{RCB73: \protect\citet{1973AJ.....78.1030R} }\\
\multicolumn{7}{l}{SH12: \protect\citet{2012MNRAS.423L..30S}}\\
\end{tabular}
\vspace{0.2cm}
Table notes:
\begin{tablenotes}
	\item[a] The given flux density for 8C~1020+675 from the Miyun survey is the peak flux density, rather than integrated flux density. We exclude this value from Figure~\ref{fig:uss_spix}, but it is quoted here for completeness. 
	\item[b] We have assumed a 10\% uncertainty in this measurement.
	\item[c] Flux scale is consistent with the VLA-CLASS survey.
	\item[d] Flux scale is within 2\% of the B77 scale.
	\item[e] This source was observed on four occasions in 2002 -- 2003, recovering integrated flux densities of 104, 106, 107 and 105 mJy (see \protect\citealt{2008ApJ...689..108L}). We quote the mean flux density here.
	\end{tablenotes}
\end{threeparttable}
}
\end{center}
\end{table*}

\subsubsection{SCG\_J102419+671742}
SCG\_J102419+671742 (also known as 8C~1020+675) has been identified in a number of previous radio surveys between 38 MHz and 1.4 GHz. From Figure~\ref{fig:uss_src} (bottom panel) it is extended at both the resolution of the GMRT and NVSS. Historic values of the flux density for 8C~1020+675 are presented in Table \ref{tab:ussc}, with their associated flux scale and the conversion factor necessary to make them consistent with the \citetalias{2012MNRAS.423L..30S} flux scale. 

Figure~\ref{fig:uss_spix} (bottom panel) also presents the flux density as a function of frequency for this source, along with the multi-frequency spectral index fit. We model the spectral index behaviour of 8C~1020+675 as a power-law between 38 MHz and 1.4 GHz, deriving a spectral index of $\alpha = -1.31\pm0.26$. While this is marginally flatter than the two-point fit listed in Table \ref{tab:uss_src} $(\alpha = -1.41\pm0.07)$ it is sufficiently steep to qualify as a USS radio source. The flux density recovered by the GMRT is in good agreement with the multi-frequency fit to measurements from the literature. 

The postage stamp image of 8C~1020+675 presented in Figure~\ref{fig:uss_src} (lower panel) reveals a complex morphology, with extended radio emission on angular scales up to $\sim120$ arcsec. A pair of potential optical hosts are identified in the DSS, although they lie toward the Southern tail of the radio emission. This source also appears to be coincident with the object GALEXASC J102418.81+671744.1. No redshift measurements are available in the literature so we cannot confirm its size. Detailed optical analysis of this field is underway and may shed further light on the nature of this source.

\subsubsection{Other steep-spectrum candidates}
As can be seen in Figure~\ref{fig:uss_src}, SCG\_J101808+665632 and SCG\_J102253+672121 (panels two and three, respectively) appear reasonably compact at the resolution of the GMRT. SCG\_J101808+665632 has the steepest spectrum of all sources with NVSS counterparts $(\alpha = -1.53\pm0.06)$ and is reasonably bright at 325 MHz (an integrated flux density $S = 59.30\pm4.23$ mJy). These sources have not been detected in any other historic radio surveys, so their steep spectra cannot yet be independently confirmed; likewise no redshift measurements exist, so we cannot yet investigate whether these may be HzRGs. However, being both bright and compact, they present themselves as promising candidates.

It should be noted that we have only examined sources which have counterparts in the NVSS. As such, our current sample of USS sources is severely limited, and many more steep-spectrum objects may be present in the field. The highly-sensitive e-MERLIN and JVLA data being taken may reveal an increased population of very steep spectrum objects, and will allow us to study the ones identified here in more detail. Additionally, a deep optical study of this region is underway as part of ancillary science work, which may provide insight into whether these sources may be HzRGs.

\subsection{Flat-/rising-spectrum Sources}
A source may possess a flat/rising radio spectrum for a number of reasons -- bright sources with rising spectra may be high-redshift quasars or blazars, whereas faint flat-spectrum sources may be star-forming galaxies. A rising low-frequency radio spectrum is usually associated with synchrotron emission from a relativistic electron population as modified by absorption processes. This may be for example free-free absorption (FFA) by a warm gas environment, or synchrotron self-absorption (SSA) of the radio-emitting electrons. 

Given the sensitivity limits of the high-frequency reference, it is unlikely that any of these sources are SFG. Instead, this sample better fits the description of gigahertz-peaked spectrum (GPS) sources. GPS radio sources are a type of AGN whose radio emission peaks at GHz frequencies, usually exhibiting a sharp drop-off either side of the turnover frequency. These sources are typically compact in nature ($\sim0.1-1$ kpc in extent) with low polarization fraction and generally exhibit little variability over long timescales \citep{1991ApJ...380...66O}. For comprehensive reviews of GPS radio sources, see \citet{1998PASP..110..493O}, \citet{2008ApJ...680..911S} or \citet{2014ApJ...780..178M}.

Of the 335 sources common to both the SCG and NVSS catalogues, 42 sources have spectral index values $\alpha > -0.16$ (the $3\sigma_1$ limit of the steep-spectrum population). By inspection, no spatial trend in their distribution is apparent -- these sources appear to be fairly evenly scattered across the survey area. From Figure~\ref{fig:fluxvspix} we identify sources with reliably steep, rising spectra for further analysis, as well as a number of reasonably bright sources with positive spectral indices, for a total of four GPS candidates. The identifiers for these sources, their right ascension and declination, as well as the flux densities at 325 and 1400 MHz, and the two-point spectral index, are also listed in Table \ref{tab:uss_src}. These selected sources all appear compact at the resolution of the GMRT; as a result, we do not present postage stamp images in this work. The first of these, SCG\_J101418+683826 does not appear in any other catalogues besides the NVSS, and cannot yet be investigated further. In this section, we investigate the remainder of this GPS sample.

\subsubsection{SCG\_J101538+672844}
We find a single source with a spectral index more than $3\sigma_2$ above the mean spectral index $(\alpha_2)$ of the flat-spectrum population (i.e. a steep rising spectrum). The two-point spectral index of this source is $\alpha = 1.73\pm0.07$. This source has also been identified in the literature in a number of high-frequency radio surveys; for example, one identifier is CGRaBS J1015+6728 \citep{2008ApJS..175...97H}. We present the flux densities listed in other catalogues, alongside the conversion factor required to bring them onto the \citetalias{2012MNRAS.423L..30S} flux density scale in Table \ref{tab:ussc}.

\begin{figure}
	\centering
	\begin{tabular}{l}
	\subfloat{
		\includegraphics[width=0.46\textwidth]{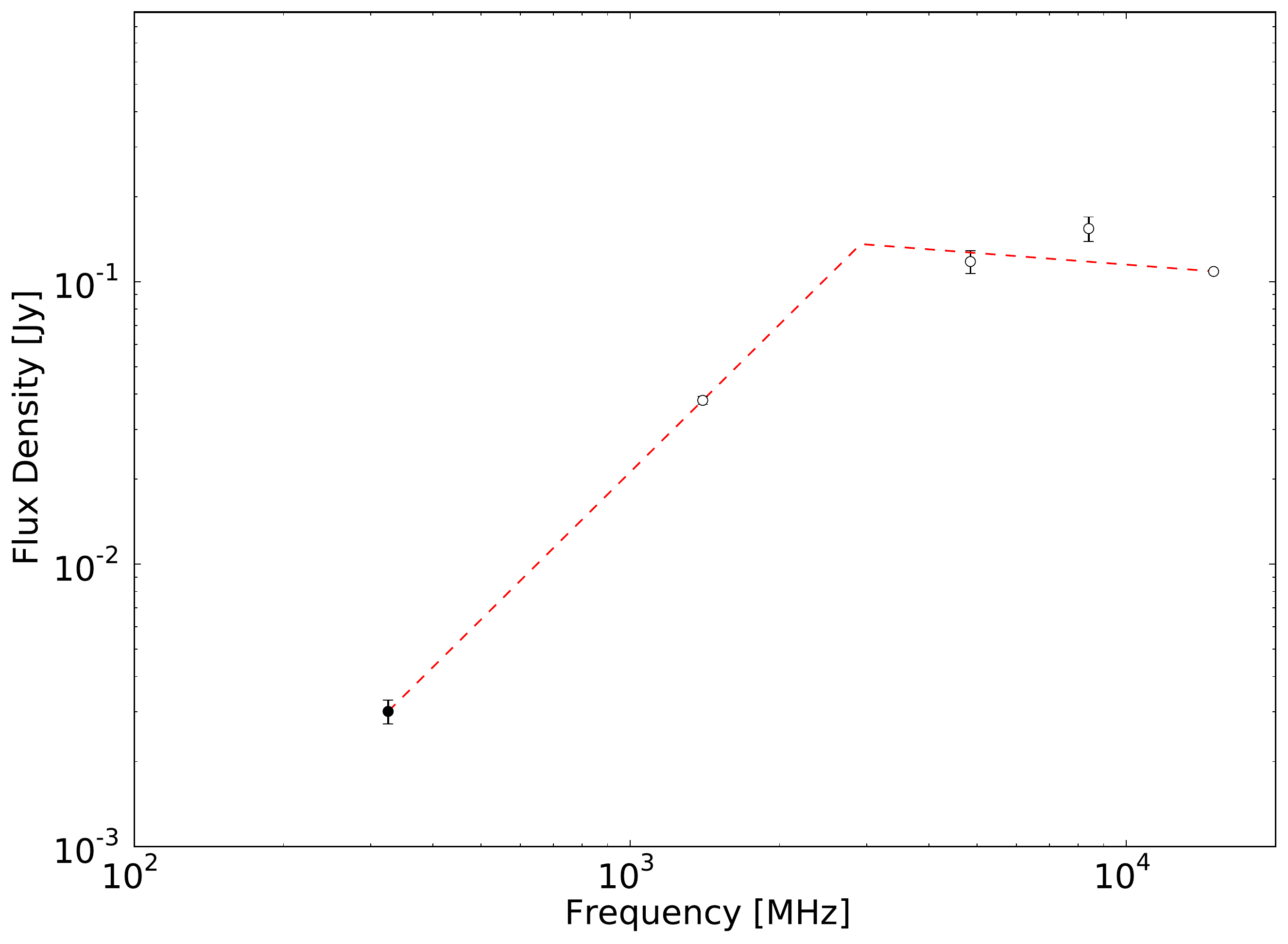}
	} \\
	\subfloat{
		\includegraphics[width=0.46\textwidth]{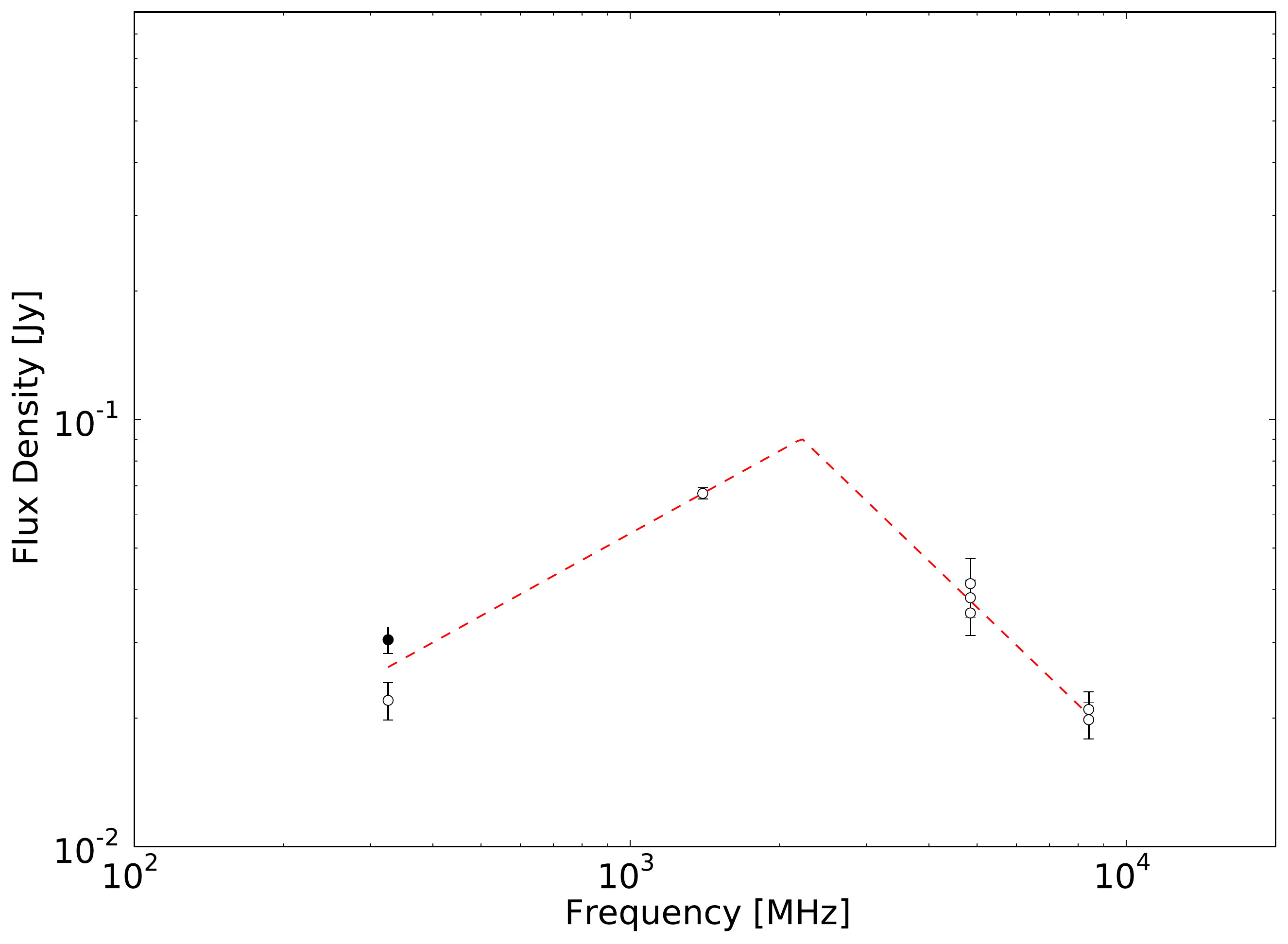}
	} \\
	\subfloat{
		\includegraphics[width=0.46\textwidth]{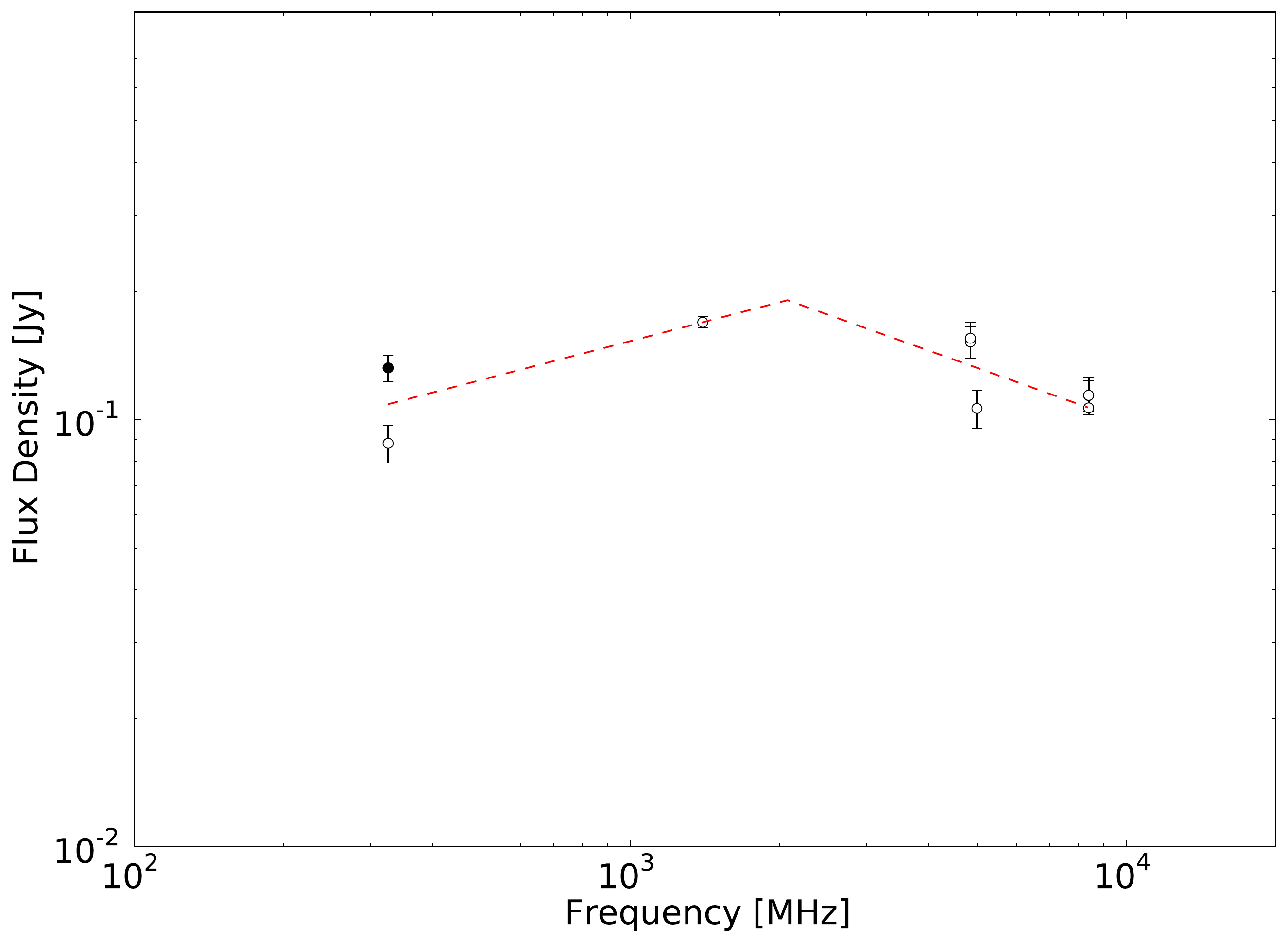}
	} \\
	\end{tabular}
\caption{SEDs for GPS sources identified from the SCG catalogue. \emph{Top:} SCG\_J101538+672844 (CGRaBS J1015+6728). \emph{Centre:} SCG\_J101723+673633 (87GB 101339.1+675144). \emph{Bottom:} SCG\_J103401+683226 (87GB 103023.1+684757). Filled (empty) circles mark data from this work (the literature) as presented in Table \ref{tab:ussc}, with appropriate correction factors applied. Dashed lines denote the broken power law fits to the data, from Table \ref{tab:issc_fits}.}
\label{fig:gps_src}
\end{figure}

Previous work suggests that CGRaBS J1015+6728 is a blazar; flux density measurements from the literature between 1.4 and 15 GHz indicate a rising spectrum with a turnover at higher frequencies (see Table \ref{tab:ussc}). We recover an integrated flux density of $3.09\pm0.31$ mJy; a value that appears consistent with this behaviour. To our knowledge this is the first flux density measurement for this source below 1 GHz. Whilst the SED of CGRaBS J1015+6728 does not show any obvious signs of variability (see the top panel of Figure~\ref{fig:gps_src}) the OVRO database\footnote{Available online at \url{http://www.astro.caltech.edu/ovroblazars/}} indicates that this source has been increasing in flux density steadily since monitoring observations began in 2009.

\subsubsection{SCG\_J101723+673633}
This source is also known in the literature as 87GB 101339.1+675144. Observations of this source have been performed over the course of several decades. From Figure~\ref{fig:gps_src} (centre panel) the flux density measurements appear to be reasonably stable, at least over a period of several years. However, all measurements of the flux density at 4.85 and 8.4 GHz were taken in the period 1986--1999; more recent observations at higher frequencies are required to test the variability on longer timescales. At low frequencies, there is a discrepancy between the flux densities from this work and from WENSS; this may be due to intrinsic variability, or alternatively may be due to the general scatter in flux density ratio between the SCG and WENSS catalogues at low flux densities.

\subsubsection{SCG\_J103401+683226}
This source is also known in the literature as 87GB 103023.1+684757, and has been the subject of a number of observations since the 1980s. From Figure~\ref{fig:gps_src} (lower panel) it appears that the spectrum of SCG\_J103401+683226 is approximately flat between 325 MHz and 8.4 GHz, with signs of variation at high frequencies. Flux density measurements in the 5--8 GHz range appear to be approximately consistent for data taken between 1986--1990 (\citealt{1991ApJS...75.1011G}, \citealt{1992MNRAS.254..655P}, \citealt{1996ApJS..103..427G}) and 1994--1999 (\citealt{2007ApJS..171...61H}). More recent observations from the 2002--2003 period perhaps suggest variation on longer timescales, as the flux density has been observed to decrease at 5 GHz \citep{2008ApJ...689..108L}. 

At lower frequencies, there is a discrepancy in the flux density recovered by the WENSS and this work. For this source, the GMRT recovers flux density in excess of the WENSS by approximately 50 per cent. This may be explained by intrinsic variability, although the general scatter in the flux density ratio between the SCG and WENSS catalogues at low flux densities may have some contribution to this difference.

\subsubsection{Modelling}
Previous work on GPS sources has often attempted to discriminate between single-/multiple-component SSA/FFA models in cases where large numbers of flux density measurements across multiple frequencies exist (for example \citealt{2015ApJ...809..168C}). For the GPS sources discussed in this work, however, there are insufficient measurements in the literature to conduct a thorough investigation. Hence, using all available data from the literature, as presented in Table \ref{tab:ussc}, as well as the new flux density measurements from the SCG catalogue, we model these GPS sources using a simple broken power-law fit of the form:
\[
	S \,\, [\rm{mJy}] = \left\{
	\begin{array}{lr}
		A_0 \nu^{\alpha_{\rm{lo}}} & : \nu < \nu_{\rm{crit}} \\
		B_0 \nu^{\alpha_{\rm{hi}}} & : \nu \geq \nu_{\rm{crit}}
	\end{array}
	\right.
\]
where $\alpha_{\rm{lo}}$ and $\alpha_{\rm{hi}}$ are the low-/high-frequency spectral index, respectively, and $\nu_{\rm{crit}}$ is the turnover frequency. The fitted spectral index values $(\alpha_{\rm{lo}} \,\, {\rm{and}} \,\, \alpha_{\rm{hi}})$ are listed in Table \ref{tab:issc_fits}, as well as the turnover frequency. In future, when the wide-band JVLA and LOFAR data become available, we should be able to investigate the validity of single-/multiple-component SSA/FFA models as the mechanism responsible for the shape of the SED.

\begin{table}
\begin{center}
\caption{Fitted spectral index models for rising-spectrum sources.}
\label{tab:issc_fits}
\begin{tabular}{ccccc}
\hline
 & & \\ 
Source & $\alpha_{\rm{lo}}$ & $\alpha_{\rm{hi}}$ & $\nu_{\rm{crit}}$ \\
& & & [GHz] \\
\hline
SCG\_J101538+672844 & $1.74\pm0.17$ & $-0.14\pm0.18$ & $2.92\pm0.17$ \\
SCG\_J101723+673633 & $0.64\pm0.07$ & $-1.12\pm0.26$ & $2.22\pm0.16$ \\
SCG\_J103401+683226 & $0.30\pm0.10$ & $-0.41\pm0.22$ & $2.07\pm0.35$ \\
\hline
\end{tabular}
\end{center}
\end{table}

We present SEDs for these sources in Figure~\ref{fig:gps_src} as well as the fitted broken power-law spectral index models. For SCG\_J101538+672844, $\alpha_{\rm{hi}}$ is consistent with an essentially flat spectral index above around 3 GHz, whereas $\alpha_{\rm{lo}}$ is strongly indicative of absorption processes, perhaps FFA or SSA, which presents with a typical spectral index of around $\alpha \simeq 2.5$. However, from previous work (for example \citealt{2005MNRAS.358.1159M}, \citealt{2006ApJS..167..103F}, \citealt{2008AJ....136.1889O}) the size of quasars and beamed radio sources is typically $\sim1^{\prime\prime}$, whereas SSA requires scale sizes of $\ll$ mas size (\citealt{2009AJ....137.4846O} and references therein). The resolution of these GMRT data do not allow us to differentiate between these scenarios; the higher-resolution data being taken with other instruments may allow us to better investigate the mechanism responsible for the shape of these spectra.

\section{Source Counts}\label{sec:src}
Source counts as a function of flux density have been extensively studied at 1.4 GHz (for example \citealt{1985ApJ...289..494W}, \citealt{1985AJ.....90.1957M}, \citealt{1997ApJ...475..479W}, \citealt{2003AJ....125..465H}, \citealt{2005AJ....130.1373H}, \citealt{2008ApJ...681.1129B}) and a number of frequencies below 1 GHz (for example \citealt{1991PhDT.......241W}, \citealt{2008MNRAS.383...75G}, \citealt{2008MNRAS.387.1037G}, \citealt{2009MNRAS.395..269S}, \citealt{2009AJ....137.4846O}, \citealt{2013MNRAS.435..650M}, \citealt{2014MNRAS.443.2590S}).

It is well-established that the Euclidean-normalised source counts distribution at 1.4 GHz exhibits a flattening at around 1 mJy; this has been seen in observations at 610 MHz and 325 MHz (see \citealt{2010A&ARv..18....1D} for a review of radio surveys across a wide range of frequencies). A subsequent downturn toward fainter flux densities ($100-150\, \umu$Jy at 1.4 GHz) has also been suggested by \citet{2008ApJ...681.1129B}. Assuming a typical synchrotron spectral index of $\alpha = -0.7$, this corresponds to a flux density of around $180-250 \, \umu$Jy at 610 MHz, or $300-400\, \umu$Jy at 325 MHz. This suggests that the majority of previous surveys at low frequencies have had insufficient sensitivity to confirm/refute this feature.

However, in this work, we recover sources with flux densities down to $183\,\umu$Jy, which suggests we possess sufficient sensitivity to investigate this further. In this section, we discuss the process by which we derive our source counts alongside our treatment of resolution bias and Eddington bias, as well as cosmic variance. Subsequently we analyse the Euclidean-normalised differential source counts distribution derived from the SCG catalogue. 

\subsection{Construction of Source Counts}\label{sec:construction}
Sources were binned according to their integrated flux density, adopting the binning strategy of \cite{hales2014b}. As such, bin widths were set at 0.07 dex for $S < 1 \, {\rm{mJy}}$, 0.13 dex for $1\,{\rm{mJy}} \leq S < 10\,{\rm{mJy}}$, and 0.2 dex for $S \geq 10$ mJy. Subsequently, bins were optimised to achieve raw counts ($N$) of at least 9, and hence maintain a signal-to-noise ratio of at least 3. Given that all sources in our catalogue exceed a flux density of $183 \, \umu$Jy, all our bins are in excess of $5\times$ the nominal image noise. 

\begin{figure}
	\centering
	\includegraphics[width=0.48\textwidth]{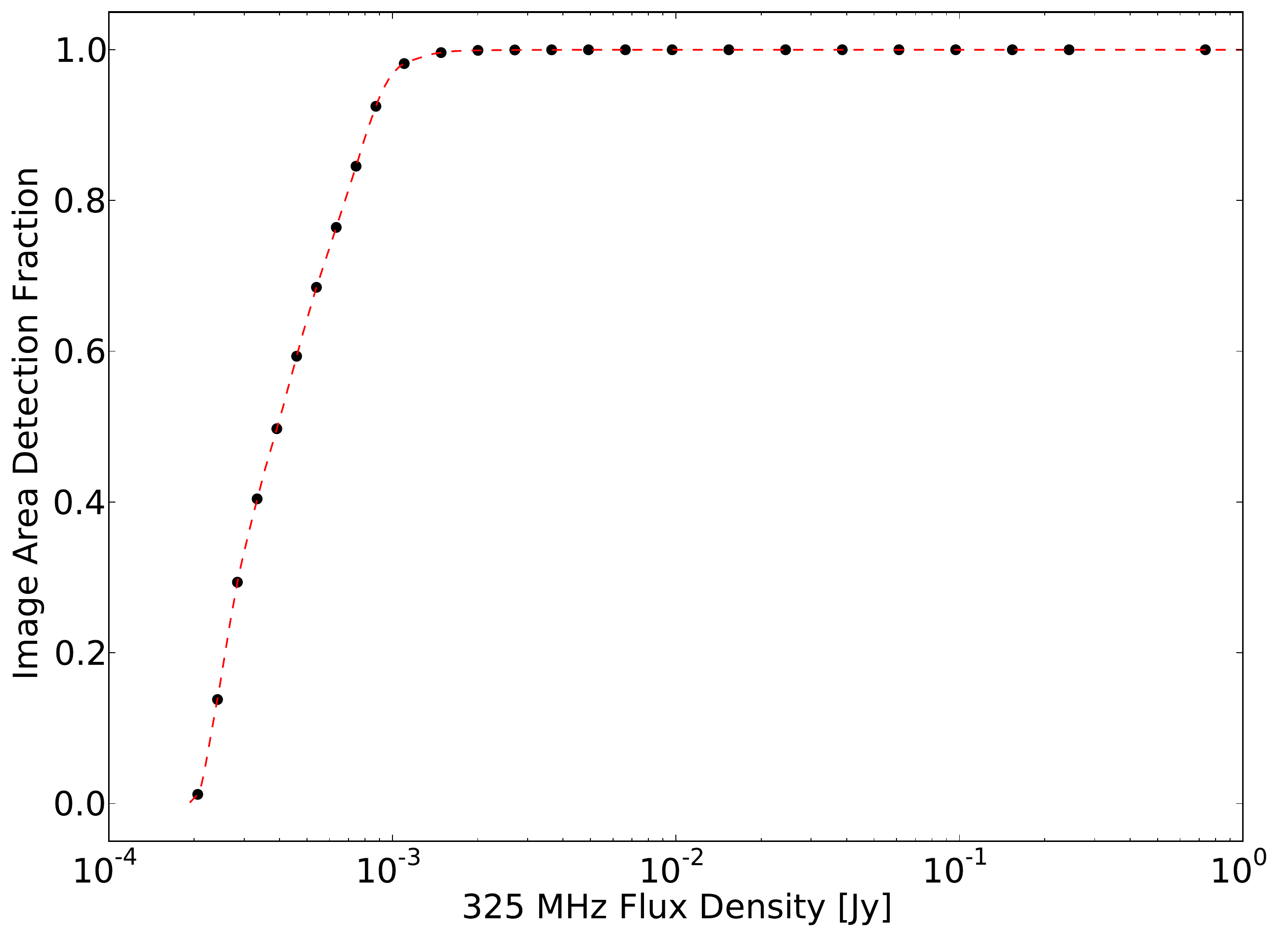}
	\caption{Image area correction factor as a function of flux density at 325 MHz. Black points mark the mean flux density of bins, red dashed line marks the per-source image detection fraction.}
	\label{fig:src_cor}
\end{figure}

The corrected number of sources in a given flux density bin, $N_c$, was found by first deriving the image area detection fraction for each source in our catalogue. This was done using the effective noise image, which is derived from the measured noise (presented in Figure~\ref{fig:mos_sens}) divided by the local bandwidth smearing ratio (see Equation \ref{eq:6} later). The detection fraction as a function of source flux density is presented in Figure~\ref{fig:src_cor}, where the black points indicate the detection fraction for the mean flux density in a given bin. Subsequently, $N_c$ is derived by:
\begin{equation}
	N_c = \sum_{\rm{nsrc}} \left( V_{\rm{area}} \right)^{-1}
\end{equation}
where $V_{\rm{area}}$ is the image detection area for a given source. Note that from Figure \ref{fig:src_cor} the faintest bin $(183 - 223 \,  \umu{\rm{Jy}})$ has a very small image area detection fraction (and therefore a high area correction factor). As such, this bin was excluded from further analysis, as the sources in this bin would only be detected over a very small (and perhaps unrepresentative) region of our survey.

\begin{table*}
\begin{center}
\caption{Differential source counts at 325 MHz. The differential source counts and Euclidean-normalised differential source counts quoted here have all been corrected for image area detection fraction, resolution bias and Eddington bias.}
\label{tab:src_counts}
\begin{tabular}{cccccc}
\hline
& & & & & \\
Flux bin &	$ S_c $ & $N$ &  $N_c$ & $\frac{{\rm{d}}N}{{\rm{d}}S}$ & $\frac{{\rm{d}}N}{{\rm{d}}S} S_c^{2.5}$ \\
$[$mJy$]$ & $[$mJy$]$ & & & $[$Jy$^{-1}$ sr$^{-1}]$ & $[$Jy$^{1.5}$ sr$^{-1}]$\\
\hline
$0.223-0.262$ &  0.242 & 114 & 877.5 & $\,\,1.37\times10^{10}$ & $12.49\pm0.42$ \\
$0.262-0.308$ &  0.284 & 198 & 697.7 & $9.01\times10^{9}$ & $12.26\pm0.46$ \\
$0.308-0.361$ &  0.333 & 242 & 605.6 & $7.06\times10^{9}$ & $14.33\pm0.58$\\
$0.361-0.424$ &  0.391 & 255 & 511.0 & $5.11\times10^{9}$ & $15.48\pm0.68$ \\
$0.424-0.498$ &  0.460 & 237 & 401.5 & $3.37\times10^{9}$ & $15.24\pm0.76$ \\
$0.498-0.585$ &  0.540 & 224 & 329.6 & $2.32\times10^{9}$ & $15.72\pm0.87$ \\
$0.585-0.687$ &  0.634 & 200 & 262.8 & $1.55\times10^{9}$ & $15.73\pm0.97$ \\
$0.687-0.807$ &  0.745 & 206 & 244.3 & $1.21\times10^{9}$ & $18.37\pm1.18$ \\
$0.807-0.948$ &  0.875 & 183 & 198.1 & $8.35\times10^{8}$ & $18.88\pm1.34$ \\
$0.948-1.279$ &  1.10 & 309 & 316.0 & $5.60\times10^{8}$ & $22.55\pm1.27$ \\
$1.279-1.725$ &  1.49 & 229 & 230.1 & $3.01\times10^{8}$ & $25.58\pm1.69$ \\
$1.725-2.327$ &  2.00 & 166 & 166.2 & $1.56\times10^{8}$ & $28.09\pm2.18$ \\
$2.327-3.139$ &  2.70 & 122 & 122.0 & $8.32\times10^{7}$ & $31.61\pm2.86$ \\
$3.139-4.234$ &  3.65 & 105 & 105.0 & $5.22\times10^{7}$ & $41.89\pm4.09$ \\
$4.234-5.712$ &  4.92 & 85 & 85.0 & $3.09\times10^{7}$ & $52.37\pm5.68$ \\
$5.712-7.705$ &  6.63 & 61 & 61.0 & $1.63\times10^{7}$ & $58.26\pm7.46$ \\
$7.705-12.21$ &  9.70 & 76 & 76.0 & $8.86\times10^{6}$ & $82.08\pm9.42$ \\
$12.21-19.35$ &  15.41 & 79 & 79.0 & $5.75\times10^{6}$ & $168.5\pm19.0$ \\
$19.35-30.67$ &  24.36 & 51 & 51.0 & $2.32\times10^{6}$ & $215.2\pm30.1$ \\
$30.67-48.61$ &  38.61 & 30 & 30.0 & $8.57\times10^{5}$ & $251.1\pm45.8$ \\
$48.61-77.04$ &  61.20 & 20 & 20.0 & $3.52\times10^{5}$ & $325.8\pm72.9$ \\
$77.04-122.1$ &  97.0 & 16 & 16.0 & $1.78\times10^{5}$ & $\,\,\,\,520.0\pm130.0$ \\
$122.1-193.5$ &  153.7 & 14 & 14.0 & $9.80\times10^{4}$ & $\,\,\,\,908.0\pm242.7$ \\
$193.5-306.7$ &  243.6 & 9 & 9.0 & $3.97\times10^{4}$ & $1164.3\pm388.1$ \\
$\,\,\,\,\,306.7-1770.8$ & 737.0 & 11 & 11.0 & $3.76\times10^{3}$ & $1751.2\pm528.0$ \\
\hline
\end{tabular}

\end{center}
\end{table*}

The differential source counts $n(S) = {\rm{d}}N / {\rm{d}}S$ were then calculated by dividing the corrected number of sources in each bin, $N_c$, by $A\Delta S$, where $A$ is the image area (in steradians) and $\Delta S$ is the bin width (in Jy). We derived the source counts using the full image area, approximately 6.5 square degrees (0.002 sr). The bins, raw counts, corrected counts, differential source counts, and Euclidean-normalised differential source counts $n(S)\, S^{2.5}$, are presented in Table \ref{tab:src_counts}. The Euclidean normalisation is performed using the geometric mean flux density, $S_c$, of sources in each bin. We have assumed Poisson statistics in deriving the errors on the source counts. The differential source counts presented in Table \ref{tab:src_counts} have been corrected for both resolution bias and Eddington bias, as described in the following sections.

\subsection{Cosmic Variance}
We can estimate the effect of cosmic variance using the work of \citet{2013MNRAS.432.2625H}. With a survey area of $\sim6.5$ square degrees, this is the widest deep-field yet studied at 325 MHz; as such we would expect the least uncertainty due to cosmic variance. Assuming a spectral index $\alpha = -0.7$, and given the field of view over which the source counts were derived we would expect the uncertainty introduced by cosmic variance effects to be at the order of $1-3$ per cent at the flux density limit of this survey; a value small compared to other uncertainties.

\subsection{Resolution Bias}
Given that source catalogues are compiled using detection algorithms are largely based on identification of sources at a given significance above the noise level -- i.e. based on peak flux density -- whereas source counts are derived using the integrated flux density, the resolution of a survey can significantly impact the recovered source counts. At increasingly high resolution, surveys may resolve out sources that possess low surface brightness, but are sufficiently large that their integrated flux density would contribute to the source counts distribution. 

\defcitealias{hales2014a}{H14}

Recent work by \cite{hales2014a} (hereafter \citetalias{hales2014a}) has provided a robust formalism for estimating the correction required to account for resolution bias. \citetalias{hales2014a} include two effects under the resolution bias umbrella: firstly the incompleteness due to lack of sensitivity to resolved low surface brightness sources, and secondly the redistribution of source counts between bins as a result of underestimating flux densities for unresolved sources. This second component of resolution bias was identified by \cite{2008ApJ...681.1129B}. In this section we will closely follow the method of \citetalias{hales2014a} to derive the correction for resolution bias.

\subsubsection{Effect 1: Sensitivity to Resolved Sources}
In our work, we possess modest resolution (13 arcsec) and use a weighting that achieves compromise between sensitivity and resolution (\texttt{AIPS} \texttt{ROBUST} $-1$). Additionally, with a well-sampled \emph{uv}-plane that has good coverage on short baselines, at 325 MHz the GMRT is sensitive to structures up to around 32 arcminutes in extent. Given that no mJy or sub-mJy sources are expected to exhibit angular sizes on these scales, we would expect no limitation on the size of source that can be recovered due to our \emph{uv}-coverage.

The other aspect of this first effect is that sources with sufficiently high integrated flux density to contribute to the source counts may be resolved to the extent that the peak falls below the detection limit of the source detection algorithm. In order to correct for this effect, we compare the maximum recoverable angular scale with the underlying size distribution to estimate the fraction of sources missed as a function of flux density. For a source of a given flux density, there exists a maximum angular size that can still be recovered by our catalogue. This is given by: 
\begin{multline} \label{eq:4}
	[ \Theta_{\rm{max}}(S) ]^2 = \Bigg\{ \int_0 ^{S/A_S} \sqrt{ \frac{S \, B_{\rm{maj}} \, B_{\rm{min}}}{A_S \, z} } f_{\tilde{\sigma}}(z) {\rm{d}}z \\ 
	\times \left[\int_0 ^{S/A_S}  f_{\tilde{\sigma}}(z^{\prime}) {\rm{d}}z^{\prime} \right]^{-1} \Bigg\}^2 - B_{\rm{maj}} \, B_{\rm{min}} 
\end{multline}
where $\Theta_{\rm{max}}(S)$ is the maximum deconvolved angular size at flux density $S$, $A_S$ is the signal-to-noise ratio threshold (in this work, 5.0) and $B_{\rm{maj}}$ and $B_{\rm{min}}$ are the major and minor axis of the restoring beam, respectively. The \emph{effective} noise at a given position, $\tilde{\sigma}(x,y)$ is defined as the local rms noise divided by the local bandwidth smearing ratio, i.e. $\tilde{\sigma}(x,y) = \sigma(x,y) / \overline{\omega}(x,y)$.  By definition, $f_{\tilde{\sigma}}$ is a probability distribution function for $\tilde{\sigma}$; in practice this takes the form of a normalised histogram of $\tilde{\sigma}$ values.

In their work, \citetalias{hales2014a} present the formalism for deriving the bandwidth smearing correction in the case of a non-circular restoring beam. For a source at a given position angle, $\zeta$ (in degrees East of North) with respect to the phase centre of observations, the projected beam is given by:
\begin{equation} \label{eq:5}
	B_{\rm{proj}}(\zeta) = \frac{ B_{\rm{maj}} \, B_{\rm{min}} }{ \sqrt{\left[ B_{\rm{maj}} \sin( \zeta - \psi ) \right]^2 + \left[ B_{\rm{min}} \sin( \zeta - \psi ) \right]^2} }
\end{equation}
where $\psi$ is the position angle of the restoring beam \citepalias{hales2014a}. In this work, our mosaic was created using a common circular restoring beam of FWHM 13 arcsec, hence $ B_{\rm{maj}} = B_{\rm{min}} = 13$ arcsec, and $\psi = 0$. Equation \ref{eq:5} was used to derive the effect of bandwidth smearing using the formalism from \citetalias{hales2014a} via
\begin{equation} \label{eq:6}
	\overline{\omega}(x,y) = \frac{S_{\rm{peak}}}{S^0_{\rm{peak}}} = \left\{ 1 + \frac{2\,\ln 2}{3} \left[ \frac{\delta\nu_{\rm{eff}}}{\nu_{\rm{ref}}} \frac{d}{B_{\rm{proj}}} \right] \right\}^{-1/2}
\end{equation}
where $S_{\rm{peak}}$ and $S^0_{\rm{peak}}$ are the measured and true peak flux densities, respectively, and $d$ is the distance from the pointing centre. In this work, the effective channel width $\delta\nu_{\rm{eff}} = 520.8$ kHz. Additionally, the reference frequency at which calibration solutions were derived $\nu_{\rm{ref}} = 322.9$ MHz. Equation \ref{eq:6} was evaluated for each pixel of the individual pointing images, and subsequently mosaicked using the same procedure and weighting as for the field images. The resulting bandwidth smearing mosaic yields $S_{\rm{peak}} \, / \, S^0_{\rm{peak}} > 0.96$ everywhere, with a typical smearing of less than 2 per cent. 

\citetalias{hales2014a} model the true size distribution for total intensity \emph{components} using a modified version of the integral angular size distribution presented by \citet{1990ASPC...10..389W} for \emph{sources} at 1.4 GHz. From \cite{1990ASPC...10..389W} the fraction of sources with a largest angular size (LAS) larger than $\Theta$ is given by
\begin{equation} \label{eq:7}
	h (> \Theta, S) = 2^{-(\Theta/\Theta_{\rm{median}})^{0.62}}
\end{equation}
where $\Theta_{\rm{median}}$ is the median LAS as a function of flux density. For single component sources, \citetalias{hales2014a} find their data is better fit by a model that predicts a median LAS that is half the size predicted by \cite{1990ASPC...10..389W}. \citetalias{hales2014a} give the density function associated with Equation \ref{eq:7} as 
\begin{equation} \label{eq:8}
	f_{\Theta}(\Theta, S) = \frac{0.62 \, \ln 2}{\Theta_{\rm{med}}} \left( \frac{\Theta}{\Theta_{\rm{med}}} \right)^{-0.38} \, h (> \Theta, S)
\end{equation}
The angular size distribution for sources at 325 MHz is less well explored in the literature than at higher frequency, particularly at increasingly faint flux densities. \cite{2009AJ....137.4846O} report higher resolution data ($6.37 \times 5.90$ arcsec) at the same frequency considered in this work; they find a median size distribution consistent with the 1.4\,GHz work by \cite{2003NewAR..47..357W} scaled to 325\,MHz. Recent work with LOFAR by \cite{2016arXiv160501531W} also evaluated the effect of resolution bias using the \cite{2003NewAR..47..357W} distribution scaled to 150 MHz.

With a resolution of 13 arcsec, our observations are unable to probe the angular size distribution of faint sources further than \cite{2009AJ....137.4846O}. The resolution of our survey is far more comparable to the resolution of \citetalias{hales2014a}. For this derivation we adopt the same formalism for the angular size -- flux density relation as \citetalias{hales2014a}:
\begin{equation} \label{eq:9}
	\Theta_{\rm{median}} = X \, \rm{arcsec} \, \left( \frac{ S_{\rm{1.4\,GHz}} }{1\,\rm{mJy}}  \right)^{0.30}
\end{equation}
where $S$ denotes the source flux density. In their work, \citetalias{hales2014a} use $X = 1$ arcsec, whereas \cite{2003NewAR..47..357W} use $X = 2$ arcsec. From theory we would expect that sources become more extended toward lower frequencies, as the lower-energy electrons have longer radiative lifetimes, and may travel further from their place of origin. As such, we tested two versions of this distribution, using $X = 1 \, (2)$ arcsec as per \citetalias{hales2014a} \citep{2003NewAR..47..357W} with flux densities scaled to 325 MHz using a spectral index $\alpha = -0.8$, following \cite{2016arXiv160501531W}. We find that using $X = 2$ arcsec provided a correction effect that brought our counts more in line with what is predicted by models and expected from other observations with similar sensitivity.

From \citetalias{hales2014a} therefore, the overall correction for incompleteness due to highly-extended sources is given by
\begin{equation} \label{eq:10}
	\frac{{\rm{d}}N_{\rm{detectable}}}{{\rm{d}}S} = \frac{{\rm{d}}N_{\rm{true}}}{{\rm{d}}S} \times \left\{1 - h (> \Theta_{\rm{max}}(S), S) \right\}
\end{equation}
where ${\rm{d}}N_{\rm{true}}/{\rm{d}}S$ is the \emph{true} source counts distribution, and ${\rm{d}}N_{\rm{detectable}}/{\rm{d}}S$ is the observable source counts (i.e. sources with angular size $\Theta \leq \Theta_{\rm{max}}$). Consequently, the overall correction factor for this first effect, $r_{\rm{effect\,1}}$, is given by

\begin{equation} \label{eq:11}
	r_{\rm{effect\,1}}(S) = \frac{{\rm{d}}N_{\rm{detectable}}}{{\rm{d}}S}(S) \div \frac{{\rm{d}}N_{\rm{true}}}{{\rm{d}}S}(S)
\end{equation}

\subsubsection{Effect 2: Underestimation of Flux Density for Unresolved Sources}
The second form of resolution bias arises as a result of underestimation of the flux density of sources classified as unresolved. From simulations, \cite{2008ApJ...681.1129B} found that a significant fraction of sources with flux densities below $150\,\umu$Jy were redistributed to \emph{lower} flux densities as a result of poor signal-to-noise prohibiting full deconvolution. \citetalias{hales2014a} derive the formalism for correcting for this effect.

In this work, we have used a different function (Equation \ref{eq:locus}) than that of \citetalias{hales2014a} to describe the locus used to differentiate between resolved and unresolved sources. For a given flux density $S$, we find the \emph{minimum} angular size $ \Theta_{\rm{min}}$ required for a source to be classified as resolved is given by
\begin{multline} \label{eq:12}
	[ \Theta_{\rm{min}}(S) ]^2 = \Bigg\{ \int_0 ^{S/A_S} \sqrt{ B_{\rm{maj}} \, B_{\rm{min}} \, k^{(S/z)^{-c}} } f_{\tilde{\sigma}}(z) {\rm{d}}z \\ 
	\times \left[\int_0 ^{S/A_S}  f_{\tilde{\sigma}}(z^{\prime}) {\rm{d}}z^{\prime} \right]^{-1} \Bigg\}^2 - B_{\rm{maj}} \, B_{\rm{min}} 
\end{multline}
where $k$ and $c$ are the fit parameters of the locus that allows us to determine which sources are unresolved (see Equation \ref{eq:locus} in \S\ref{sec:size}) and the remaining parameters are defined as for Equation \ref{eq:4}. The differential number counts for detectable resolved sources are given by
\begin{align} \label{eq:13}
	\frac{{\rm{d}}N_{\rm{resolved}}}{{\rm{d}}S}(S) = & \frac{{\rm{d}}N_{\rm{detectable}}}{{\rm{d}}S}(S) \nonumber \\
	& \times [ \, h (> \{ {\rm{min}}(\Theta_{\rm{min}}(S), \Theta_{\rm{max}}(S) )\}, S)  \nonumber \\
	& - h (> \Theta_{\rm{max}}(S), S ) \, ] \nonumber \\
	& \div [ \, 1 - h (> \Theta_{\rm{max}}(S), S ) \, ] 
\end{align}
Assuming the flux density recovered for resolved sources is equal to their true flux density, we can predict the differential source counts for detectable, unresolved components via
\begin{equation} \label{eq:14}
	\frac{{\rm{d}}N_{\rm{unresolved}}}{{\rm{d}}S}(S) = \frac{{\rm{d}}N_{\rm{detectable}}}{{\rm{d}}S}(S) - \frac{{\rm{d}}N_{\rm{resolved}}}{{\rm{d}}S}(S)
\end{equation}
In the absence of any measurement error, for an unresolved source the integrated flux density is equal to the integrated surface brightness recovered by our instrument. However, from Figure~\ref{fig:peakvint} it can be seen that some sources have been recovered which appear to have an integrated flux density $S_{\rm{int}}$ less than the peak flux density $S_{\rm{peak}}$. These sources have clearly been affected by noise, and we have assumed $S_{\rm{int}} = S_{\rm{peak}}$. However, this is based on the assumption of zero intrinsic angular size, whereas sources always have non-zero physical size. As such, sources are pushed toward fainter flux density bins \citepalias{hales2014a}. This effect can be modelled using the formalism of \citetalias{hales2014a} via
\begin{align} \label{eq:15}
	\frac{{\rm{d}}N_{\rm{unresolved, observed}}}{{\rm{d}}S}(S) = &  \int_S^{\infty} \frac{{\rm{d}}N_{\rm{unresolved}}}{{\rm{d}}S^{\prime}}(S^{\prime}) \,\, H\left(\tilde{\Theta} - \Theta^{\prime} \right) \nonumber \\
	& \times \frac{f_{\Theta^{\prime}}(\Theta^{\prime}, S^{\prime}) }{ \int_0^{\tilde{\Theta}} f_{\Theta}(\Theta^{\prime\prime}, S^{\prime}) {\rm{d}} \Theta^{\prime\prime}} {\rm{d}}S^{\prime}
\end{align}
where $H(x)$ is a step function, given by
\begin{equation} \label{eq:16}
	H(x) = \left\{
	\begin{array}{lr}
		1 & : x \geq 0 \\
		0 & : x < 0
	\end{array}
	\right.
\end{equation}
and $\Theta^{\prime}$ and $\tilde{\Theta}$ are defined by
\begin{align}\label{eq:17}
	& \Theta^{\prime} = \sqrt{ B_{\rm{maj}} \, B_{\rm{min}} \left( \frac{S^{\prime}}{S} -1 \right) }
\end{align}
and
\begin{align}\label{eq:18}
	& \tilde{\Theta} = {\rm{min}} \left[ \Theta_{\rm{min}}, \Theta_{\rm{max}} \right]
\end{align}
To derive the overall correction factor for this second effect, we first derive the minimum angular size using Equation \ref{eq:12} for test flux densities across the entire range covered by our survey. Subsequently we use this minimum size as well as the maximum size given by Equation \ref{eq:4} to evaluate Equations \ref{eq:13} and \ref{eq:15}, and use these to derive the correction for this second effect via
\begin{align}\label{eq:19}
	r_{\rm{effect\,2}}(S) =  & \frac{{\rm{d}}N_{\rm{detectable}}}{{\rm{d}}S}(S) \nonumber \\
			& \div \left[ \frac{{\rm{d}}N_{\rm{resolved}}}{{\rm{d}}S}(S) + \frac{{\rm{d}}N_{\rm{unresolved, observed}}}{{\rm{d}}S}(S) \right]
\end{align}

\subsubsection{Overall Resolution Bias Correction}
Following \citetalias{hales2014a} therefore the overall correction factor which accounts for both forms of resolution bias, $R(S)$ is given by 
\begin{align}\label{eq:20}
	R(S) = r_{\rm{effect\,1}}(S) \times r_{\rm{effect\,2}}(S)
\end{align}
Many previous low-frequency radio surveys have not achieved the depth where the effect of resolution bias becomes significant, or have typically only applied corrections for image area detection fraction. 

\defcitealias{massardi2010}{M10}

\begin{figure*}
	\centering
	\subfloat{
		\includegraphics[width=0.5\textwidth]{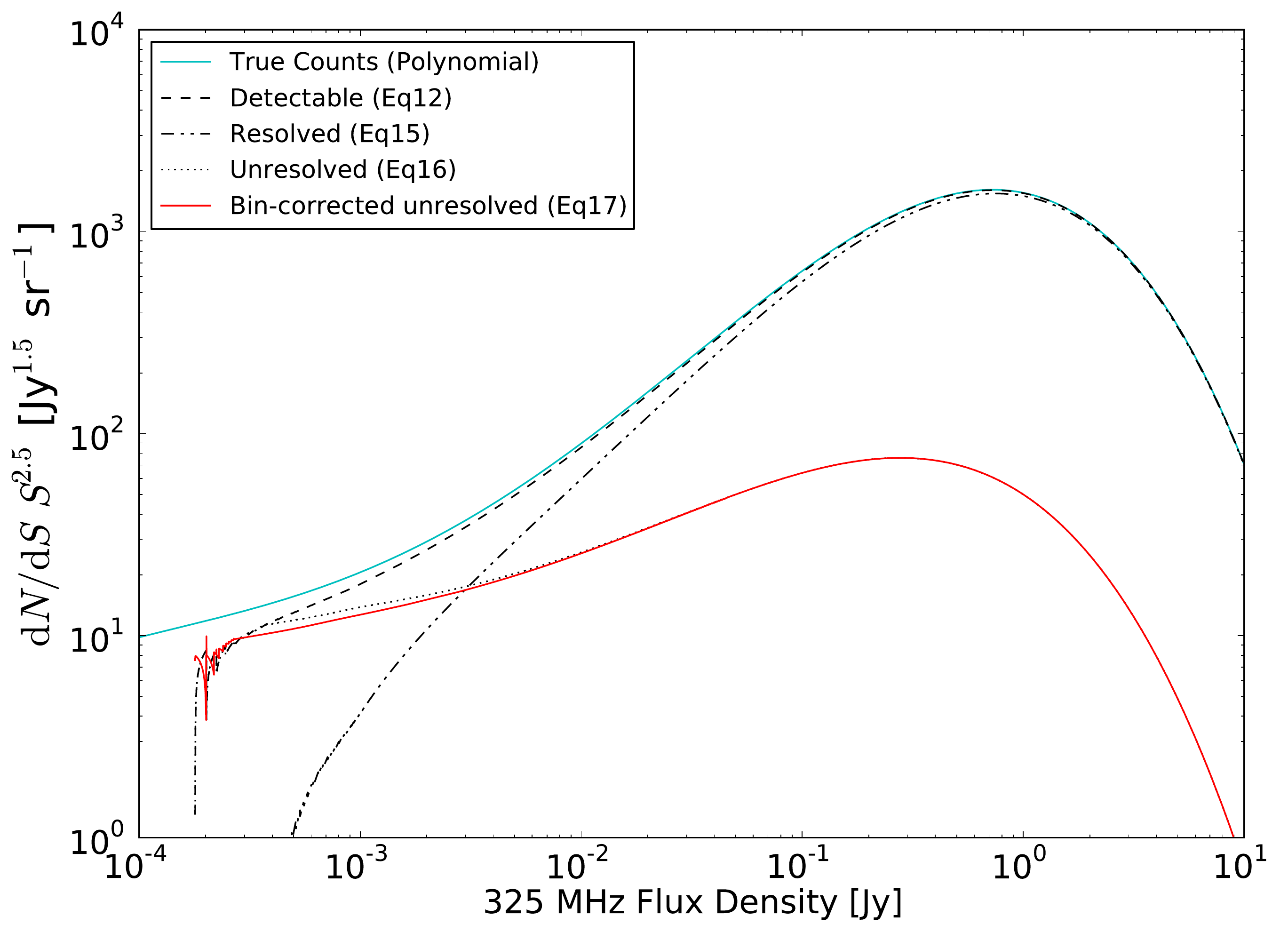}
	} 
	\subfloat{
		\includegraphics[width=0.486\textwidth]{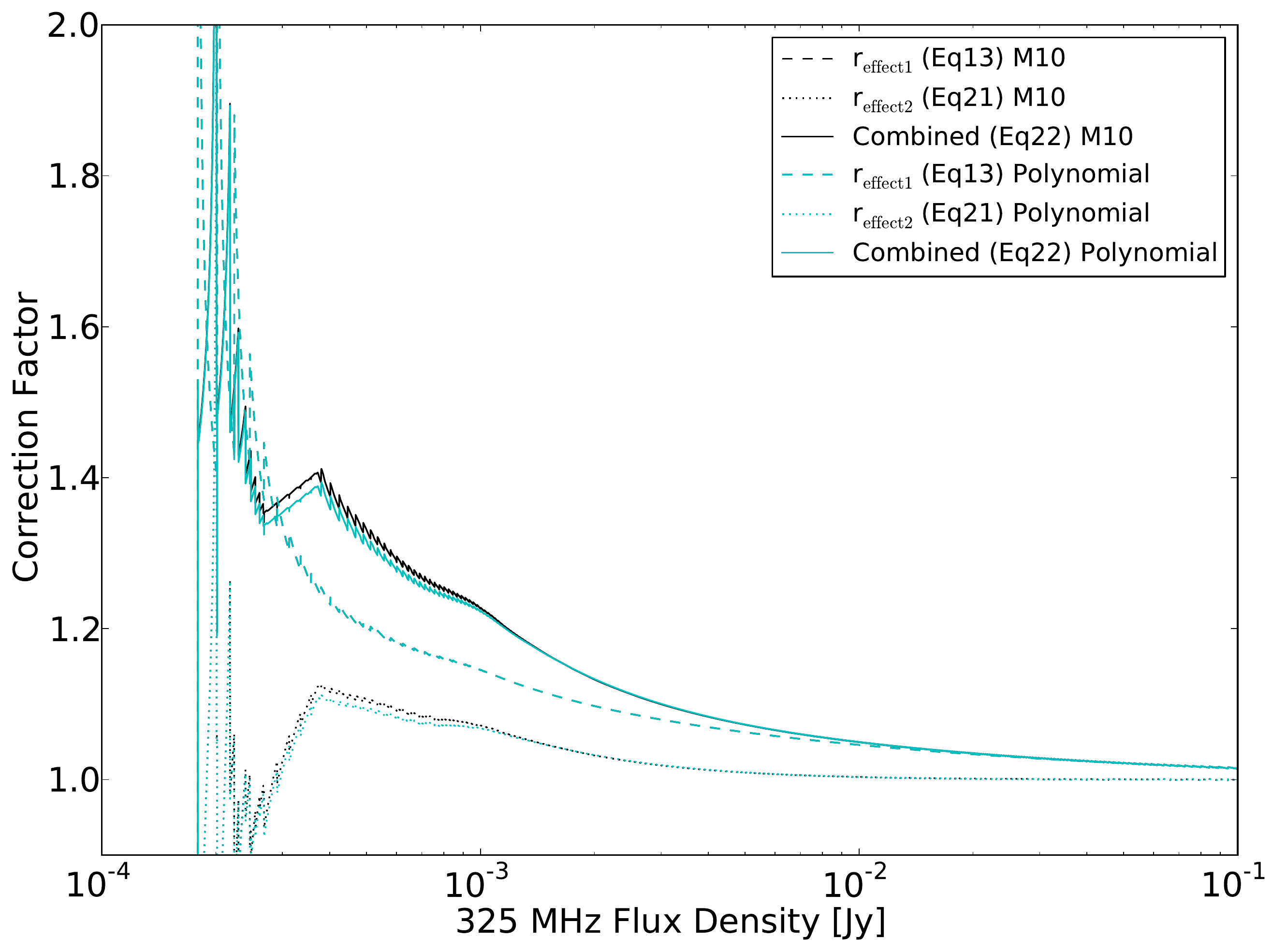}
	}
	\caption{\emph{Left panel:} the effect of resolution bias on differential source counts, showing the following results: true source counts model from the polynomial model (cyan), detectable source counts (Equation \ref{eq:10}, dashed), resolved sources (Equation \ref{eq:13}, dot-dashed), unresolved sources assuming no systematics (Equation \ref{eq:14}, dotted) and unresolved sources corrected for flux density underestimation (Equation \ref{eq:15}, red). \emph{Right panel:} resolution bias correction factors, to account for incompleteness due to highly-extended sources (Equation \ref{eq:11}, dashed line), flux density underestimation for resolved sources (Equation \ref{eq:19}, dotted line) and the overall combined correction factor (Equation \ref{eq:20}, solid line). Black (cyan) curves denote the corrections derived assuming the \citetalias{massardi2010} (polynomial fit) model.}
	\label{fig:res_bias}
\end{figure*}

\subsubsection{Evaluating the Resolution Bias Correction}
\cite{massardi2010} (hereafter \citetalias{massardi2010}) present a comprehensive evaluation of the various source populations recovered by radio surveys across a wide range of frequencies. The \citetalias{massardi2010} model includes the contribution from AGN-type sources (BL-Lac objects, flat-spectrum radio quasars and steep-spectrum objects) as well as star-forming galaxies (SFG) and starburst galaxies (SBG). The 325 MHz model suggests a source counts distribution that flattens below 1 mJy, as has been detected in many previous surveys. However, the raw counts from our work (presented later) suggest a source counts distribution that drops off significantly at the faintest flux densities. In order to investigate the effect of resolution bias, we derive the corrections assuming two models. Firstly, we derive the corrections assuming the model of \citetalias{massardi2010}. Following \cite{2003AJ....125..465H} we also derive a fourth-order polynomial fit of the form:

\begin{equation} \label{eq:poly}
	\log \left[ \left( {\rm{d}}N / {\rm{d}}S \right) \, S^{2.5} \right] = \sum_{i = 0}^4 \, a_i \, \left[ \log(S) \right]^i
\end{equation}
where $S$ is the flux density, measured in Jy. In order that our corrections converge to a stable result, we iterated through the polynomial fitting / bias correction derivation process three times. The coefficients of the final fit are as follows: $a_0 = 3.192\pm0.172$, $a_1 = -0.223\pm0.428$, $a_2 = -0.846\pm0.361$, $a_3=-0.261\pm0.120$, $a_4 = -0.024\pm0.013$. This fit is valid for flux densities between $242 \, \umu$Jy and 0.74 Jy; additional data at higher flux densities would be required to improve the fit in bright source regime.

The equations presented in this section were evaluated with the chosen model in place of ${\rm{d}}N_{\rm{true}}/{\rm{d}}S$. The modelled effects of resolution bias on the differential source counts (assuming the polynomial fit describes the underlying counts distribution) are presented in the left-hand panel of Figure~\ref{fig:res_bias}, and the overall correction factor for resolution bias (for both the polynomial model and the \citetalias{massardi2010} model) in the right-hand panel of Figure~\ref{fig:res_bias}.

From Figure~\ref{fig:res_bias}, it appears that incompleteness due to sensitivity to resolved sources is in excess of the five per cent level at around 10 mJy, and becomes more significant below around 1 mJy. It also appears that the redistribution of sources to different flux density bins is most significant for flux densities between a few mJy and a few hundred $\umu$Jy. This is exemplified by the correction factor in the right-hand panel of Figure~\ref{fig:res_bias}, where the overall correction factor rises with decreasing flux density, before falling, and then subsequently rising again as incompleteness dominates at the very faintest flux densities. 

The resolution bias correction profile exhibited in Figure~\ref{fig:res_bias} is similar to that derived by \citetalias{hales2014a}, including the oscillations toward the very faintest flux densities. These are due to the sensitivity across the field, exhibited in Figure \ref{fig:mos_sens}, which appears to vary more significantly than the sensitivity in \citetalias{hales2014a}. However, our results should not be affected by the large oscillations toward very faint flux densities in Figure \ref{fig:res_bias} as our faintest bin has a typical flux density of $242 \, \umu$Jy, a regime where the gradient across the bin should not be too significant. Resolution bias corrections were applied per-bin when the effect was at the two per cent level or higher.


\subsection{Eddington Bias}
As source number counts drop off rapidly with increasing flux density, it is more likely that noise will scatter fainter sources to higher flux densities rather than vice versa. This is known as Eddington bias, and we naturally expect this effect to be most significant at the faint flux density limit of our survey, where number counts are higher. In general, previous work at higher frequency has quantified Eddington bias using semi-empirical methods. For example, \citet{2007MNRAS.378..995M} use a source counts model fitted to their data to generate a population of sources below the detection limit. Subsequently, \cite{2007MNRAS.378..995M} derive counts from this population and use the difference between the recovered population and input model to quantify the Eddington bias. 

\citetalias{hales2014a} also present a robust formalism for deriving the correction due to Eddington bias. We computed the effect of Eddington bias using the formalism of \citetalias{hales2014a} by assuming the differential source counts model from \citetalias{massardi2010}, including contributions from AGN, SFG and SBG. As with the resolution bias, we also derived the corrections using our sixth order polynomial fit, iterating through the fit / correction process three times. From \citetalias{hales2014a}, the proportion of components with a \emph{true} flux density $S + \epsilon$ that are observed to have a flux density $S$ resulting from a Gaussian measurement error $-\epsilon$ is given by:

\begin{figure*}
	\centering
	\subfloat{
		\includegraphics[width=0.51\textwidth]{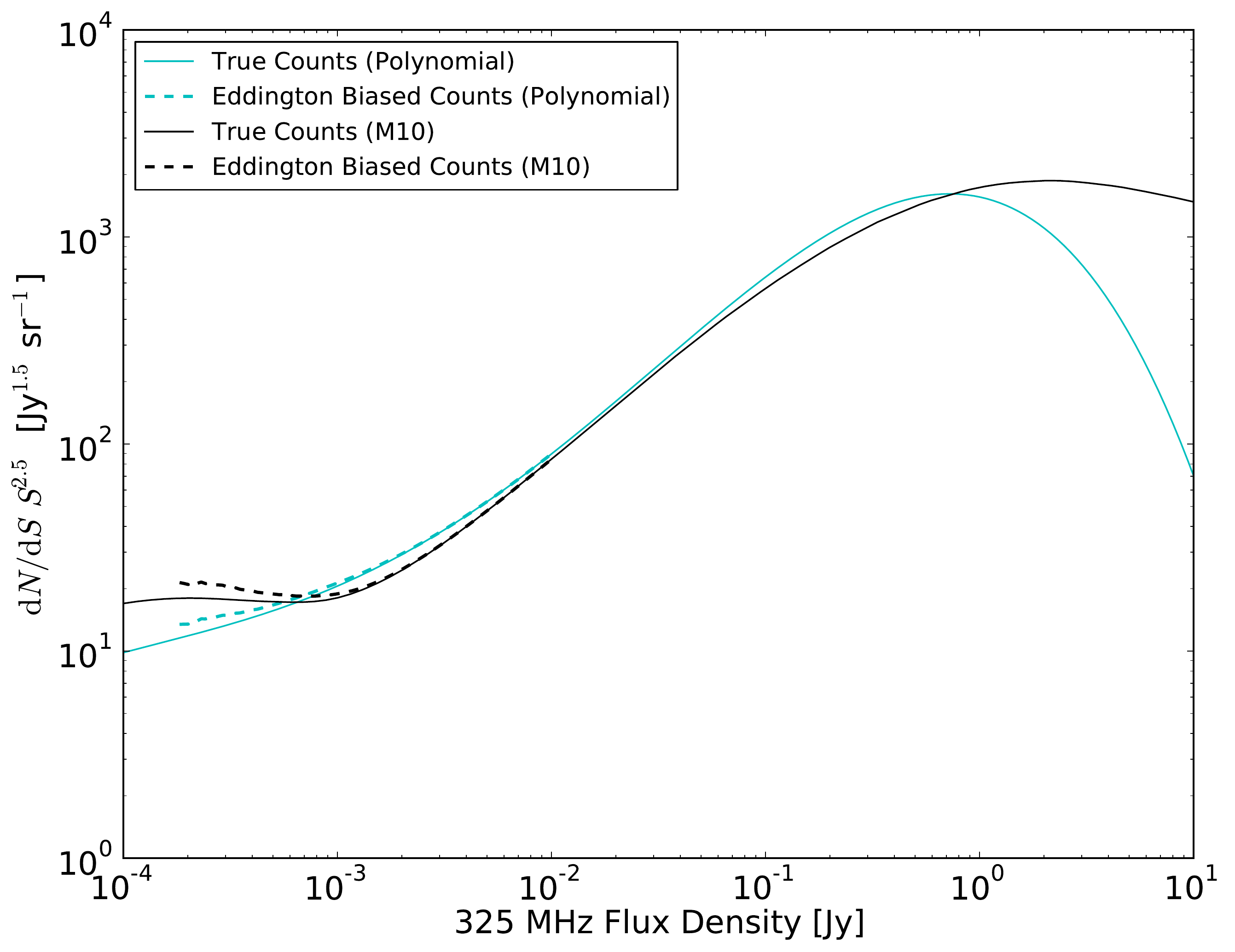}
	} 
	\subfloat{
		\includegraphics[width=0.485\textwidth]{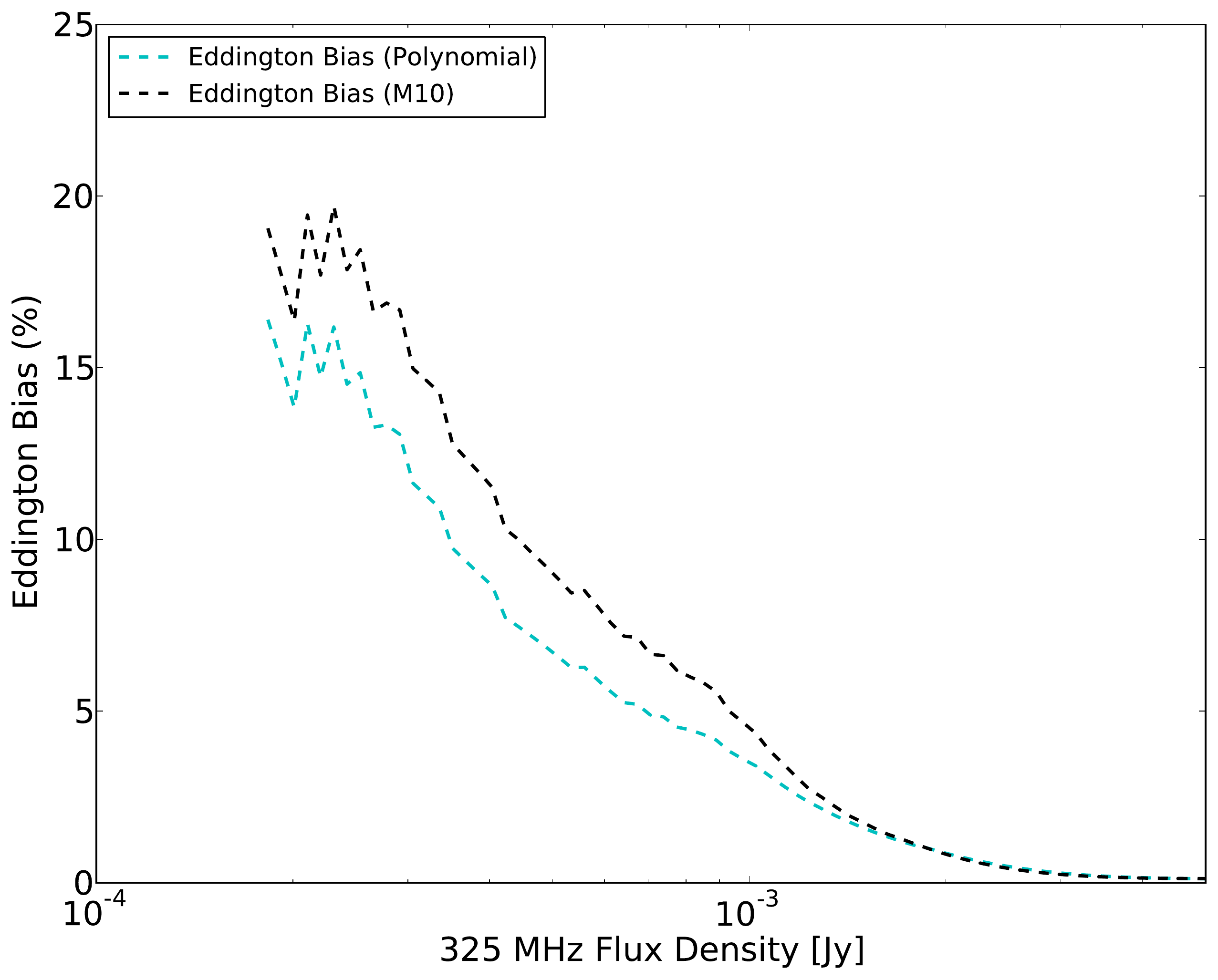}
	}
	\caption{\emph{Left panel:} the effect of Eddington bias on differential source counts. The cyan (black) curves indicate the source counts for the polynomial fit \citepalias{massardi2010} model. Dashed curves indicate the modelled effect of Eddington bias, solid curves indicate the underlying distribution. \emph{Right panel:} the percentage effect of Eddington bias on the source counts model for the region below 5 mJy, where Eddington bias becomes significant. Cyan (black) curves again indicate the polynomial \citepalias{massardi2010} model described by Equation \ref{eq:poly}.}
	\label{fig:edd_bias}
\end{figure*}

\begin{align}\label{eq:21}
	\frac{ {\rm{d}}N_{\rm{Edd}} }{ {\rm{d}}S }\left(S\right) & = \int_{-\infty}^{\infty} \int_{0}^{z^{\prime\prime}} \frac{1}{\sqrt{2\pi}} \exp \left( \frac{-\xi^2}{2} \right)  \nonumber \\
		& \times \frac{{\rm{d}}N_{\rm{True}}}{{\rm{d}}S} \left(S+\xi z \right) \, \frac{ f_{\tilde{\sigma}}(z) }{ \int_0 ^{z^{\prime\prime}}  f_{\tilde{\sigma}}(z^{\prime}) {\rm{d}}z^{\prime} } \, {\rm{d}}z \, {\rm{d}} \xi
\end{align}
where ${\rm{d}}N_{\rm{Edd}} / {\rm{d}}S$ denotes the biased source counts. The limit of the integral over $z$ is given by:
\begin{equation} \label{eq:22}
	z^{\prime\prime} = \left\{
	\begin{array}{lr}
		S/A_S & : \xi \geq 0 \\
		{\rm{min}} \left(-S+\xi, \, S/A_S \right) & : \xi < 0
	\end{array}
	\right.
\end{equation}

In Equation \ref{eq:21} and \ref{eq:22}, $A_S$ and $f_{\tilde{\sigma}}(z)$ are defined as for Equation \ref{eq:4}. Note that Equation \ref{eq:22} differs from the equivalent equation in \citetalias{hales2014a} (Equation 44). This modification was necessary to account for the variable sensitivity across the FOV, ensuring that for a given flux density bin we do not integrate over regions where the noise level would prevent sources being detected.

Eddington bias corrections were derived only for flux densities below a few mJy, as these flux densities begin to approach the noise level in our survey. This correction relies on the assumption that the sources are unresolved. For the flux density range where Eddington bias is likely to be most significant, this is an appropriate assumption -- see \S\ref{sec:size}.

Figure \ref{fig:edd_bias} presents the effect of Eddington bias on the assumed underlying source counts, as well as the absolute value of the Eddington bias. In Figure \ref{fig:edd_bias}, the cyan (black) curves denote the polynomial fit described by Equation \ref{eq:poly} (\citetalias{massardi2010} model). From Figure \ref{fig:edd_bias}, Eddington bias exceeds the five per cent level for flux densities below around 1 mJy. However, from Figure \ref{fig:edd_bias} there is slight discrepancy in the predicted effect of Eddington bias: for all flux density bins below 1 mJy, the \citetalias{massardi2010} model predicts an excess Eddington bias at around the two per cent level. 

The difference between the two curves in the right hand panel of Figure \ref{fig:edd_bias} arises naturally as a result of the gradient of the source counts distribution. The \citetalias{massardi2010} models suggest that the Euclidean-normalised source counts are approximately flat from 1 mJy to below the flux density limit of this survey; as such the non-Euclidean differential differential source counts continue to rise, so the likelihood of sources being upscattered by noise is greater, and Eddington bias rises. Equation \ref{eq:poly} predicts a slightly less significant flattening in the Euclidean-normalised counts, meaning the gradient of the non-Euclidean source counts is shallower, and the effect of Eddington bias is reduced. Eddington bias corrections were subsequently applied per-bin when the effect was at the two per cent level or greater.

\subsection{A Note Regarding Model Selection}
\begin{figure*}
	\centering
	\subfloat{
		\includegraphics[width=0.49\textwidth]{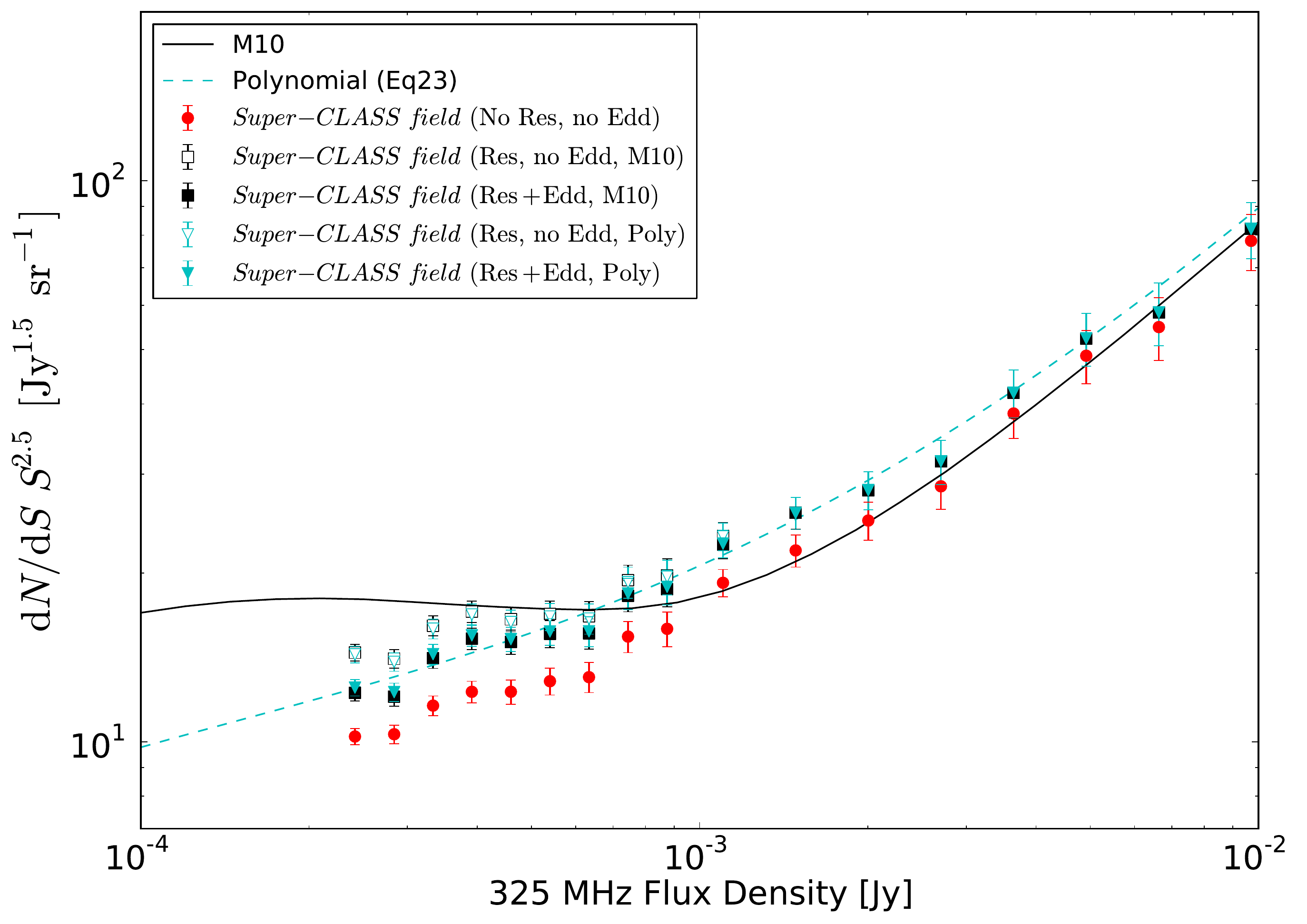}
	} 
	\subfloat{
		\includegraphics[width=0.49\textwidth]{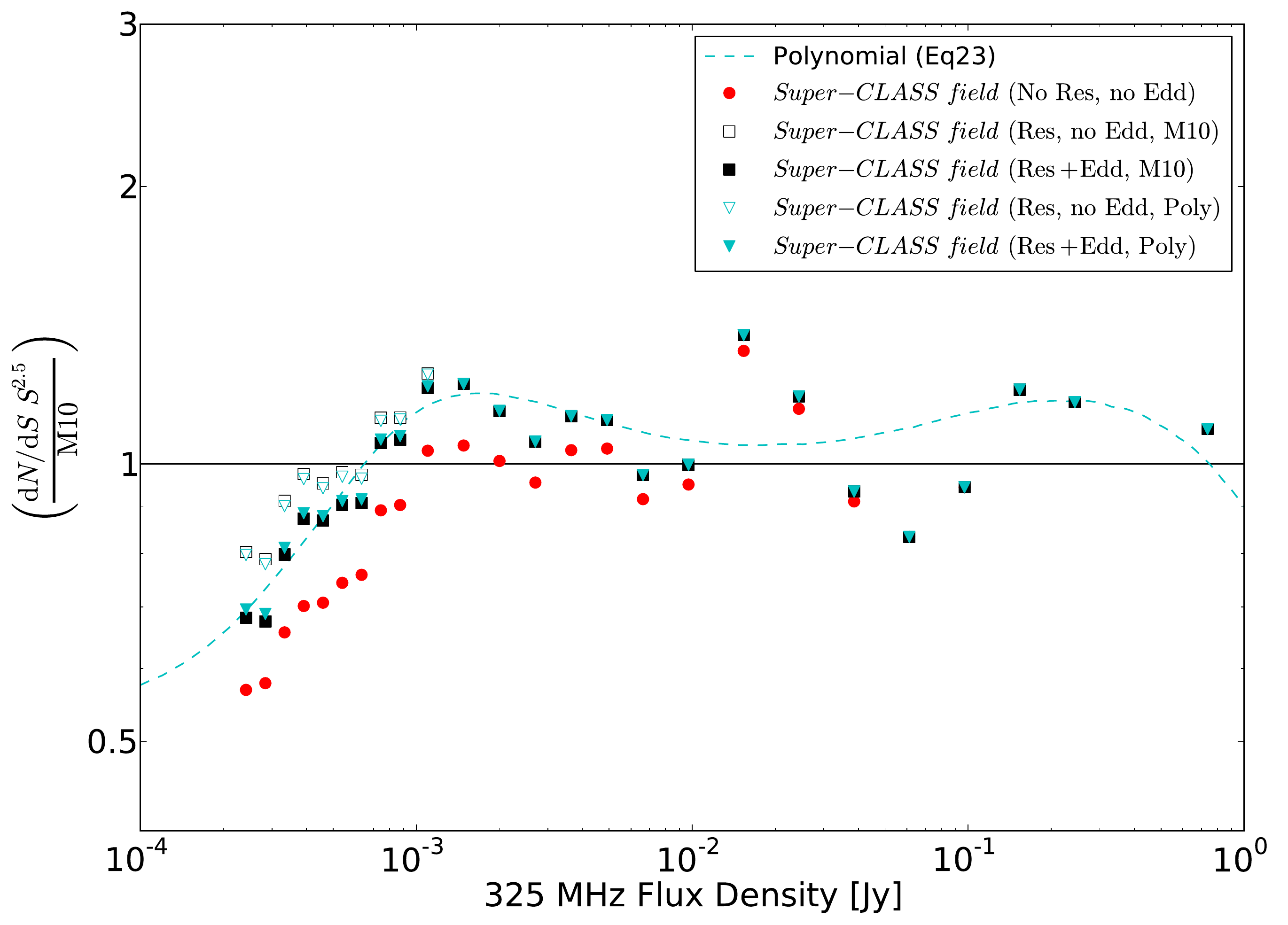}
	}
	\caption{Euclidean-normalised differential source counts at 325 MHz derived from the SCG catalogue, showing the effect of the corrections for resolution and Eddington bias below 10 mJy. Black squares (purple triangles) indicate the effect of bias corrections derived assuming the \citetalias{massardi2010} (polynomial, Equation \ref{eq:poly}) model. Filled (empty) symbols mark the differential source counts corrected for resolution and Eddington (only Eddington) bias. Filled red points mark the Euclidean-normalised differential source counts prior to resolution and Eddington bias correction. Left panel presents a zoom on the region below 10 mJy, right panel presents the residual compared to the \citetalias{massardi2010} model.}
	\label{fig:scg_cnts}
\end{figure*}

Throughout this section, we have derived the corrections for resolution and Eddington bias for two cases -- the polynomial fit to our data as well as the \citetalias{2012MNRAS.423L..30S} model. Figure \ref{fig:scg_cnts} presents the Euclidean-normalised differential source counts derived from the SCG catalogue, with corrections for resolution bias and Eddington bias, assuming both models, as well as the residual between our differential source counts and the \citetalias{massardi2010} model.

From Figure \ref{fig:scg_cnts}, it appears that the choice of model has a relatively small effect on the overall differential source counts. For the faintest flux density bin, the corrections derived from the polynomial fit model yield differential source counts that are around eight per cent higher than those derived assuming the \citetalias{massardi2010} model. This is to be expected given the profiles exhibited in the right-hand panels of Figures \ref{fig:res_bias} and \ref{fig:edd_bias}, as the resolution bias and Eddington bias corrections have the opposite effects. The difference decreases rapidly with increasing flux density, becoming negligible above around 0.5 mJy. Given that our corrected source counts exhibit a discrepancy with the \citetalias{2012MNRAS.423L..30S} model toward the faintest flux densities, we use our polynomial model described by Equation \ref{eq:poly} to derive the final corrections, which were then applied to the source counts we present here.

\subsection{Source Counts Profile}
\begin{figure*}
	\centering
	\begin{tabular}{r}
	\subfloat{
		\includegraphics[width=0.78\textwidth]{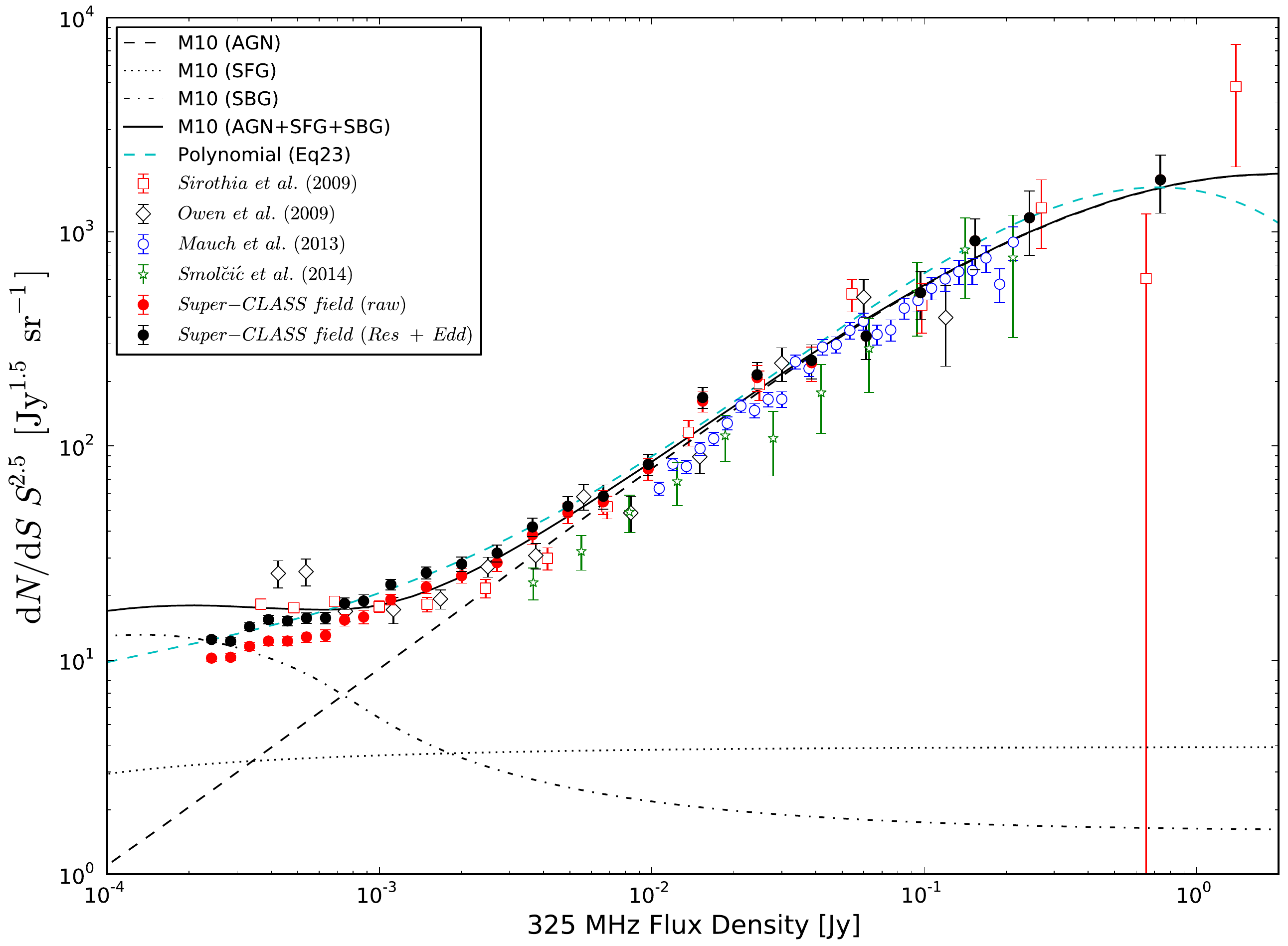}
	} \\
	\subfloat{
		\includegraphics[width=0.78\textwidth]{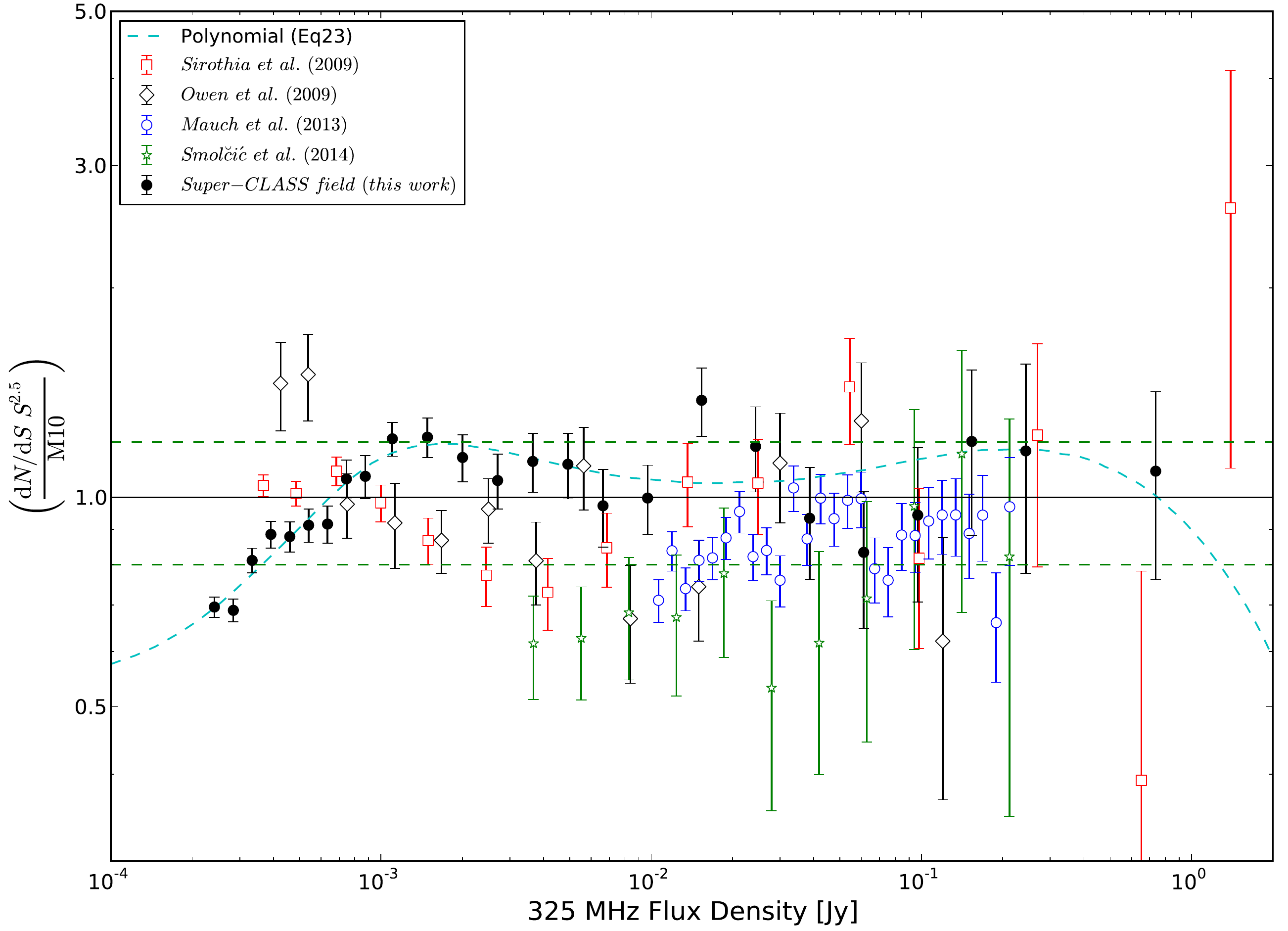}
	}
	\end{tabular}
	\caption{\emph{Top panel:} Euclidean-normalised differential source counts at 325 MHz for sources detected in this work, corrected for resolution bias and Eddington bias (filled black circles). Note that filled red circles show the source counts from this field \emph{prior} to resolution and Eddington bias correction, for comparison. We also show the Euclidean-normalised differential source counts from \protect\citet{2009MNRAS.395..269S} for the ELAIS-N1 field, those of \protect\citet{2013MNRAS.435..650M} for the \emph{Herschel}-ATLAS/GAMA field, those derived by \protect\citet{2014MNRAS.443.2590S} for the VLA-COSMOS field, and those from \protect\citet{2009AJ....137.4846O} for the SWIRE field. Dashed / dot-dashed / dotted curves denote the expected contribution at 325 MHz from AGN / starburst galaxies (SBG) / spiral galaxies (SFG) respectively, from the model of \citetalias{massardi2010}; the solid line represents the sum of the populations. The dashed cyan curve denotes the fourth-order polynomial fit to the source counts derived in this work, from Equation \ref{eq:poly}. \emph{Bottom panel:} Observed counts from the literature relative to the total \citetalias{massardi2010} model (AGN+SBG+SFG) illustrating the large scatter in source counts. Dashed cyan curve is the fourth-order polynomial fit relative to the \citetalias{massardi2010} model. Dashed green lines indicate the limit of $\pm20\%$ compared to the \citetalias{massardi2010} model.}
	 \label{fig:src_full}
\end{figure*}

We present the Euclidean-normalised differential source counts in Figure~\ref{fig:src_full}, with the numerical models from \citet{massardi2010} for AGN, SFG and SBG, shown respectively by the dashed, dotted and dot-dashed curves. The sum of contributions from AGN, SFG and SBG is denoted by the solid black curve. The polynomial described by Equation \ref{eq:poly} is denoted by the dashed cyan curve. Also shown in Figure~\ref{fig:src_full} are the Euclidean-normalised source counts derived in previous works with the GMRT (\citealt{2009MNRAS.395..269S}, \citealt{2013MNRAS.435..650M}) as well as the 324 MHz VLA-COSMOS survey \citep{2014MNRAS.443.2590S} and the SWIRE field \citep{2009AJ....137.4846O}.

From Figure~\ref{fig:src_full}, our source counts profile is consistent with models that suggest that steep-spectrum objects dominate above a few mJy. The contribution from SFG becomes increasingly important at flux densities below a few mJy. The functional form of the normalised source counts derived in this work closely follows the models derived by \citetalias{massardi2010}. This is exemplified in the lower panel of Figure~\ref{fig:src_full}, where we present the ratio of the observed source counts to the \citetalias{massardi2010} model; for all bins above 0.33 mJy, the fractional difference is typically less than 20 per cent. 

It can be seen in Figure~\ref{fig:src_full} that we detect the same flattening in the Euclidean-normalised source counts distribution that has been seen in previous surveys at higher frequencies (for example \citealt{2008MNRAS.387.1037G}, and references therein) and deep GMRT surveys at 325 MHz (\citealt{2009MNRAS.395..269S}, \citealt{2009AJ....137.4846O}, \citealt{2010iska.meetE..51S}). Additionally, the observations presented in this work are among the first low-frequency studies with sufficient sensitivity to probe the source counts distribution at flux densities where a secondary drop has been seen at higher frequency (1.4\,GHz flux density less than around $100 - 150\,\umu$Jy; \citealt{2008ApJ...681.1129B}). 

However, some works at 1.4 GHz have not seen this feature (for example \citealt{2008AJ....136.1889O}) and this has not yet been reported at 325 MHz or 610 MHz. Given a typical synchrotron spectral index of $\alpha=-0.7$, this secondary drop would appear at a 325 MHz flux density of around $280-416\,\umu$Jy, well within the range of flux densities recovered in this work (our source counts are derived down to a limiting flux density of $242\,\umu$Jy). 

From Figure~\ref{fig:src_full} there is marginal evidence of this feature at faint flux densities $(S < 308 \, \umu$Jy$)$, as our source counts profile drops significantly below the predictions of \citetalias{massardi2010}. However, given that the resolution bias correction is heavily influenced by the assumed underlying source size distribution model, we cannot discount the possibility that this may be due to a differences between the \emph{true} size distribution of low-frequency radio sources and the model we have assumed.

\section{Conclusions}\label{sec:CONC}
We report deep 325 MHz GMRT observations of the Super-CLASS field, a region of sky known to contain 5 Abell clusters. We achieve a nominal sensitivity of $34\,\umu$Jy beam$^{-1}$ toward the centre of the field, the deepest study conducted at this frequency to-date. From our mosaicked image, which covers approximately 6.5 square degrees, we recover a catalogue of 3257 sources down to a flux density limit of $183 \, \umu$Jy. 

We have compared with available catalogues from other surveys, and identified 335 sources in common with the NVSS, after accounting for sources which become resolved in the higher resolution GMRT data. The spectral index distribution indicates two populations within the data: a more numerous population of sources centred around a spectral index of $\alpha = -0.81$, which dominate at higher flux densities, and a less numerous, fainter population of sources with flat- and rising spectra, which appear to dominate the fainter flux densities. However, this spectral index sample is currently severely limited by the sensitivity of the high-frequency reference.

Adopting the definition of an ultra-steep spectrum object as a radio source with $\alpha < -1.3$ (as is typically the case in the literature) we find a total of four ultra-steep spectrum radio sources which have counterparts in the NVSS. Of these, we identify two AGN-type sources whose steep spectra may be the result of our superior sensitivity to faint diffuse emission, and two candidate HzRGs.

Additionally, we have identified three GPS sources, and a further candidate GPS source which is not identified elsewhere in the literature aside from the NVSS. One of these GPS sources is the blazar candidate CGRaBS J1015+6728; we recover an integrated flux density of $3.09\pm0.31$ mJy, to our knowledge the first flux density measurement for this object below 1 GHz. We model the spectra of these GPS sources using a broken power-law; our fits suggest break frequencies in the $2-3$\,GHz range.

Finally, we derive the Euclidean-normalised differential source counts for sources with flux densities in excess of $223\,\umu$Jy. We have performed a rigorous mathematical treatment of the various biases in the differential source counts, including Eddington bias and resolution bias. The differential source counts appear to be consistent with predictions from numerical models which account for steep-spectrum sources (AGN, FSRQ and BL Lac objects) and star-forming galaxies (spiral galaxies and starburst galaxies). We have also provided a new empirical model for describing the Euclidean-normalised differential source counts that is valid between $242 \, \umu$Jy and 0.74 Jy.

The source counts distribution derived in this work exhibits marginal evidence of a secondary downturn at flux densities below $308\,\umu$Jy. This corresponds well with the flux density regime where this feature has been detected at 1.4 GHz. To our knowledge, this is the first detection of this feature in the differential source counts at 325 MHz.

\subsection{Acknowledgements}
We thank our anonymous referee for their comments, which have helped improve the quality of the scientific output of this paper. We also thank the operators and engineers of the GMRT who made these observations possible. The GMRT is operated by the National Centre for Radio Astrophysics (NCRA) of the Tata Institute of Fundamental Research (TIFR), India. CJR wishes to thank H.~Intema for many helpful conversations during the data reduction process. 

This work has made use of the NASA/IPAC Extragalactic Database (NED), operated by JPL under contract with NASA, as well as NASA's Astrophysics Data System (ADS) and the Cosmological Calculator developed by \citet{2006PASP..118.1711W}. This work has also used data from the Owens Valley Radio Observatory (OVRO) blazar monitoring database, available at \url{http://www.astro.caltech.edu/ovroblazars/}. 

CJR gratefully acknowledges funding support from the United Kingdom Science \& Technology Facilities Council (STFC). AMS gratefully acknowledges support from the European Research Council under grant ERC-2012-StG-307215 LODESTONE. IH, MLB and CD are grateful to the European Research Council for support through EC FP7 grant number 280127. MLB also thanks the STFC for the award of Advanced and Halliday fellowships (grant number ST/I005129/1).

\bibliographystyle{mnras}
\bibliography{SuperCLASS_GMRT_survey}

\end{document}